\documentclass[twocolumn,amsmath,showkeys,amssymb,superscriptaddress,longbibliography,floatfix, pre]{revtex4-2}

\pdfoutput=1

\usepackage{bm}
\usepackage{graphicx,epsfig}
\usepackage{amsmath,amssymb,mathtools}
\usepackage{url}
\usepackage[rgb]{xcolor}
\usepackage{hyperref}
\def\be{\begin{equation}}
\def\ee{\end{equation}}

\def\bc{\begin{center}}
\def\ec{\end{center}}
\def\bea{\begin{eqnarray}}
\def\eea{\end{eqnarray}}
\newcommand{\avg}[1]{\langle{#1}\rangle}

\newcommand{\hanlin}[1]{{\bf\color{black}#1}}

\begin{document}

\title{Higher-order triadic percolation on random hypergraphs}

\author{Hanlin Sun}
\email{hanlin.sun@su.se}
\affiliation{Nordita, KTH Royal Institute of Technology and Stockholm University, Hannes Alfvéns väg 12, SE-106 91 Stockholm, Sweden}

\author{Ginestra Bianconi}
\email{ginestra.bianconi@gmail.com}
\affiliation{School of Mathematical Sciences, Queen Mary University of London, London, E1 4NS, United Kingdom}
\affiliation{The  Alan  Turing  Institute,  96  Euston  Road,  London,  NW1  2DB,  United  Kingdom}

\begin{abstract}
In this work, we propose a comprehensive theoretical framework combining percolation theory with nonlinear dynamics to study hypergraphs with a time-varying giant component. We consider in particular hypergraphs with higher-order triadic interactions.  Higher-order triadic interactions occur when one or more nodes upregulate or downregulate a hyperedge. For instance, enzymes regulate chemical reactions involving multiple reactants. Here we propose and investigate higher-order triadic percolation on hypergraphs showing that the giant component can have a non-trivial dynamics. Specifically, we show that the fraction of nodes in the giant component undergoes a route to chaos in the universality class of the logistic map. In hierarchical higher-order triadic percolation, we extend this paradigm in order to treat hierarchically nested higher-order triadic interactions. We demonstrate the non-trivial effects of their increased combinatorial complexity on the critical phenomena and the dynamical properties of the process. Finally, we consider other generalizations of the model studying the effect of {\color{black} adopting} interdependencies and node regulation instead of hyperedge regulation.
The comprehensive theoretical framework presented here sheds light on possible scenarios for climate networks, biological networks, and brain networks, where the hypergraph connectivity changes over time.
\end{abstract}
\maketitle

\section{Introduction}

Higher-order networks \cite{battiston2021physics, bianconi2021higher,boccaletti2023structure,battiston2020networks,salnikov2018simplicial,torres2021and} capture the many-body interactions among two or more nodes present in a large variety of complex systems, ranging from the brain to biochemical reaction networks.
Higher-order networks display a very rich interplay between topology and dynamics \cite{battiston2021physics,bianconi2021higher,majhi2022dynamics} which is leading to significant progress in network theory.   
Notably, higher-order interactions affect synchronization \cite{millan2020explosive,ghorbanchian2021higher,skardal2019abrupt,zhang2021unified,gambuzza2016inhomogeneity}, percolation \cite{sun2023dynamic,sun2021higher, liu2023higher, kim2024higher,di2024percolation,bianconi2018topological, bianconi2024theory,pan2024robustness, chen2024cascading}, epidemic spreading \cite{landry2020effect, st2021universal, iacopini2019simplicial},  and random walks and diffusion \cite{carletti2020random,torres2020simplicial}.

Triadic interactions are a general type of higher-order interaction in which one or more nodes regulate the activity of a link between two other nodes. This type of interaction is widely observed in nature such as brain networks, climate networks, ecological networks, and biochemical reaction networks \cite{cho2016optogenetic, grilli2017higher,bairey2016high, chen2023teasing, boers2019complex}. In the brain, the glia inhibit or facilitate the synaptic interactions between pairs of neurons. 
{\color{black} In climate networks of extreme rainfall events, triadic interactions can be used to explain the situations in which the long-range synchronized rainfall events, represented by links of the network, are modulated by large-scale patterns, such as Rossby waves \cite{boers2019complex}.
In biochemical reaction networks, chemical reactions are controlled by enzymes, whose activity can be regulated by other reactions. }
Recently, there has been a growing interest in triadic interactions from the theoretical side. Importantly, triadic interactions have been recently shown to give rise to a paradigmatic change in the theory of percolation, according to the recently proposed triadic percolation \cite{sun2023dynamic,millan2023triadic}  model. Furthermore, it has been shown that triadic interactions affect learning \cite{kozachkov2023neuron,herron2023robust}, signal processing, and network dynamics~\cite{nicoletti2024information,baptista2024mining}.

Triadic interactions can be generalized to hypergraphs, where one or more nodes can regulate the presence of a hyperedge or its strength. For instance,  in chemical reaction networks, enzymes can regulate chemical reactions which can be represented as hyperedges \cite{chen2023teasing,jost2019hypergraph} including the set of their reactants. In this work, we propose higher-order triadic percolation (HOTP) and its variations showing the important effects arising when triadic percolation is formulated on hypergraphs.

Percolation \cite{dorogovtsev2008critical,bianconi2018multilayer,lee2018recent,li2021percolation} is one of the most fundamental critical phenomena defined on networks and has been extensively used to characterize the robustness of networks~\cite{artime2024robustness}.  Percolation on simple networks is characterized by a second-order phase transition at which the emergence of the giant component can be observed when nodes or links are randomly damaged. In the last decades important progress in network theory has demonstrated that more general percolation problems, including k-core percolation \cite{dorogovtsev2006k}, and interdependent percolation on multiplex networks \cite{buldyrev2010catastrophic, baxter2012avalanche,shekhtman2018percolation,del2018finding,son2012percolation} can display discontinuous hybrid transitions, higher-order critical points \cite{cellai2013percolation} and exotic phase diagrams \cite{baxter2016correlated}. In all these problems the size of the giant component is monitored after a (random) damage is inflicted to the nodes or to the links of the network, and after the eventual cascade of failure events triggered by the initial perturbation reaches a steady state.

Triadic percolation \cite{sun2023dynamic,millan2023triadic}, however, demonstrates that percolation can become a fully-fledged dynamical process in which the giant component never reaches a steady state.  If the triadic interactions are defined on top of a random network~\cite{sun2023dynamic},  the giant component becomes in general time-dependent, and its relative size, which defines the order parameter of the model, can undergo a route to chaos in the logistic map universality class \cite{feigenbaum1978quantitative,strogatz2018nonlinear}. Moreover, triadic percolation on spatial networks leads to topological changes in the structure of the giant component generating non-trivial spatio-temporal patterns, as shown in Ref. \cite{millan2023triadic}.

In this work, we investigate higher-order triadic percolation (HOTP) on hypergraphs. We reveal that on hypergraphs, the HOTP is a fully-fledged dynamical process whose order parameter displays period-doubling and a route to chaos when both positive and negative regulations are present with non-zero probability.  Thus the phase diagram of HOTP reduces in this scenario to an orbit diagram that we hereby prove to be in the universality class of the logistic map. 

The critical properties of  HOTP are highly non-trivial. Here we characterize them in important limiting cases in which there are exclusively positive or exclusively negative regulatory interactions, characterizing their discontinuous hybrid transitions, period-2 bifurcations, and continuous transitions. {\color{black} The critical conditions are theoretically predicted and numerically validated on random hypergraphs with triadic interactions.}
When both positive and negative regulations are present,  we indicate the conditions for observing unusual orbit diagrams with a reentrant collapsed phase. In this case, we observe a phase transition between a collapsed state with a null giant component and an active phase with a non-zero and time-varying giant component while the probability of down-regulating the hyperedges increases.  {\color{black} By numerically testing  HOTP  on real-world hypergraph data we observe that this dynamical behavior of HOTP remains qualitatively unchanged when HOTP  occurs on non-random hypergraph architectures.} 

Triadic interactions can also be hierarchically nested as recognized in the context of ecological networks in Ref. \cite{bairey2016high}. In order to study the effects of hierarchical regulation, we propose and study hierarchical higher-order triadic percolation (HHOTP). In this scenario, the regulatory interaction between a node and a hyperedge is controlled by other regulatory interactions and so on. We reveal that the dynamics of HHOTP is significantly different from the dynamics of HOTP due to the combinatorial complexity of their HHOTIs. Indeed as long as the hypergraphs contain negative HHOTIs with a non-vanishing probability, HHOTP displays a much richer dynamics than HOTP  which can lead to period-doubling and a route to chaos also if no positive HHOTIs are present. Moreover, our theoretical derivation reveals that the route to chaos of HHOTP is no longer in the logistic map universality class.

{\color{black} Note that here, due to space limitation, we focus exclusively on the scenario in which nodes regulate hyperedges. However, in the future, it would be also interesting to consider HOTP in which hyperedges regulate nodes. This scenario is in line with recent results obtained in Ref. \cite{chen2024cascading} where the authors consider only positive regulations}.

This work includes an in-depth discussion of two other interesting variations of HOTP: the interdependent higher-order triadic percolation (IHOTP), and the higher-order node dynamical percolation model (HONDP). In IHOTP,  we adopt the notion of interdependent hypergraph giant component introduced in Ref. \cite{sun2021higher} which in the absence of regulation is already known to display discontinuous hybrid transition. This different nature of the underlying percolation process has dramatic effects on the dynamics of IHOTP.
In particular, here we indicate the conditions under which the route-to-chaos in IHOTP is impeded.
The study of HONDP allows us to discuss in detail the effect of regulating the nodes instead of the hyperedges revealing important differences with HOTP.

The paper is structured as follows. In Section II, we introduce random hypergraphs with higher-order triadic interactions and with hierarchical triadic interactions; in Sections III and IV we define HOTP and HHOTP and we investigate their critical properties; in Section V, we discuss other generalizations of HOTP, including IHOTP and HONDP; finally, in Section VI we provide the concluding remarks.
The paper also includes two Appendices. In the Appendix $\ref{ApA}$  we prove that the order parameter of HOTP undergoes a route-to-chaos in the universality class of the logistic map. In Appendix $\ref{ApB}$ we discuss HOTP and HONDP in the presence of partial regulation, i.e. where hyperedges (for HOTP) or nodes (for HONDP) are regulated with probability $1-\rho_0<1$.
\section{Random hypergraphs with higher-order triadic interactions}
We consider random hypergraphs with higher-order triadic interactions. These higher-order networks can be modeled as a multilayer structure \cite{bianconi2018multilayer} formed by a structural hypergraph and a signed bipartite network encoding the triadic interactions.

\subsection{Structural hypergraphs}
\label{sec:structural_hypergraphs}
We consider a structural hypergraph $\mathcal{H}(V,E_H)$  formed by a set of nodes $V$ and a set of hyperedges $E_H$. Each hyperedge $\alpha$ of cardinality $m_{\alpha}=m$ is characterized by the set of nodes $v$ it contains, i.e.,
\bea
\alpha = \left[v_1, v_2, \cdots, v_m\right].
\eea
 The hyperdegree $k$ of a node is defined as the number of its incident hyperedges. 

\begin{figure*}[!htb]
  \includegraphics[width=0.9\textwidth]{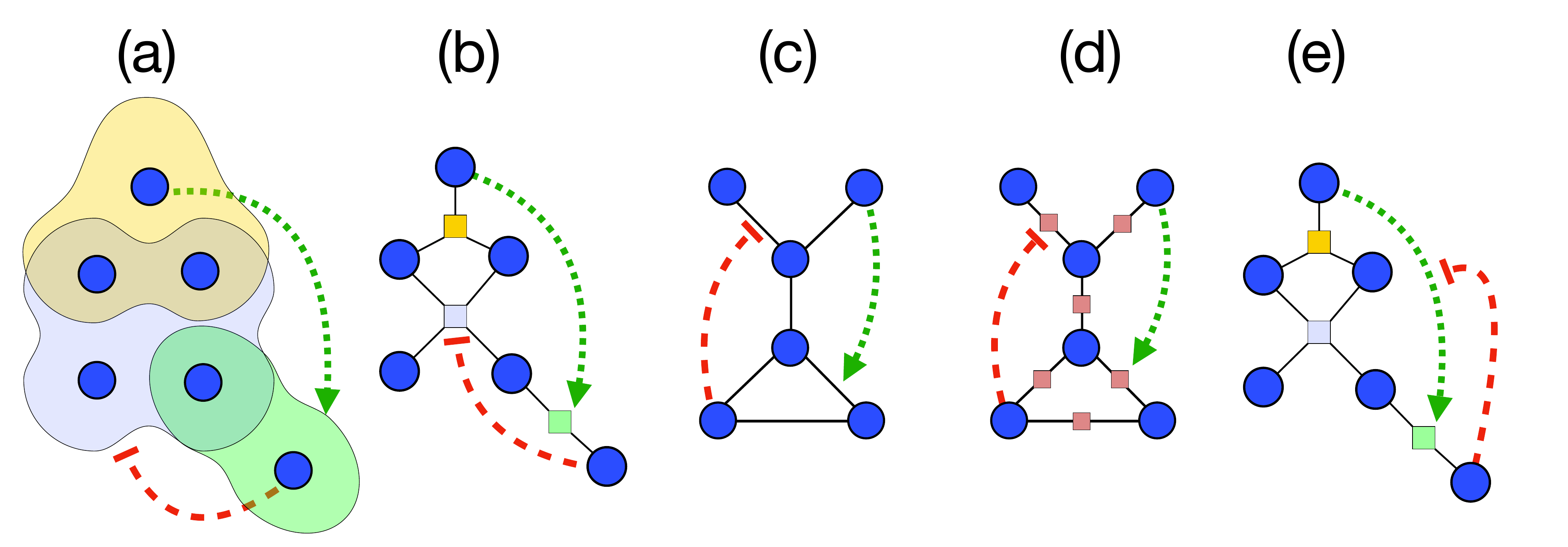}
  \caption{Schematic diagrams of hypergraphs with higher-order triadic interactions (HOTI) and hierarchical higher-order triadic interactions(HHOTI). In panel (a), the shaded areas represent hyperedges with cardinality 3 (yellow), 4 (purple) and 2 (green). The green arrow denotes positive regulation and the red arrow denotes negative regulation between nodes and hyperedges. Panel (b) shows the corresponding factor graph representation of the hypergraph in (a). Panel (c) shows an example of a network with triadic interactions and panel (d) shows the corresponding factor graph representation. Panel (e) shows an example of hierarchical triadic interactions. The positive regulatory interaction between a node and a hyperedge (the green arrow) is further regulated by a fourth node (the red arrow).}
  \label{fig:figure1}
 \end{figure*}

 Hypergraphs can always be represented as factor graphs, i.e. 
 bipartite networks $G=(V,U,E)$ formed by a set of nodes $V$, a set of factor nodes $U$, and a set of links $E_F$ between nodes and factor nodes.  The factor graph $G=(V,U,E)$ corresponding to the hypergraph $H(V,E_H)$ can be constructed by mapping each hyperedge $\alpha \in E_H$ uniquely to factor nodes $u_\alpha \in U$ and connecting node $v \in V$ to the factor node $u_\alpha \in U$ if the node $v$ belongs to hyperedge $\alpha$ in the hypergraph $H$. Thus, the hyperdegree of a node in the hypergraph is indicated by the degree of its corresponding node in the factor graph and the cardinality of a hyperedge is indicated by the degree of its corresponding factor node (see Fig. \ref{fig:figure1}).
 
In this work, we consider exclusively random hypergraphs with hyperdegree distribution $P(k)$  and hyperedge cardinality distribution $Q(m)$.
We denote with $\avg{k}=\sum_kkP(k)$ the average hyperdegree of the nodes of the hypergraph and with $\avg{m}=\sum_mmQ(m)$ the average cardinality of the hyperedges of the hypergraph.  We furthermore assume that their factor graph representation is locally tree-like (see for a detailed discussion of this assumption Ref. \cite{sun2021higher}). This assumption is generally met by the considered random hypergraphs under very general conditions over the distributions $P(k)$ and $Q(m)$, and applies to the case in which $P(k)$ and $Q(m)$ have finite moments as for Poisson distributions.
In  the following we make use of the  hypergraph generating functions  $G_0(x),G_1(x)$ and $G_{1,m}(x)$ defined as
\bea
G_1(x)&=&\sum_k \frac{k P(k)}{\avg{k}} x^{k-1},\nonumber \\ 
G_0(x)&=&\sum_k P(k) x^{k}, \nonumber\\
G_{1,m}(x)&=&\sum_m \frac{m Q(m)} {\avg{m}} x^{m-1}.
\eea 
The considered random hypergraphs reduce to networks when all the hyperedges have cardinality $m=2$, (i.e. edges). In terms of the cardinality distribution $Q(m)$,  this happens when $Q(m)=\delta_{m,2}$ where here and in the following $\delta_{x,y}$ indicates the Kronecker delta.
In this work, we provide general results for arbitrary hypergraphs but if not otherwise explicitly stated, we will perform the numerical analysis and the Monte Carlo simulations on hypergraphs having hyperedges with constant cardinality $m$ and  Poisson hyperdegree distribution $P(k)$ with average hyperdegree $c$.
\subsection{Higher-order triadic interactions}

\subsubsection{Higher-order triadic interactions}
\label{sec:higher_order_triadic_interactions}
Higher-order triadic interactions (HOTIs) occur when one or more nodes regulate a hyperedge. The HOTIs are signed, i.e. they can be associated with a positive or negative regulatory role. For instance, an enzyme can speed up or inhibit a chemical reaction.  Here we consider the simple scenario in which the HOTIs occur between the same type of nodes  (i.e. we do not distinguish between "enzyme" and "metabolite" in the previous example), generalizations in this direction will be considered in future works.  
In this framework, triadic interactions are encoded by a regulatory bipartite network $\mathcal{G}_T(V,E_H, W)$ formed by the set of nodes $V$ and the set of factor nodes $E_H$ where each factor node represents a hyperedge of the structural hypergraph. The signed HOTIs are represented by the edges $W$ of the bipartite network $\mathcal{G}_T$. Specifically, each edge in $W$  indicates the existence of a (triadic) regulatory interaction from a node in $V$  to a hyperedge in $E_H$ (see Fig. \ref{fig:figure1} (a) and (b)). We call a node a positive (or negative) regulator of a hyperedge if the node regulates the hyperedge positively (or negatively). Note that the sign of regulation is a property of the regulatory interaction but not of a node. 

We observe that HOTIs reduce to the triadic interactions investigated in Ref. \cite{sun2023dynamic} when the structural hypergraph reduces to a network, i.e. has all hyperedges of cardinality $m=2$ (see Fig. \ref{fig:figure1} (c) and (d)).

In this work, we consider an ensemble of hypergraphs with random HOTIs. While the structural hypergraph is sampled from the ensemble of random hypergraphs defined in the previous paragraph, the HOTIs are described by a random bipartite regulatory network. 
Specifically, we assume that the regulatory interactions between nodes and hyperedges are random and each hyperedge is regulated by $\kappa^+$ positive regulators and $\kappa^-$ negative regulators. The regulatory degree distributions are given by $\hat{P}(\kappa^+)$ and $\hat{P}(\kappa^-)$, respectively. We also assume that there are no correlations among the regulatory degree $\kappa^\pm_{\alpha}$ of a hyperedge $\alpha$, its cardinality  $m_{\alpha}$ and the hyperdegree $k_i$ of its regulatory nodes.
Finally, we assume that positive and negative regulatory degrees are independent, which is a valid assumption since the regulatory networks we consider are sparse. 

As a reference for the future derivations, we observe that we indicate with $G_0^{\pm}(x)$ the  generating function of the degree distribution $\hat{P}(\kappa^\pm)$ of the positive and negative regulatory interactions of a random hyperedge, given by 
\bea
G_0^{\pm}(x) = \sum_{\kappa^\pm} \hat{P}(\kappa^\pm) x^{\kappa^\pm}.
\label{G0pm}
\eea
In this work we will discuss the general theory for arbitrary distribution $\hat{P}(\kappa^\pm)$, however, if not explicitly stated,  our numerical and Monte Carlo results will be conducted always for Poisson $\hat{P}(\kappa^\pm)$ distribution with average degree $c^{\pm}$.
\subsubsection{Hierarchical higher-order triadic interactions}
\label{Sec:HHOTI}
Triadic interactions can be nested hierarchically, leading to significant effects as it has been recognized in the context of ecological networks \cite{bairey2016high}.

Hierarchical higher-order triadic interactions (HHOTI) can be described by a multilayer regulatory network formed by $L$ layers, where each layer $\mu\in \{1,2,\ldots L\}$ is a bipartite network $\mathcal{G}_{T}^{[\mu]}$.
The first layer is formed by nodes in $V$ connected to factor nodes in $E_H$, i.e. $\mathcal{G}_{T}^{[1]}=(V,E_H,W^{[1]})$. Thus the first layer reduces to the fundamental HOTI defined in the previous paragraph. The second layer indicates the set of regulatory interactions $W^{[2]}$ of the regulatory interactions in $\mathcal{G}_T^{[1]}$, the third layer indicates the regulatory interactions $W^{3}$ of the regulatory interactions in layer 2 and so on (see Fig. \ref{fig:figure1} (e)).
Thus we can describe these further layers with $\mu>1$ as  bipartite networks $\mathcal{G}_T^{[\mu]}=(V,W^{[\mu-1]},W^{[\mu]})$ fully determining the regulatory interactions $W^{[\mu]}$ between the nodes in $V$ and the factor nodes representing the regulatory interactions $W^{[\mu-1]}$ at the previous layer. 

The bipartite networks $\mathcal{G}_T^{[\mu]}$ that will be considered will be random bipartite networks. Specifically, we assume that the regulatory interactions between nodes and factor nodes in $W^{[\mu-1]}$ are random and that each factor node in $W^{[\mu-1]}$ has degree $\kappa^{\pm}$ indicating the number of positive and negative regulatory interactions, respectively each drawn from the corresponding distribution $\hat{P}_{\mu}(\kappa^\pm)$.

As a reference for the future derivations, we observe that we indicate with $G_{0,[\mu]}^{\pm}(x)$ the  generating function of the degree distribution $\hat{P}_{[\mu]}(\kappa^\pm)$ of the positive and negative regulatory interactions of the factor nodes in $W^{[\mu]}$ 
\bea
G_{0, [\mu]}^{\pm}(x) = \sum_{\kappa^\pm} \hat{P}_{[\mu]}(\kappa^\pm) x^{\kappa^\pm}.
\label{G0mupm}
\eea
In this work, we will focus mostly on numerical results obtained for distributions $\hat{P}_{[\mu]}(\kappa^\pm)$ independent of the layer $L$, i.e. $\hat{P}_{[\mu]}(\kappa^\pm)=\hat{P}(\kappa^\pm)$ with $\hat{P}(\kappa^\pm)$ being a Poisson distribution with average $c^{\pm}$.
 
\section{Higher-order triadic percolation on random hypergraphs}

\subsection{Higher-order triadic percolation }

Here we formulate higher-order triadic percolation (HOTP) that generalizes triadic percolation, proposed for networks in Ref. \cite{sun2023dynamic}, to hypergraphs.
In this framework,  nodes are active if they belong to the hypergraph giant component, while hyperedges are upregulated or downregulated according to the activity of their regulator nodes and stochastic noise. 
The hypergraph giant component considered here is the largest extensive connected component whose nodes and hyperedges satisfy the following self-consistent and recursive conditions~\cite{sun2021higher}.
\begin{itemize}
\item
A node is in the giant component if it belongs to at least one hyperedge that is in the hypergraph giant component. 
\item A hyperedge is in the giant component if 
\begin{itemize}\item[(i)] it is not down-regulated, \item[(ii)] it includes at least one node that is in the hypergraph giant component.
\end{itemize}
\end{itemize}

Higher-order triadic percolation is a dynamic process determined by a simple 2-step iterative algorithm. At time $t=0$, every hyperedge is active with probability $p_H^0$. For $t \geq 1$:
\begin{itemize}
    \item Step 1: Given the set of active hyperedges at time $t-1$, a node is considered active at time $t$ if it belongs to a least a hyperedge in the hypergraph giant component.
    \item Step 2: A hyperedge is deactivated if it is regulated by at least one active negative regulator and/or is not regulated by any active positive regulator and it is considered active otherwise. All other hyperedges are deactivated with probability $1-p$.
\end{itemize}

Given a random hypergraph {\color{black} with hyperdegree distribution $P(k)$ and hyperedge cardinality distribution $Q(m)$ defined in Sec. \ref{sec:structural_hypergraphs}}, at each time $t$, Step 1 implements hyperedge percolation~\cite{sun2021higher} where the probability that a hyperedge is intact is given by $p_H^{(t-1)}$.
This process determines the fraction of nodes in the hypergraph giant component $R^{(t)}$ at time $t$, given by 
\bea
R^{(t)} = 1-G_0(1-\hat{S}^{(t)}). \label{eq:hypergraph_triadic_R}
\eea
where $\hat{S}^{(t)}$ is the probability that starting from a random node and choosing one of its hyperedges at random we reach a hyperedge in the hypergraph giant component. 
The probability $\hat{S}^{(t)}$ together with  the probability $S^{(t)}$ that starting from a random hyperedge and choosing one of its nodes at random we reach a node in the hypergraph giant component, obey the following self-consistent set of equations
\bea
\hat{S}^{(t)}&=& p_H^{(t-1)}\left[1-G_{1,m}(1-S^{(t)})\right] \label{eq:hypergraph_triadic_S_hat}, \\
S^{(t)} &=& 1-G_1(1-\hat{S}^{(t)}).\label{eq:hypergraph_triadic_S}
\eea
Thus Eqs. (\ref{eq:hypergraph_triadic_S_hat}) and (\ref{eq:hypergraph_triadic_S}) together with Eq. (\ref{eq:hypergraph_triadic_R}) determine $R^{(t)}$ starting from the knowledge of $p_H^{(t)}$ and implements Step 1.
Note that this step differs from Step 1 of standard triadic percolation~\cite{sun2023dynamic}, as here we define a node active only if it belongs to the hypergraph giant component while in the standard triadic percolation we define a node active only if it belongs to the standard giant component of the network under consideration.

Step 2 requires the formulation of an additional equation determining the probability $p_H^{(t)}$ that the hyperedges are intact given the probability $R^{(t)}$ that the nodes are active, (i.e. they belong to the hypergraph giant component at Step 1). {\color{black} On the hypergraphs with random interactions defined in Sec. \ref{sec:higher_order_triadic_interactions}, $p_H^{(t)}$ obeys the following dynamical equations:}
\bea
p_H^{(t)} &=& p G_0^{-}\left(1-R^{(t)}\right) \left(1-G_0^{+}\left(1-R^{(t)}\right)\right),
\label{eq:hypergraph_triadic_pL}
\eea
where $G_0^{\pm}(x)$ are the generating functions defined in Eq. (\ref{G0pm}). Note that Eq. (\ref{eq:hypergraph_triadic_pL}) is the same equation used in Ref. \cite{sun2023dynamic} to define Step 2 of standard triadic percolation, however here $p_H^{(t)}$ indicates the probability that a hyperedge is intact while in standard triadic percolation it indicates the probability that an edge is intact.

From this definition of higher-order triadic percolation, it is clear that this dynamical process reduces to standard triadic percolation  \cite{sun2023dynamic} if $P(m)=\delta_{m,2}$, i.e. if the hypergraph reduces to a network. 
\begin{figure*}[!htb!]
  \includegraphics[width=\textwidth]{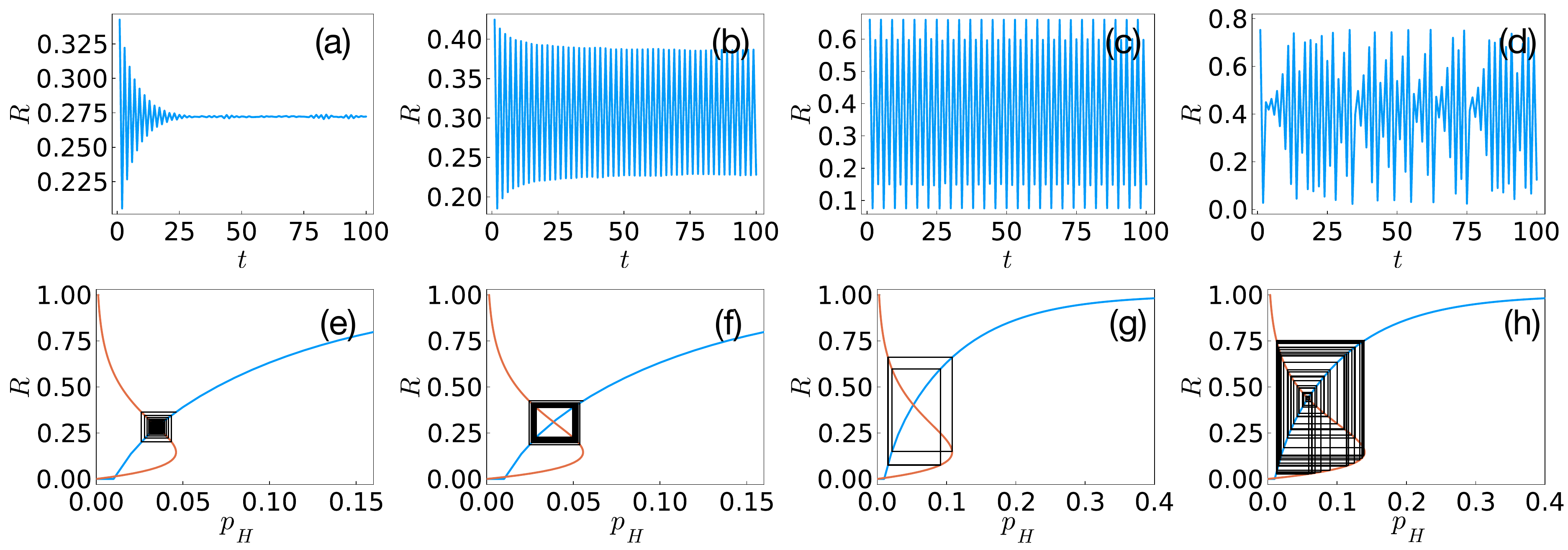}
  \caption{Time series (panels (a)-(d)) of the order parameter $R$ of HOTP and their corresponding Cobwebs (panels (e)-(h)) are shown to display a variety of dynamical behavior:  a steady state (panels (a) and (e)),  period-2 oscillations (panels (b) and (f)),  period-4 oscillations (panel (c) and (g)) and chaotic dynamics (panel (d) and (h)).
The time series are obtained by Monte Carlo simulations on hypergraphs of $N=10^6$ nodes with HOTIs. The structural hypergraph has hyperedges of fixed cardinality $m=10$ and Poisson hyperdegree distribution $P(k)$ with an average $c=10$. The degree distributions $\hat{P}(\kappa^\pm)$ are Poisson with an average regulatory positive and negative degree $c^+=3$ and $c^-=5.5$ respectively. In the Cobweb, the blue curves denote function $R=f_m(p_H)$ and the orange curves denote function $p_H=g_p(R)$ as defined in Eq. (\ref{eq:fm}) and Eq. (\ref{eq:gp}).  The time series and the Cobweb plots are obtained for   $p=0.27$ (panels (a) and (e)),  $p=0.35$ (panels (b) and (f)),  $p=0.68$  (panels (c) and (g)) and  $p=0.88$ (panels (d) and (h)).}
  \label{fig:cobweb}
 \end{figure*}

 \begin{figure*}[!htb!]
  \includegraphics[width=\textwidth]{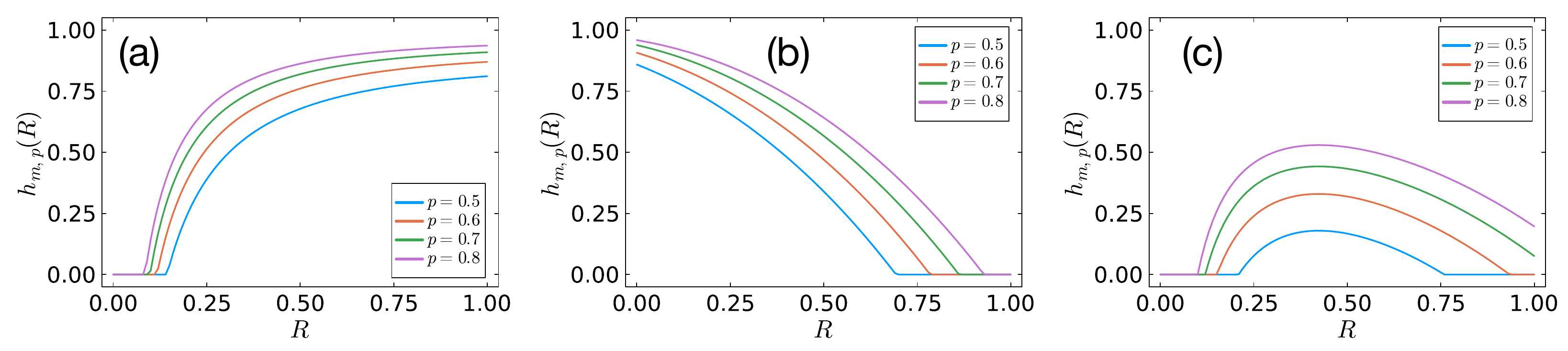}
  \caption{Examples of the map $h_{m,p}(R)$ encoding for the dynamics of HOTP when the HOTIs are exclusively positive  (panel (a)); when the HOTIs are exclusively negative (panel (b)) and when the hypergraphs include both positive and negative HOTIs with non-vanishing probability (panel (c)).  The map is unimodal only the scenario (c), thus we can observe a route to chaos of the order parameter only in the presence of both positive and negative HOTIs.   For all three panels, the structural networks have constant hyperedge cardinality $m=3$ and  Poisson hyperdegree distribution with an average $c=4$. Moreover the data ar obtained for $\hat{P}^{\pm}(\kappa)$ with average degree $c^{\pm}$ given by $c^+=2, c^{-}=0$ (panel (a));  $c^{+}=0,c^-=2$ (panel (b)).; $c^+=2,$ $c^-=1.5$ (panel (c)).}
  \label{fig:hmp}
 \end{figure*}

\subsection{Higher-order triadic percolation as a fully-fledged dynamical process }

Higher-order triadic percolation defines a fully-fledged dynamical process in which the fraction of nodes in the giant component $R^{(t)}$ is in general a non-trivial function of time.
In order to see this we can encode the dynamical Eqs. (\ref{eq:hypergraph_triadic_R})-(\ref{eq:hypergraph_triadic_S}) implementing Step 1 and determining $R^{(t)}$ once $p_H^{(t-1)}$ is known as a function
\bea
R^{(t)}=f_m\left(p_H^{(t-1)}\right).
\label{eq:fm}
\eea
Note that the function $f_m(p_H^{(t-1)})$ already encodes for the result of the percolation process, thus encodes already for the percolation phase transition of hyperedge percolation.
On the other side, Step 2 is determined by the Eq. (\ref{eq:hypergraph_triadic_pL}) that evidently can be expressed as a function determining $p_H^{(t)}$ once $R^{(t)}$ is known, i.e. 
\bea 
p_H^{(t)}=g_p\left(R^{(t)}\right),
\label{eq:gp}
\eea
The iteration of Step 1 and Step 2 occurring at time $t$ is thus enclosed in the one-dimensional map 
\bea
R^{(t)} = h_{m, p}\left(R^{(t-1)}\right) = f_m\left(g_p\left(R^{(t-1)}\right)\right).
\label{eq:one_dimensional_map}
\eea
This one-dimensional map defines the evolution of the percolation order parameter $R$ of HOTP as a function of time, i.e. $R=R^{(t)}$ for an infinite random hypergraph with HOTI.
This one-dimensional map thus encodes the dynamical nature of HOTP and can be used to predict the dynamical behavior of the order parameter.
Indeed, by iterating the map we can build a Cobweb plot that accurately predicts the dynamics of the order parameter on large hypergraphs (see Fig. (\ref{fig:cobweb})).
As we will discuss below and prove in Appendix \ref{ApA}, this map undergoes a route to chaos in the universality class of the logistic map.
In particular, the dynamics of the order parameter can either go to a fixed point $R=R^{\star}$, corresponding to a static asymptotic state of HOTP, or it can display periodic oscillations or even display a chaotic dynamics in the thermodynamic limit  (see Fig. \ref{fig:cobweb}). 

The critical behavior of the dynamics can be studied by analyzing its derivative $J$. A stable  stationary solution $R^{\star}$ of Eq. (\ref{eq:one_dimensional_map}) satisfy the fix point equation
\bea
R^{\star} = h_{m,p}(R^{\star}),
\eea
together with the stability conditions on the derivative  $J$ of the map,
\bea
\left|J\right|= \left|h_{m,p}^\prime(R^{\star})\right|<1.
\eea
Thus the stationary solution loses its stability for $|J|=1$.
When $J=1$, as long as  $R=R^{\star}>0$, we observe a discontinuous hybrid transition with a square root singularity.
When $J=-1$ we observe instead a bifurcation, resulting in period-2 oscillations. Similarly, the critical point of the onset of period-4 oscillations can be obtained by analyzing the derivative $J_2$ of the second-iterate function $h^2_{m,p}=h_{m,p}\circ h_{m,p}$. The bifurcation of period-2 oscillations takes place at $J_2=-1$, resulting in period-4 oscillations.

In the next paragraphs, we will discuss how these theoretical insights shed light on the critical properties of HOTP.

\subsection{Critical properties of higher-order triadic percolation}

In order to investigate the critical properties of HOTP we consider the effect induced by the sign of the HOTI. Thus we will discuss the scenarios in which only positive HOTIs are present,  the case in which only negative HOTIs are present, and the case in which both positive and negative HOTIs are present with non-vanishing probability.
\subsubsection{Higher-order triadic percolation in the absence of negative HOTIs}
If only positive HOTIs are present, the Eq. (\ref{eq:hypergraph_triadic_pL}) determining the regulatory step (Step 2) reduces to
\bea
p_H^{(t)} =g_p\left(R^{(t)}\right) = p \left(1-G_0^{+}\left(1-R^{(t)}\right)\right).
\label{eq:hypergraph_triadic_pL_only_positive}
\eea
In this case, we cannot have a route to chaos because the map 
$h_{m,p}(R)=f_m(g_p(R))$ is monotonically increasing with $R$ \cite{strogatz2018nonlinear} (see Fig. $\ref{fig:hmp}$ for an example).
Indeed we have 
\bea
h_{m,p_c}^\prime(R)=f_m^{\prime}(g_p(R))g_p^{\prime}(R)>0,
\eea
as it can be easily checked that
\bea
\frac{\partial f_m}{\partial p_H^{(t)}} \geq 0, \quad \frac{\partial g_p}{\partial R^{(t)}}>0.
\eea
Specifically, we have that the dynamics reaches a stationary state $R=R^{\star}$ such that 
\bea
R^{\star}=h_{m,p}(R^{\star})=f_m\left(g_p\Big(R^{\star}\Big)\right).
\label{eq:stat1}
\eea
This stationary state, $R^{\star}$ displays a discontinuous hybrid phase transition between a zero value $R^{\star}=0$ to a non-zero value $R=R^{\star}>0$ at the critical point $p=p_c$.

The phase transition takes place at $p=p_c$ when the stationary solution $R=R^{\star}=0$ loses its stability, i.e. when 
\bea
\left|J\right| = \left|h_{m,p_c}^\prime(R^{\star})\right|_{R^{\star}=0}=1.
\label{Jaus}
\eea

\begin{figure}[!htb]
  \includegraphics[width=0.5\textwidth]{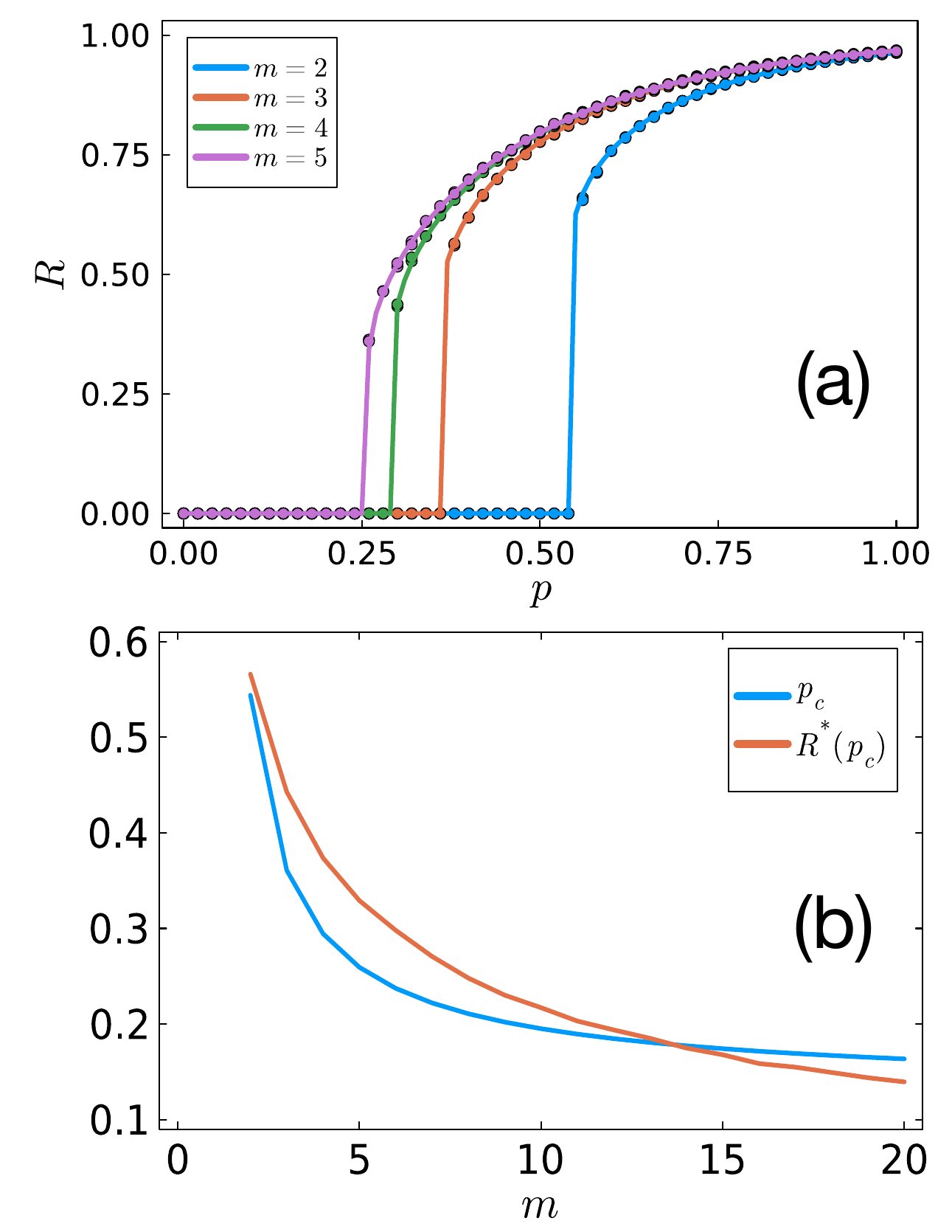}
  \caption{Phase diagram of random hyperedge percolation in the absence of negative HOTIs.  Panel (a) shows the size of the giant component $R$ as a function of $p_H$. Solid lines indicate theoretical predictions (Eq. (\ref{eq:hypergraph_triadic_S_hat}) - Eq. (\ref{eq:hypergraph_triadic_pL})) symbols indicate the results of  Monte Carlo simulations. Panel (b) shows the numerical calculation of critical threshold $p_c$ and the non-trivial fixed point $R^{\star}(p_c)$ at $p=p_c$ as a function of the hyperedge cardinality $m$. The Monte Carlo simulations are performed on hypergraphs with HOTI  having $N=10^5$ nodes. The structural hypergraph has hyperedges of fixed cardinality $m$ and has Poisson hyperdegree distribution $P(k)$ with an average hyperdegree $c=4$. The degree distribution $\hat{P}^+(\kappa)$ of is Poisson with an average positive regulatory degree $c^+=2$. }
  \label{fig:figure4}
 \end{figure}

In order to show that the transition is discontinuous and hybrid \cite{sun2023dynamic, bianconi2018multilayer}, consider expanding the stationary point equation around the critical point $p=p_c$. Let us indicate with $\delta p$ the distance from the critical point, i.e. $0<\delta p= p-p_c \ll 1$ and with $\delta R$ the corresponding {\color{black} variation of the order parameter} $0<\delta R=R^{\star}(p)-R^{\star}(p_c) \ll 1$. {\color{black} Using Eq.(\ref{eq:stat1}), and assuming that $h_{m,p}(R)$ is  differentiable up to the second order at $R^{\star}(p_c)$, we obtain
\bea
\delta R = \frac{\partial h_m(R^{\star}(p_c), p_c)}{\partial R} \delta R&+&\frac{1}{2} \frac{\partial^2 h_m(R^{\star}(p_c), p_c)}{\partial R^2} (\delta R)^2 \nonumber\\
&+&\frac{\partial h_{m}\left(R^{\star}(p_c), p_c\right)}{\partial p} \delta p.\nonumber
\eea
Since Eq. (\ref{Jaus}) holds, $\partial h_{m}\left(R^{\star}(p_c), p_c\right)/\partial R=1$. Thus inserting this in the above equation we obtain
\bea
\frac{1}{2} \frac{\partial^2 h_m(R_c, p_c)}{\partial R^2}(\delta R)^2+\frac{\partial h_{m}\left(R_c, p_c\right)}{\partial p} \delta p=0.
\eea
Given that $\partial^2 h_m(R^{\star}(p_c), p_c)/\partial R^2$ and ${\partial h_{m}\left(R^{\star}(p_c), p_c\right)}/{\partial p}$ are finite and have opposite signs, the order parameter $R$ has a square-root singularity for $p\to p_c^{+}$, i.e.
}
\bea
\delta R = R^{\star}(p)-R^{\star}(p_c) \propto (p-p_c)^{1/2}
\eea
indicating that the transition is hybrid. 
Our theoretical predictions perfectly match our simulation results (see Fig. \ref{fig:figure4}).
In particular, we have considered a random hypergraph with fixed hyperedge cardinality $m$ (i.e. with $Q(m')=\delta_{m',m}$ and Poisson distribution $P(k)$ and $\hat{P}(\kappa)$.
The hypergraph with larger hyperedge cardinality $m$ is more robust and displays simultaneously a lower critical threshold $p_c$, and a smaller discontinuity $R^{\star}(p_c)$ (see Fig. \ref{fig:figure4}(b)).
This result is in line with the result on simple hyperedge percolation \cite{sun2021higher}, where hypergraphs with larger hyperedge cardinality $m$ are more robust, albeit in that case the percolation transition is continuous.

\subsubsection{Higher-order triadic percolation in the absence of positive HOTIs}
\label{sec:hotp_no_positive}
In this paragraph, we consider the other limiting case in which HOTP  includes only negative HOTIs. In this case, a hyperedge is down-regulated if at least one of is negative regulator nodes is active. This scenario can be interpreted as HOTP as the limit in which the positive regulatory \hanlin{degrees are infinite}.  Thus in Step 2 of HOTP we substitute    Eq. (\ref{eq:hypergraph_triadic_pL})  with
\bea
p_H^{(t)} = p G_0^- \left(1-R^{(t)}\right).
\eea
Indeed, according to our interpretation, this equation can be obtained from Eq. (\ref{eq:hypergraph_triadic_pL})  by putting  $G_0^+ \left(1-R^{(t)}\right)$  identically equal to zero.

\begin{figure}[!htb]
  \includegraphics[width=0.5\textwidth]{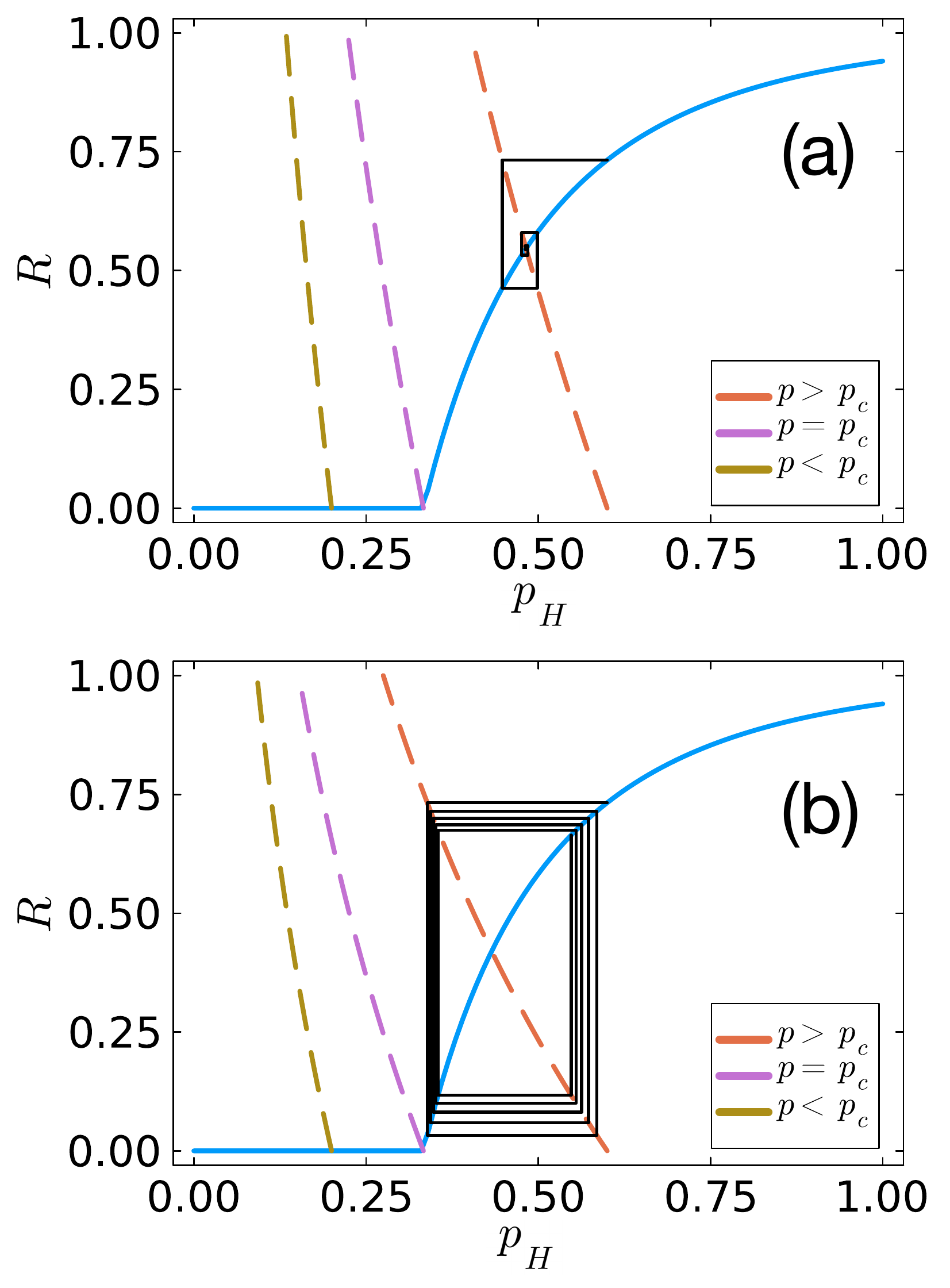}
  \caption{The function $g_p(R)$ and $f_m(p_H)$ of HOTP and their corresponding Cobweb are shown on hypergraphs including only negative HOTIs.  The function $g_p(R)$ with $p$ that is below, at, and above the critical threshold $p_c$ of continuous transition (a) and period-2 bifurcation (b) are shown. The structural hypergraphs have  Poisson hyperdegree distribution with an average $c=3$ and fixed hyperedge cardinality $m=2$. The regulatory network is formed exclusively by negative regulations with a Poisson degree distribution $\hat{P}^{-}(\kappa)$ with an average $c^-=0.4$ (panel (a)) and $c^-=0.78$ (panel (b)).  }
  \label{fig:neg_tricritical}
 \end{figure}

Also in this case, similarly to the precedent limiting case, the map $h_{m,p}(R)$ is monotonic, thus we do not observe a route to chaos (see Figure \ref{fig:hmp}). However, it can be easily shown that, differently from the precedent limiting case, in the absence of any positive interactions the map $h_{m,p}(R)$ is monotonically decreasing. In fact, if we have 
\bea
J = h_{m,p}^\prime(R) = \frac{\partial f_m}{\partial p_H^{(t)}}  \frac{\partial g_p}{\partial R^{(t)}}\leq 0
\eea
since it can be shown easily that 
\bea
\frac{\partial f_m}{\partial p_H^{(t)}} \geq 0, \quad \frac{\partial g_p}{\partial R^{(t)}}<0.
\eea
It follows that the only critical points of the dynamic equations can be a period-2 bifurcation for $J=-1$. This can occur at $R^{\star}=0$ or at $R^{\star}>0$.
Note however that we can additionally also observe continuous phase transitions from a stationary solution $R^{\star}=0$ to a stationary solution $R^{\star}>0$. This transition will not occur at a specific value of $J$, but can occur as long as $|J|<1$, the condition that ensures the stability of the stationary solution. Indeed the continuous transition occurs whereas the hyperedge percolation equation displays a phase transition, as the change of solutions between  $R^{\star}=0$ and $R^{\star}>0$ is already encoded in the function $f_m(p_H)$. 
This happens when the maximum of $p_H=g_p(R)$, which is reached at $R=0$, equals the critical threshold of hyperedge percolation $p_H^c$ provided that  $|J|<1$ (see Fig. \ref{fig:neg_tricritical}.(a)). Interestingly, in this scenario, the threshold for the continuous transitions is independent of the regulatory network and only depends on the structural hypergraph.

The period-2 oscillation emerges when the non-trivial fixed point $R^{\star}$ loses its stability. This happens when $p>p_H^c$ and $J=-1$ (see Fig. \ref{fig:neg_tricritical}(b)). Thus, the tricritical point separating the period-2 oscillation and the continuous transition is reached when
\bea
J = -1 \quad \mbox{and} \quad p=p_H^c.
\eea

Assuming the random structural hypergraph has a Poisson hyperdegree distribution with an average $c$ and a fixed hyperedge cardinality $m$, 
the critical threshold of hyperedge percolation is given by~\cite{sun2021higher}
\bea
p = p_H^c = 1/c(m-1).
\eea
Furthermore considering a  Poisson regulatory degree distribution $\hat{P}^-(\kappa)$ with an average $c^-$ we can calculate $J$ at $R=0$, $p_H=p_H^c$ obtaining
\bea
J = \left.\frac{\partial f_m}{\partial p_H^{(t)}}\right|_{p_H=p_H^c}\left.\frac{\partial g_p}{\partial R^{(t)}}\right|_{R=0} = -2cpc^-. 
\eea
Thus imposing $J=-1$ we find that the tricritical point occurs for 
\bea
c^-=\frac{m-1}{2}, \quad p = \frac{1}{c(m-1)}.
\eea
For a generalization of this approach to the more complex scenario in which hyperedges are not regulated with probability $\rho_0>0$ we refer the reader to Appendix $\ref{ApB}$.

By studying the the iteration of the map $h_{m,p}\circ h_{m,p} (R)$ it can be shown that period-4 bifurcations are not possible.
Thus the characterization of the critical properties of HOTP in the absence of positive regulation reveals that in this case only period-2 bifurcations and continuous transitions can be observed.

\begin{figure}[!htb]
  \includegraphics[width=0.33\textwidth]{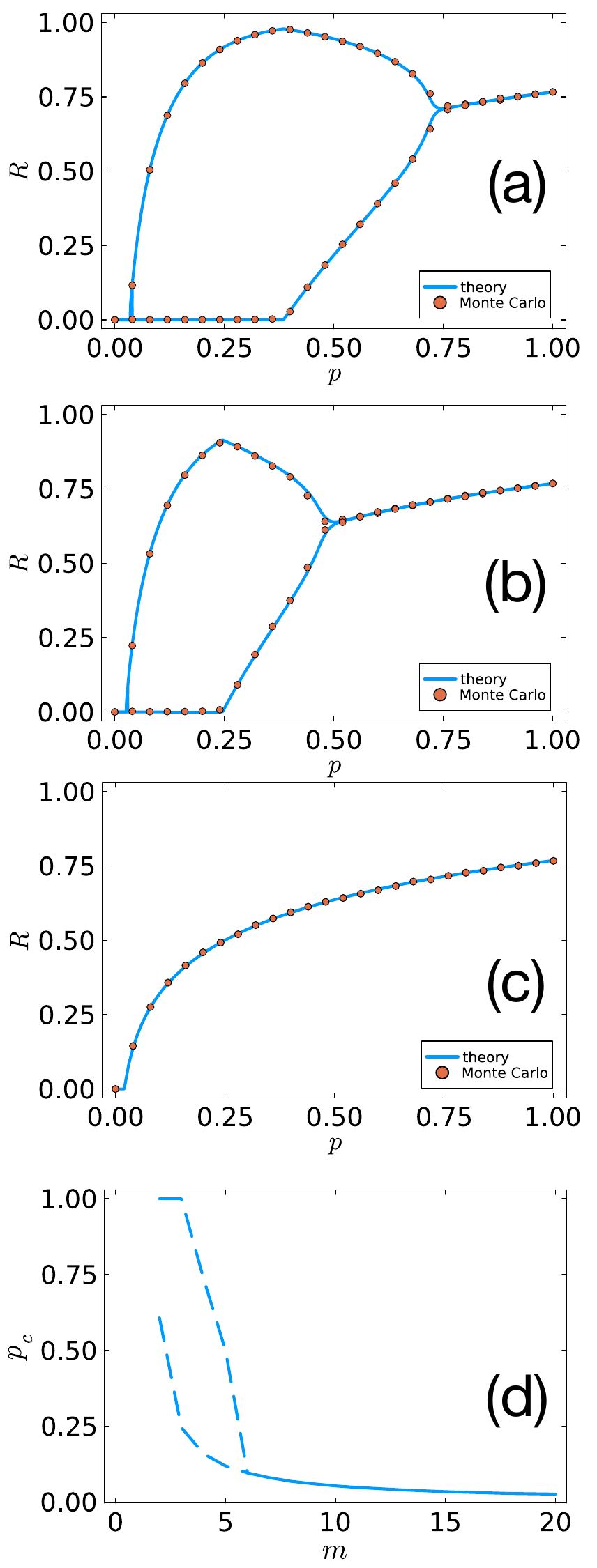}
  \caption{The theoretical phase diagrams/orbit diagrams of HOTP with exclusive negative HOTIs are compared to Monte Carlo simulations (panels (a), (b), and (c)). Panel (d) shows the critical points for period-2 bifurcation (dashed line) and continuous transition (solid line) as a function of $m$. The dashed lines join the solid line at the tricritical point. The structural hypergraph has  $N=5\times 10^5$ nodes, Poisson hyperdegree distribution $P(k)$ with an average hyperdegree $c=10$ and  hyperedges with fixed cardinality $m=4$ (panel (a)), $m=5$ (panel (b)) and $m=6$ (panel (c)). The  Poisson distribution $\hat{P}^-(\kappa)$ has average degree  $c^-=2.5$.}
  \label{fig:figure6}
 \end{figure}

In Fig \ref{fig:figure6} we compare our theoretical predictions with Monte Carlo simulations performed on hypergraphs with fixed cardinality $m$ and Poisson hyperdegree distribution $P(k)$. From this figure, we can observe the presence of two period-2 bifurcations of small value of $m$ which at the tricritical point disappear giving rise to a continuous phase transition, as predicted by our theory (see Fig. \ref{fig:figure6} (d)).
Moreover, also in this case, as in the scenario where only positive regulations are present, we observe that for hypergraphs with hyperedges of fixed cardinality $m$, hypergraphs with larger value of $m$ are more robust. In fact, for larger values of $m$, the onset of period-2 oscillation occurs at smaller values of $p$, and collapse of the network (onset of period-2 oscillations or continuous phase transition at $R=0$) also occurs at values of $p$ that decreases with $m$.

 \subsubsection{Route to chaos in higher-order triadic percolation}

\begin{figure}[!htb]
  \includegraphics[width=0.5\textwidth]{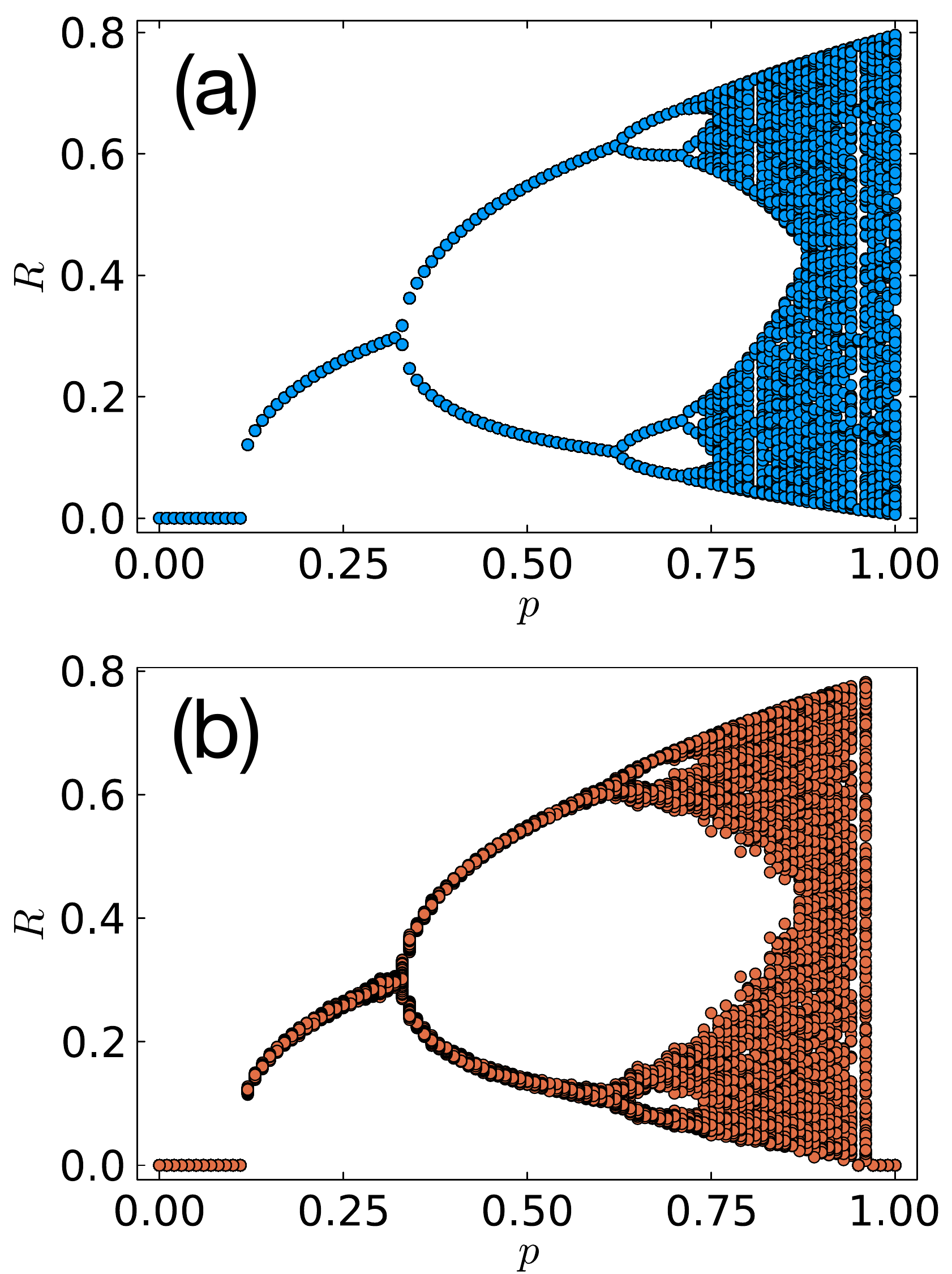}
  \caption{In the presence of both positive and negative HOTIs, the phase diagram of HOTP reduces to an orbit diagram displaying a route to chaos in the universality class of the logistic map. Panel (a) displays the theoretical prediction of this orbit diagram and panel (b) displays the results of  Monte Carlo simulations. The structural hypergraph has $N=10^6$ nodes, hyperedges of fixed cardinality $m=10$, and Poisson hyperdegree distribution $P(k)$ with an average hyperdegree $c=10$. The  Poisson degree distribution $\hat{P}^{\pm}(\kappa)$ with an average $c^+=3$ and $c^-=5.5$.}
  \label{fig:figure7}
 \end{figure}

When both positive and negative HOTIs are present, the order parameter of HOTP undergoes a route to chaos in the universality class of the logistic map, turning HOTP to a fully-fledged dynamical process (see Fig. $\ref{fig:figure7}$).  This can be demonstrated by considering the map $R^{(t)}=h_{m,p}(R^{(t-1)})$ and, using the results of Feigenbaum, \cite{feigenbaum1978quantitative,strogatz2018nonlinear}, which show that if the map is continuous, unimodal and displays a quadratic maximum, it will undergo the route to chaos in the universality class of the logistic map (see Fig. $\ref{fig:hmp}$ for an example and Appendix \ref{ApA} For a detailed discussion).
This phenomenology is in line with what happens for standard triadic percolation  \cite{sun2023dynamic}.

\begin{figure}[!htb]
  \includegraphics[width=0.4\textwidth]{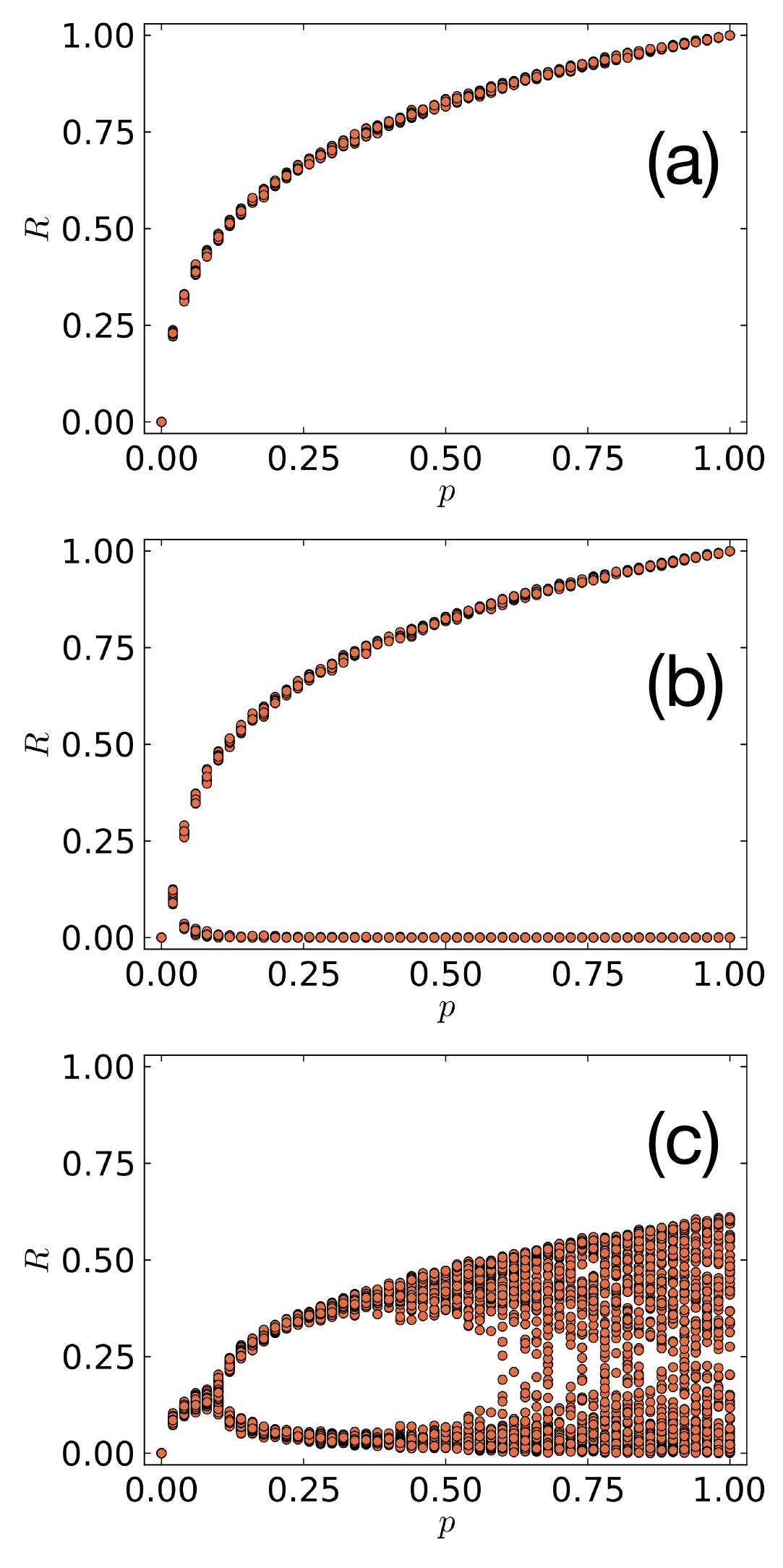}
  \caption{The phase diagram of  HOTP occurring on a real-world hypergraph displays a dynamical behavior that is qualitatively similar to the dynamical behavior of HOTP on random hypergraphs. The three panels display the phase diagram of  HOTP in the presence of exclusively positive HOTI (panel (a)); in the presence of exclusively negative HOTI (panel(b)); in the presence of both positive and negative HOTIs (panel (c)). The structural hypergraph is the recipe ingredient hypergraph of $N=6714$ nodes and $E_H=39774$ hyperedges, with the largest hyperedge cardinality of 65. The data is freely available at the repository ~\cite{landry_2023}. The HOTIs are generated randomly between the nodes and the hyperedge of the real hypergraph and the average number $c^{+}$ and $c^{-}$ of positive and negative regulatory interactions is: $c^+ =10,c^{-}=0$ (panel (a));  $c^+=\infty, c^-=15$ (panel (b)), and $c^+=10,c^-=15$ (panel (c)).}
  \label{fig:kaggle}
 \end{figure}
 
\subsubsection{HOTP on real hypergraphs}
{\color{black} An important question that arises is whether the observed critical and dynamical behavior of HOTP remains unchanged also on hypergraphs that are defined departing from the random hypergraphs that we have considered so far.
Our results show that the predicted dynamical behavior of HOTP, derived analytically for random hypergraphs, is not restricted to random hypergraphs and can remain qualitatively unchanged also if HOTP is defined on real hypergraph architectures.
In particular, to investigate the matter, we have considered a real-world hypergraph of recipe ingredients with $N=6714$ nodes representing ingredients and $E_H=39774$ hyperedges with the largest hyperedge cardinality of 65, representing recipes containing a combination of the ingredients, from the Repository \cite{landry_2023}. On top of this real-world hypergraph, we have generated random regulatory networks determining the HOTIs between the nodes and the hyperedges.  In Fig. \ref{fig:kaggle} we show numerical evidence that HOTP occurring on the considered real hypergraph and studied in the presence of only positive, only negative, and in the presence of both positive and negative HOTIs reveals the same dynamical behavior of HOTP occurring on a random hypergraph. Our course does not exclude that other real-world hypergraphs might display a different behavior in particular if they are embedded into real space where we know that triadic percolation displays spatiotemporal dynamics of the giant component (see for instance discussion in Ref. \cite{millan2023triadic}).}

 \subsubsection{Reentrant orbit diagram in HOTP}
\begin{figure}[!htb]
  \includegraphics[width=0.5\textwidth]{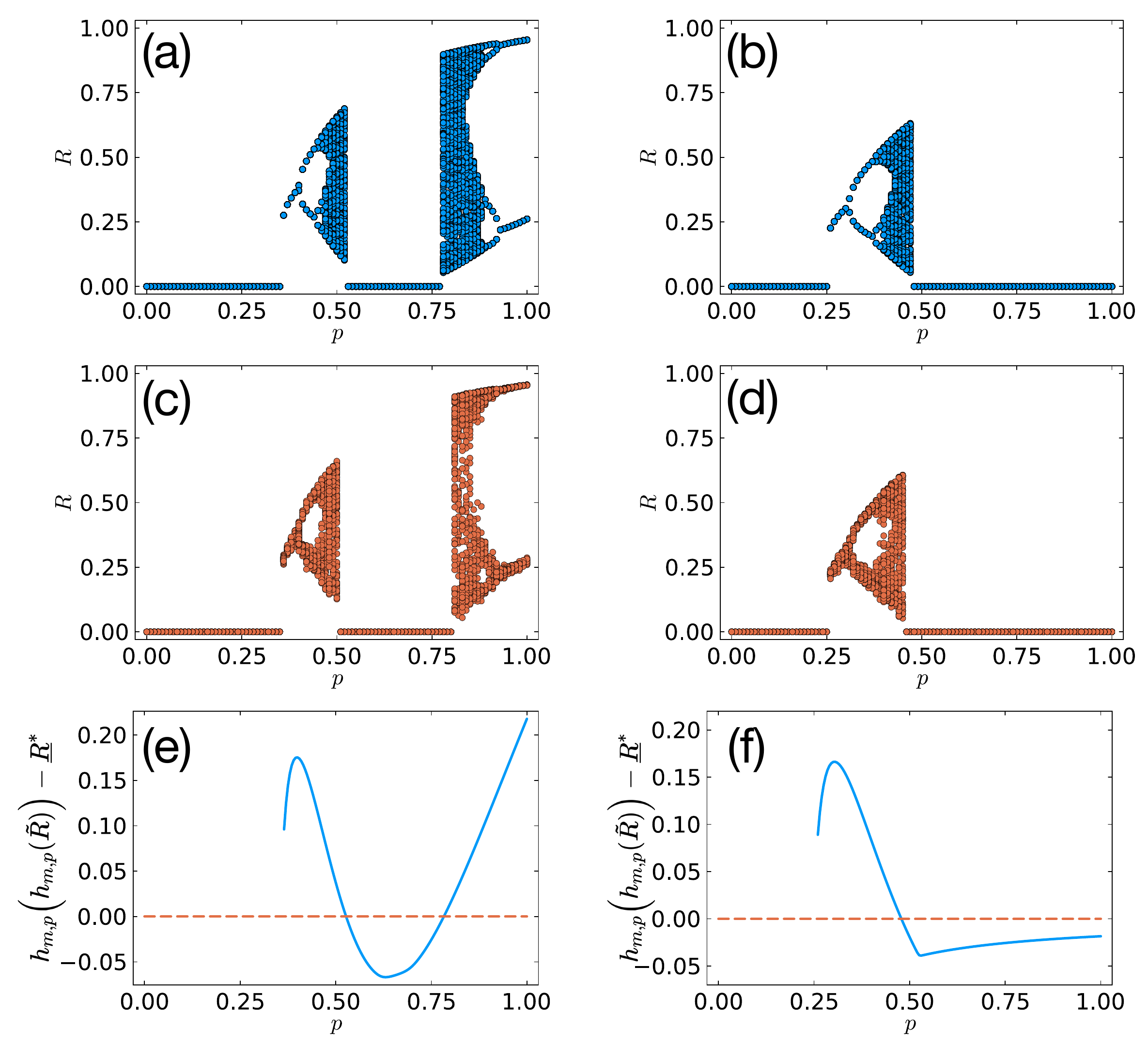}
  \caption{We provide two examples of orbit diagrams showing the reentrant phase transition on HOTP, in which the hypergraph can be re-activated by decreasing the value of $p$, i.e. increasing the probability of random deactivation of the hyperedges.  Panels (a) and (b) show two examples of theoretical orbit diagrams of HOTP displaying the reentrant phase transition. Panels (c) and (d) show the orbit diagrams obtained by implementing the corresponding Monte Carlo simulations. Panels (e) and (f) reveals that Eq. (\ref{eq:condition_nontrivial_oscillation}), i.e. the intersection between the curve $y=h_{m,p}(h_{m,p}(\tilde{R}))-R^{\star}$  and the orange dash line indicates $y=0$, provides the correct prediction of the value of $p$ at which either the hypergraph collapse of the reentrant phase transition are observed.
  In both examples, the structural hypergraph has $N=5 \times 10^5$ nodes, Poisson hyperdegree distribution $P(k)$ with average $c=30$ and  hyperedges with fixed cardinality given by $m=2$ (in panels (a),(c),(e)) and $m=3$ (in panels (b),(d),(f)).
  The   Poisson degree distributions $\hat{P}^{\pm}(\kappa)$ have average positive and negative regulatory degree given by $c^+=1, c^-=2.9$ (in panels (a),(c),(e)); and  $c^+=1,c^-=4$ (in panels (b),(d),(f)).}
  \label{fig:figure8}
 \end{figure}

Interestingly,  for some choices of the HOTp parameter values, the increase of random deactivation of hyperedges (i.e. the decrease of $p$) does not result in a monotonic suppression of the giant component. Instead, the order parameter $R$ in some cases displays a non-monotonous behavior as a function of $p$ and reentrant phase transitions (see Fig. \ref{fig:figure8}). This consists of a  {\em collapsed state} with $R^{(t)}=0$ for all values of $t$, observed in a range of values of $p\in [p_-,p_+]$, while for $p<p_-$ and for $p>p_+$ the giant component is non zero.  The reentrant nature of the phase implies the na\"ively not intuitive result that for a  higher level of random damage (smaller values of $p$)  the giant component can be restored.  Reentrant phase transitions are also observed in  correlated multilayer networks  \cite{baxter2016correlated} and in multilayer networks with extended-range percolation \cite{cirigliano2024general}, however in all these example the fraction of nodes in the giant component is unique in the thermodynamics limit, while this is an example of a reentrant orbit diagram of percolation. Finally we note that such phenomenon can be also observed on networks, (i.e. for $Q(m)=\delta_{m,2}$), under suitable conditions.

This interesting result can be physically interpreted as follows. The random deactivation of hyperedges has two opposite consequences. The decrease in network connectivity will reduce the number of nodes in the giant component. Meanwhile, the number of active negative regulators is also reduced, resulting in an increasing number of active hyperedges. The dynamics is driven by the competition between these two effects. Therefore, on hypergraphs with a larger number of negative regulators, smaller value of $p$ might result in the activation of more hyperedges and hence might lead to a transition between an inactive state without giant component to an active state with non-zero giant component and a non-trivial dynamics.

\begin{figure}[!htb]
  \includegraphics[width=0.5\textwidth]{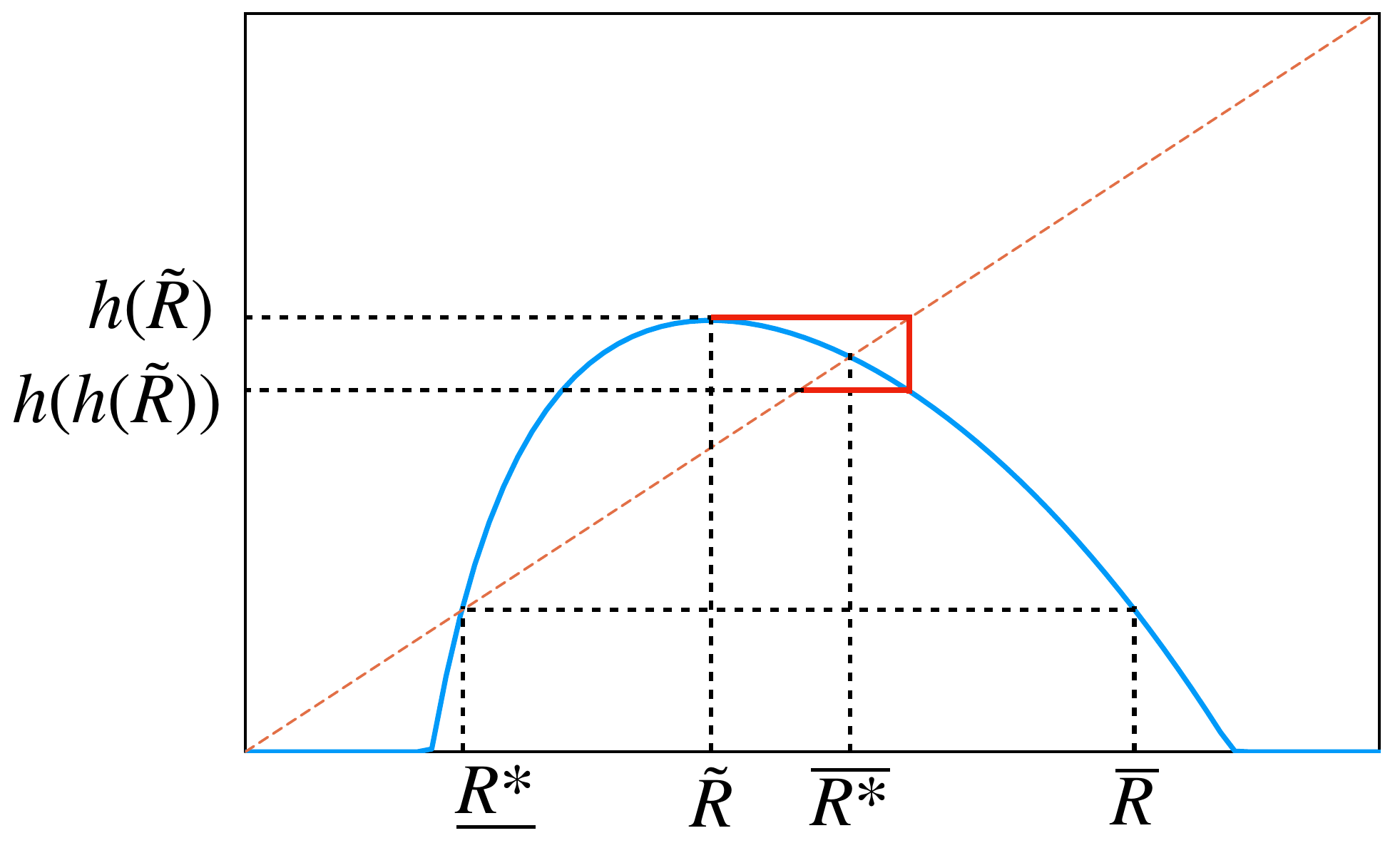}
  \caption{A graphic illustration of the critical condition for having a non-trivial giant component in HOTP. The blue curve indicates function $h_{m,p}(R)$ and the orange line indicates $f(R)=R$. The red line shows the dynamic of iterating from initial condition $R=\tilde{R}$.
  }
  \label{fig:figure9}
 \end{figure}

Let us provide the theoretical understanding of the occurrence of the reentrant collapsed phase.
Specifically, let us investigate the conditions for observing non-trivial dynamics of the order parameter. 
In order to illustrate the argument, let us consider a continuous unimodal map  $h_{m,p}(R)$ (see Fig. \ref{fig:figure9}) with maximum $\tilde{R}$ having compact support on a proper subinterval of $[0,1]$.  Furthermore, let us assume that the map has two fixed points of $R=h_{m,p}(R)$ denoted  $\underline{R^{\star}}$ and $\overline{R^{\star}}$ with $\underline{R^{\star}}<\overline{R^{\star}}$ where $\underline{R^{\star}}$ is an unstable fixed point.  Let us denote with $\overline{R}$ the value of $R$ with $R>\underline{R^{\star}}$ satisfying $h_{m,p}(\overline{R})=\underline{R^{\star}}$.
This scenario is consistent for example (see Fig. \ref{fig:figure9}) with the properties of the map $h_{m,p}(R)$ for a random hypergraph with Poisson hyperdegree distribution $P(k)$ with average hyperdegree $c$ and constant hyperedge cardinality $m$ where the regulatory degree distribution $\hat{P}(\kappa^\pm)$ are Poisson distributions with averages $c$ and $c^\pm$ respectively. In this case the function $h_{m,p}(R)$ has a unique maximum at $R=\tilde{R}$ at
\bea
\tilde{R}=-\frac{1}{c^+} \log \frac{c^-}{c^+ + c^-}.
\label{eq:tildeR}
\eea
Moreover, $\left. \partial h / \partial R \right|_{R=\underline{R^{\star}}}>1$, thus $\underline{R^{\star}}$ is an unstable fixed point, while $\left. \partial h / \partial R \right|_{R=\overline{R^{\star}}}<1$, thus $\overline{R^{\star}}$ is a stable fixed point.

Under the stated conditions of the map $h_{m,p}$, if $R>\overline{R}$ or $R<\underline{R^{\star}}$,  the dynamics will converge to the trivial collapsed state. In order to observe a non-trivial dynamics, the interval $\left(\underline{R^{\star}}, \overline{R}\right)$  must be mapped onto this itself under the action of the map $h_{m,p}$. Therefore, the condition for having non-trivial dynamics is given by
\bea
h_{m,p}(\tilde{R})<\overline{R}, 
\eea
or using $h_{m,p}(\overline{R})=\underline{R^{\star}}$, 
\bea
\quad h_{m,p}(h_{m,p}(\tilde{R})) > \underline{R^{\star}}.
\label{eq:condition_nontrivial_oscillation}
\eea
A graphic illustration can be found in Fig. \ref{fig:figure9}.

In Fig. \ref{fig:figure8} (c) and (d) we show two examples in which the reentrant collapsed state is observed. In panels (e) and (f) we examine the condition of having a non-trivial dynamics for corresponding cases. We reveal that the collapsed state emerges exactly when the mentioned condition is fulfilled.

\begin{figure}[!htb]
  \includegraphics[width=0.48\textwidth]{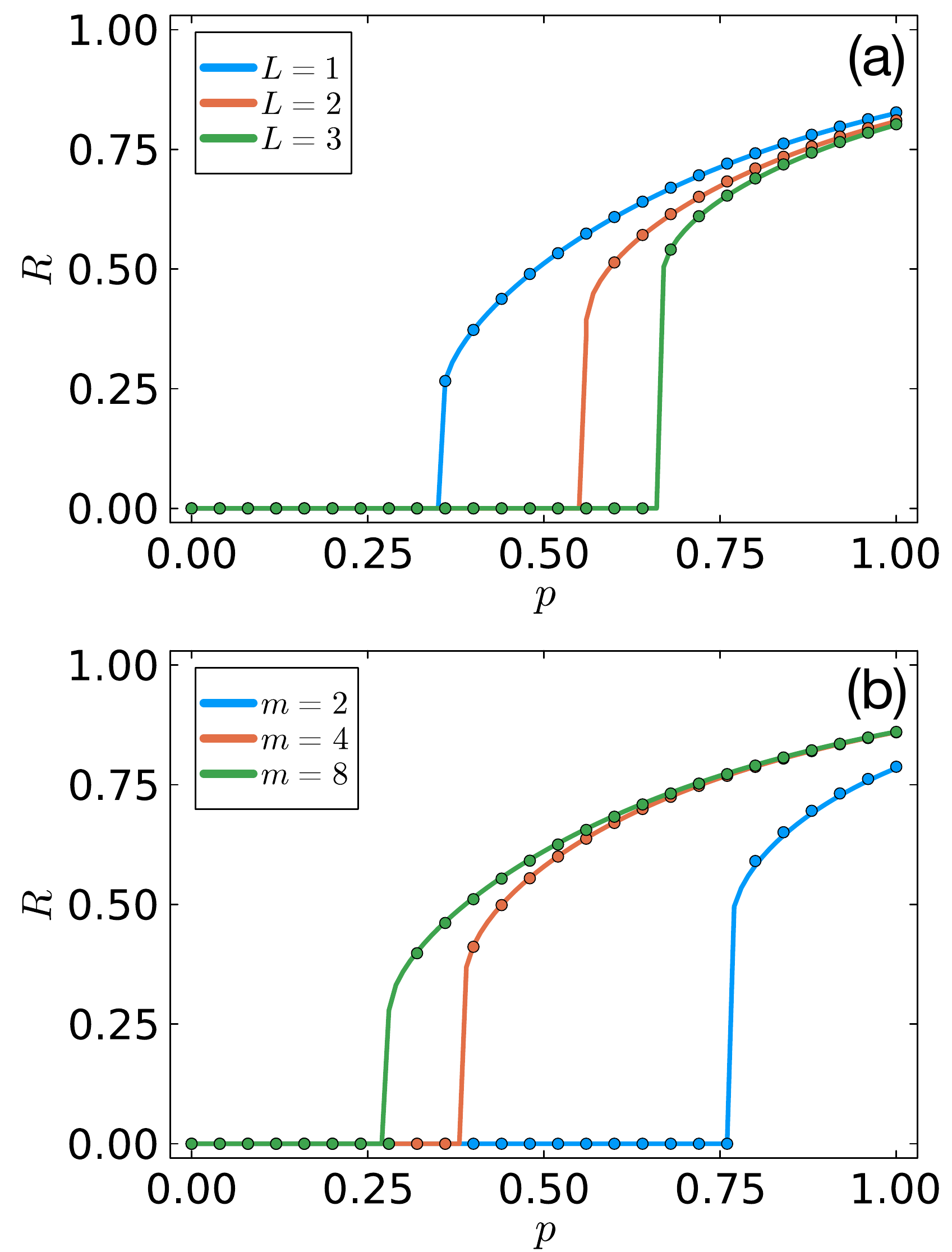}
  \caption{The dependence of the order parameter $R$ of HHOTI versus $p$ is shown for hypergraphs having exclusively positive HHOTIs.
  Panel (a) shows the dependence of the phase diagram with the number of hierarchical layers $L$, and panel (b) shows the dependence with the value $m$ of the hyperedge cardinality. Solid lines indicate theoretical predictions symbols indicate the results of Monte Carlo simulations on hypergraphs of $N=5\times10^4$ nodes.   The structural hypergraphs have a Poisson hyperdegree distribution P(k) with an average $c=2$ and hyperedges of fixed cardinality $m$ ($m=8$ in panel (a)). The regulatory network is formed exclusively by positive regulations and has a Poisson regulatory degree with an average $c^+=2.5$ (panel (a)) and $c^{+}=5$ (panel (b)).}
  \label{fig:figure10}
 \end{figure}

\begin{figure*}[!htb]
  \includegraphics[width=\textwidth]{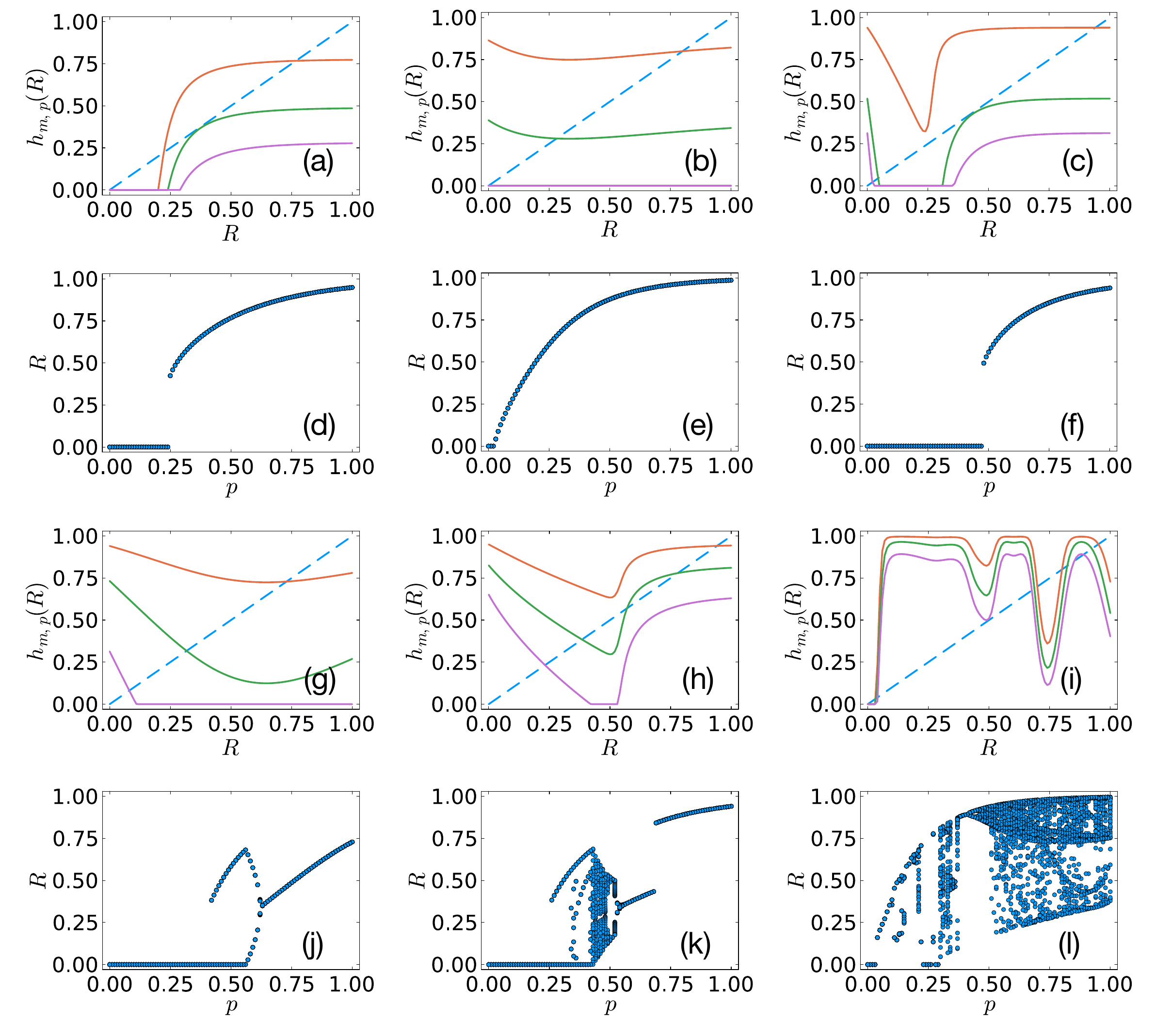}
  \caption{Illustrative examples of the map $h_{m,p}(R)$ representing the  HHOTP dynamics for specific values of $p$ are plotted in the first and third row together with the identity line (dashed) whose intersection with the map indicates the fixed point $R^{\star} = h_{m,p}(R^{\star})$. The corresponding theoretical phase diagrams/orbit diagrams of HHOTI are plotted in the second and fourth rows.  Panel (a) shows the HHOTP with exclusive positive regulations. Panels (b), (c), (g), and (h) show the HHOTP with exclusive negative regulations, and panel (i) shows an example of the general case with both positive and negative regulations. The different panels correspond to different phase transitions: a discontinuous transition (d, f), a continuous transition (e), a  period-2 bifurcation (j), and a route to chaos (k,l).  The random hypergraphs have a Poisson hyperdegree distribution with an average $c$ and a Poisson regulatory degree distribution with an average $c^+$ and $c^-$ for all layers. The hyperedges have a fixed cardinality $m$. The hierarchical regulations have $L$ layers. The parameters used in the figure are:  $c=3$, $c^+=5$, $m=5$, $L=10$ (panel (a)), $c=5$, $c^-=3$, $m=10$, $L=2$ (panel (b)), $c=3$, $c^-=10$, $m=2$, $L=50$ (panel (c)), $c=3$, $c^-=2$, $m=2$, $L=4$ (panel (j)), $c=3$, $c^-=5$, $m=3$, $L=100$ (panel (k)), $c=30$, $c^+=20$, $c^-=30$, $m=10$, $L=10$ (panel (l)). 
  }
  \label{fig:figure11}
 \end{figure*}

\section{Hierarchical higher-order triadic percolation }

Higher-order triadic interactions can be hierarchically nested, as discussed in Sec. \ref{Sec:HHOTI}. This leads to hierarchical higher-order triadic percolation (HHOTP) where the regulatory interactions of the hyperedges can be further regulated by a hierarchy triadic interactions  (see Fig. \ref{fig:figure1}). These hierarchical triadic interactions (HHOTIs) are for instance observed in ecological networks \cite{bairey2016high}. 

Assuming that the nested HHOTIs regulate the interactions simultaneously, we define the HHOTP by two-step iterative dynamics, which differs from the dynamics of HOTP only by a modification of Step 2. Indeed Step 2  becomes

\begin{itemize}
\item[Step 2$^\prime$]: A hyperedge is down-regulated if 
\item[(a)] it is regulated by at least one active negative regulator and/or is not regulated by any active positive regulator via active regulatory interactions in $W^{[1]}$, and it is considered active otherwise. All other hyperedges are deactivated with probability $1-p$.
\item[(b)]
{\color{black}An active regulatory interaction in $W^{[\mu]}$ requires that it is not regulated by any active negative regulator via active regulatory interactions in $W^{[\mu+1]}$ and is regulated by at least one active positive regulator via active regulatory interactions in $W^{[\mu+1]}$.}
\end{itemize}

According to this rule, the probability of retaining a hyperedge is given by
\bea
p_H^{(t)} &=& p G_0^- \left(1-p_{H,[2]}^{(t)}R^{(t)}\right)\nonumber \\&&\times\left(1-G_0^+ \left(1-p_{H,[2]}^{(t)}R^{(t)}\right)\right), \nonumber\\
p_{H,[\mu]}^{(t)} &=& G_{0, [\mu]}^{-}\left(1-p_{{H,[{\mu+1}]}}^{(t)}R^{(t)}\right)\nonumber \\&&\times \left(1-G_{0,[\mu]}^{+}\left(1-p_{{H,[\mu+1]}}^{(t)}R^{(t)}\right)\right)
\eea
for $\mu\in \{2,\ldots, L\}$ with $p_{H,[L+1]}=1$. Note that here the generating function  $G_{0, \ell}^{\pm}\left(x\right)$ is given by Eq. (\ref{G0mupm}).
These equations that implement Step 2$^{\prime}$ can be encoded into a map
\bea 
p_H^{(t)}=g_p\left(R^{(t)}\right),
\label{eq:gp2}
\eea
while the equations that implement Step 1 remain the same as for HOTP and obey Eq. (\ref{eq:fm}), i.e. they are given by $R^{(t)}=f_m\left(p_H^{(t-1)}\right)$.
Thus we obtain that the dynamics of HHOTP can be encoded in a one-dimensional map 
\bea
R^{(t)}=h_{m,p}(R^{(t-1)}=f_m(g_p(R^{(t-1})).
\eea
This map in the presence of negative regulations is no longer unimodal, thus
HHOTP displays a phenomenology that is significantly different from HOTP and displays a route to chaos that is no longer in the universality class of the logistic map.
Here we discuss the rich phenomenology of this complex dynamical process by considering first the simple case in which only positive regulations are present, then the case in which only negative regulations are present and finally the general case in which both positive and negative regulations are present with non-zero probability.

When only positive regulations are present, similarly to what happens for HOTP under the same conditions, HHOTP  displays a discontinuous hybrid transition. In order to illustrate this phenomenology in a simple case we consider a random hypergraph with fixed hyperedge cardinality $m$, a Poisson hyperdegree distribution $P(k)$, and Poisson regulatory degree distribution $\hat{P}_{[\mu]}(\kappa^\pm)$ of a same average degree across all layers. In this scenario, we observe that the hypergraphs with a larger hyperedge cardinality $m$ and a smaller number of regulatory layers $L$ are more robust, as indicated by their lower critical percolation threshold $p_c$, and their smaller discontinuity $R^*(p_c)$ (see Fig. \ref{fig:figure10}).

When only negative regulations are present, HHOTP displays a rich phenomenology and the nature of the dynamics changes drastically with respect to HOTP. We recall that when only negative regulations are present,  HOTP  displays either continuous transitions or period-2 oscillations. In contrast, HHOTP can display continuous transitions, discontinuous hybrid transitions, periodic oscillations, and chaos (see Fig. \ref{fig:figure11}). Specifically, as long as there are negative regulatory interactions and more than one layer, i.e. $L>1$, the map $h_{m,p}(R)$ can have multiple local maxima and minima in general, and the order parameter of HHOTP undergoes a route to chaos no longer in the universality class of the logistic map (see Fig. \ref{fig:figure11}).

\section{Generalizations}
\subsection{Interdependent higher-order triadic percolation (IHOTP)}

The connectivity of hypergraphs can be probed not only with hypergraph percolation resulting from the random deactivation of hyperedges but also with several other higher-order percolation processes \cite{sun2021higher}, in which the activation of a hyperedge requires, in addition to the regulatory rules, also cooperation among all nodes that belong to it. 

In interdependent hyperedge percolation, the interdependent hypergraph giant component is the extensive connected component whose nodes and hyperedges satisfy the following recursive self-consistent conditions. \begin{itemize}
\item
A node is in the giant component if
  it belongs to at least one hyperedge that is in the interdependent hypergraph giant component. 
\item A hyperedge is in the giant component if 
\begin{itemize}\item[(i)] it is not deactivated, \item[(ii)] {\em all its nodes} belong to the interdependent hypergraph giant component.
\end{itemize}
\end{itemize}
Note that the difference with the hypergraph giant component is in the requirement (ii) for the hyperedge to be active, which imposes that {\em all} but not {\em at least one} of the nodes belonging to the hyperedge are in the interdependent hypergraph giant component.

Here we define interdependent higher-order triadic percolation (IHOTP) which generalizes HOTP. As HOTP, IHOTP is defined in terms of a two-step iterative process, the first step implementing percolation of the structural hypergraph and the second step implementing regulation of the hyperedges. 
The difference between HOTP and IHOTP is that in Step 1 of IHOTP, we consider interdependent hyperedge percolation instead of simple hyperedge percolation.
Specifically, in IHOTP Step 1 of the iterative algorithm is modified as:
\begin{itemize}
\item Step 1': Given the set of active hyperedges at time $t-1$, a node is considered active if it is in the interdependent hypergraph giant component.
\end{itemize}
Step 2 of IHOTP is instead the same as for HOTP.
It follows that at time $t$ the dynamical equations implementing Step 1 in IHOTP are given by the equations \cite{sun2021higher} determining the size of the interdependent hypergraph giant component $R^{(t)}$ when hyperedges are intact with probability $p_H^{(t-1)}$, i.e.  
\bea
\hat{S}^{(t)}&=& p_H^{(t-1)} G_{1,m}(S^{(t)}), \nonumber\\
S^{(t)} &=& 1-G_1(1-\hat{S}^{(t)}),\nonumber\\
R^{(t)} &=& 1-G_0(1-\hat{S}^{(t)}).
\label{eq:interdependent_hyperedge}
\eea
The equation determining the regulation process occurring at Step 2 is unchanged with respect to HOTP and is given by Eq. (\ref{eq:hypergraph_triadic_pL}) that we rewrite here for convenience,
\bea
p_H^{(t)} &=& p G_0^{-}\left(1-R^{(t)}\right) \left(1-G_0^{+}\left(1-R^{(t)}\right)\right).
\label{eq:hypergraph_triadic_pL2}
\eea

\begin{figure*}[!htb]
  \includegraphics[width=\textwidth]{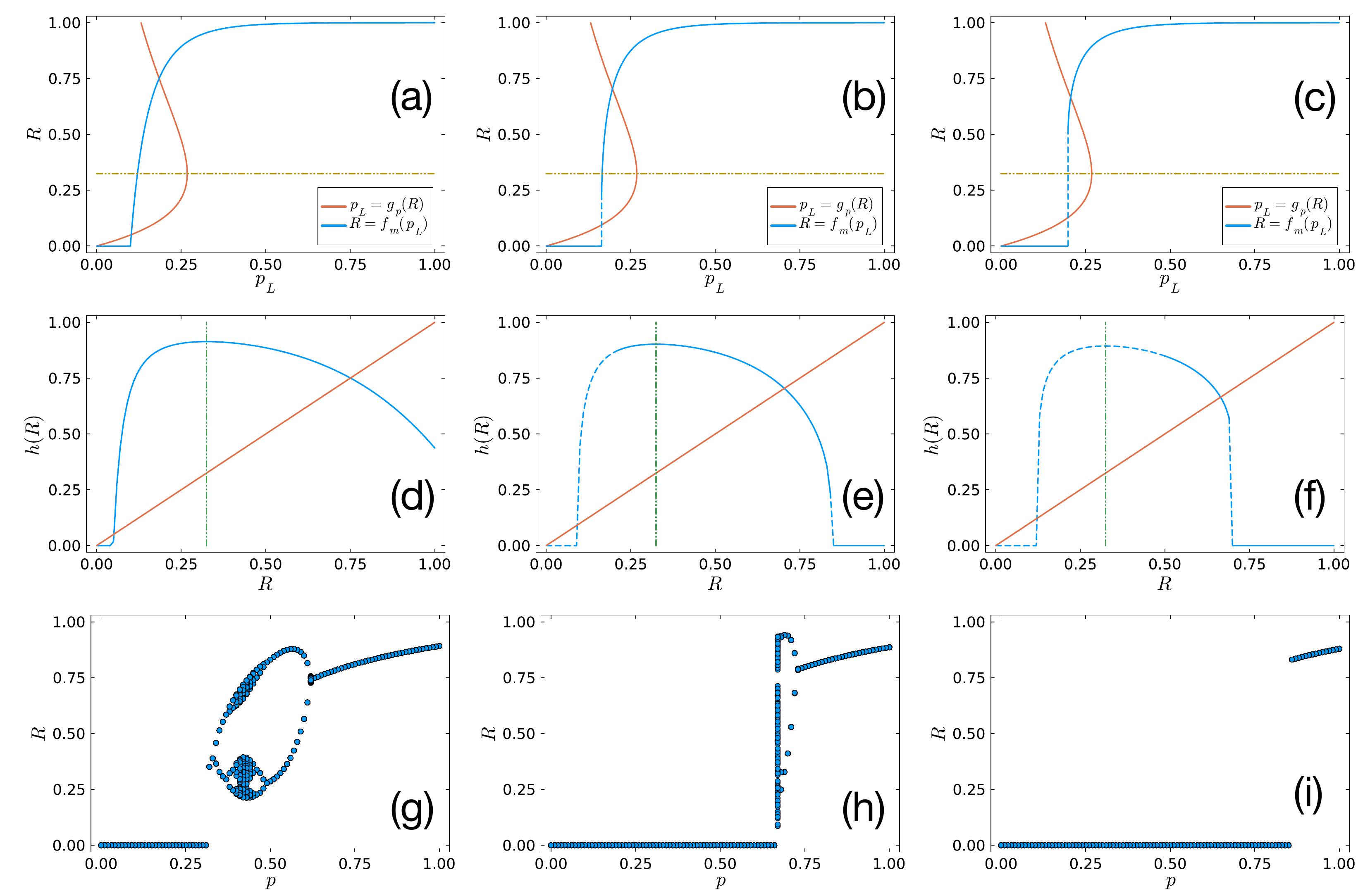}
  \caption{The effect due to the discontinuous hybrid transition of interdependent hyperedge percolation on the dynamical properties of IHOTP.
  The figure displays the maps $g_p(R)$, $f_m(p_H)$ determining Step 1 and Step 2 of the IHOTP dynamics (panels (a)-(c)),  the one-dimensional map $h_{m,p}(R)=f_m(g_p(R))$ encoding for the combined Step 1 and Step 2 dynamics (panels (d)-(f)) and their corresponding phase diagrams/orbit diagrams (panels (g)-(i)). The panel corresponds to parameter values in which the interdependent hyperedge percolation (function $R=g_p(p_H)$) displays either a continuous (panels (a)-(d)-(g)) or a discontinuous phase transition (panels (b),(c),(e),(f),(h),(i)). In panels (a)-(f) the dashed lines indicate either the discontinuous jump of interdependent percolation (panels (a)-(c)) or the values of $R$ that cannot be reached by the IHOTP dynamics (panels (d)-(f)). The horizontal line in panels (a)-(c) shows the value $R=\tilde{R}$ at which the function $p_H=g_p(R)$ reaches its local maximum. The structural hypergraphs have Poisson hyperdegree distribution with average hyperdegree $c=10$ and hyperedges with cardinality distribution $Q(m)$ given by Eq. (\ref{eq:Qm}). Here $r$ indicates the ratio between the number of hyperedges of cardinality $m=2$ and the number of hyperedges of cardinality $m=3$. The hypergraphs have regulatory degree distributions with average degrees $c^+=4$ and $c^-=4.5$. The adopted value of $r$ are $r=1$ (panels (a),(d),(g)), $r=0.6$ (panels (b),(e),(h)) and $r=0.4$ (panels (c),(e),(i)). The critical value of $r=r_H^c$ of this model is $r_H^c=2/3$ see Eq. (\ref{eq:critical_r}). Panels (a)-(f) are plotted for $p=0.6$.}
  \label{fig:figure12}
 \end{figure*}
We observe that interdependent hyperedge percolation is significantly different from hyperedge percolation, and can undergo a discontinuous phase transition \cite{sun2021higher}.  
Interestingly, the discontinuous nature of the percolation transition in interdependent hyperedge percolation dramatically changes the nature of the dynamics in IHOTP.
In particular, for certain parameter values, the IHOTP cannot admit a complete route to chaos and can only display steady states or period-2 oscillation.

In order to provide concrete evidence of the effect due to the discontinuous hybrid transition of interdependent hyperedge percolation on the dynamical properties of IHOTP, let us again write Eq. (\ref{eq:interdependent_hyperedge}) and Eq. (\ref{eq:hypergraph_triadic_pL2}) in the form of one-dimensional maps
\bea
R^{(t)} = f_m\left(p_H^{(t-1)}\right), \quad p_H^{(t)}=g_p\left(R^{(t)}\right),
\label{eq:fmgp}
\eea
and let us adopt again  the notation
\bea
R^{(t)} = h_{m,p}(R^{(t-1)}) = f_m(g_p(R^{(t-1)})).
\label{eq:h2}
\eea
We consider structural hypergraphs with a Poisson hyperdegree distribution $P(k)$ with average hyperdegree $c$ and hyperedges of cardinality given either by $m=2$ or $m=3$. Specifically, we consider the cardinality distribution $Q(m)$ given by
 \bea
 Q(m)=q_2\delta_{m,2}+q_3\delta_{m,3},
 \label{eq:Qm}
 \eea
 where we denote with $r$ the ratio between the number of hyperedges of cardinality $m=2$ and the number of hyperedges of cardinality $m=3$, i.e. $r=q_2/q_3$.
 It is known \cite{sun2021higher} that in this ensemble of hypergraphs, interdependent hypergraph percolation displays a tricritical point at $r=r_H^c$, $p=p_H^c$ given by 
 \bea
r_H^c=2/3, \quad p_H^c = 3/2c.
\label{eq:rc}
\eea 
For $r>r_H^c$  interdependent hypergraph percolation displays a continuous percolation transition while for $r<r_H^c$ the model displays a discontinuous percolation transition. 

\begin{figure}[!htb]
  \includegraphics[width=0.45\textwidth]{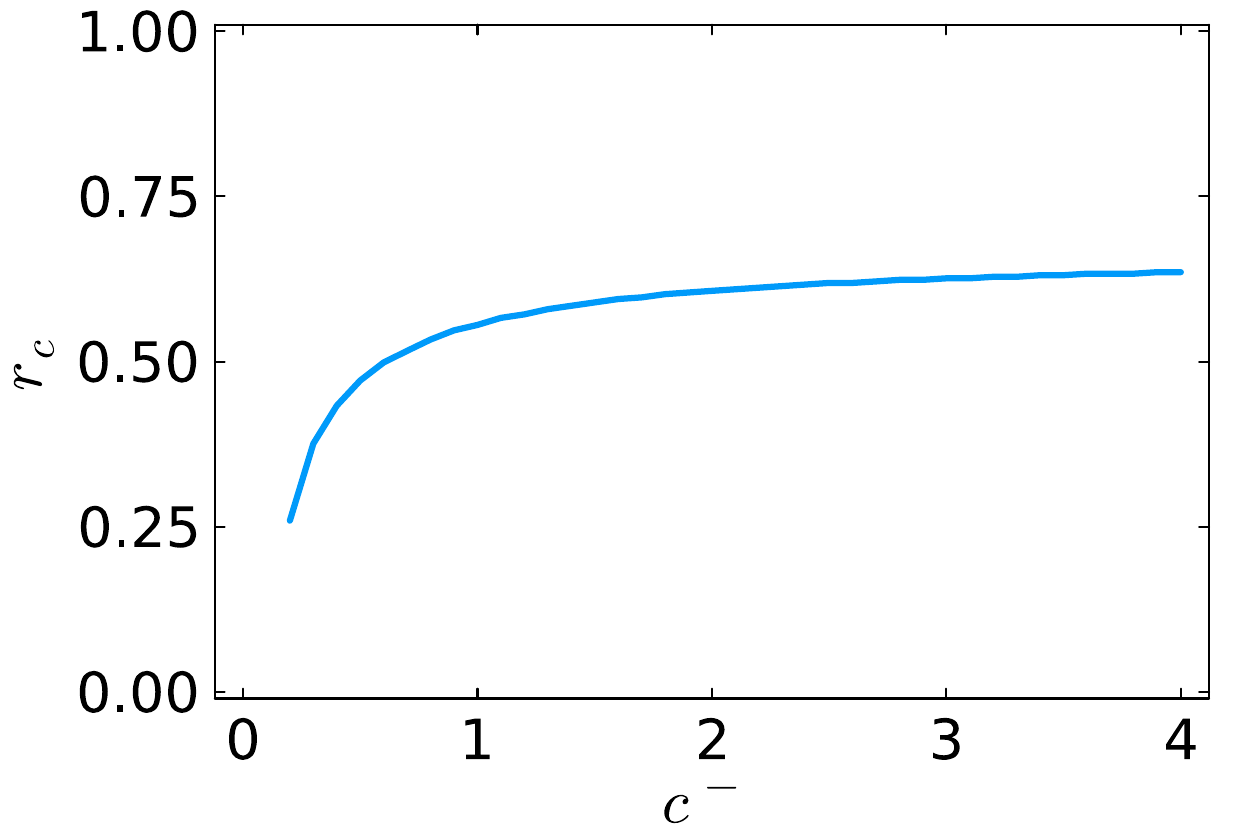}
  \caption{The critical value $r_c$ (satisfying Eq. (\ref{eq:critical_r}))  of the fraction between the number of hyperedges of cardinality $m=2$ and the number of hyperedges of cardinality $m=3$, is plotted as a function of the average negative regulatory degree $c^-$.  Thus if $r<r_c$ a route to chaos cannot be observed in IHOTP. The plot is obtained by considering a random hypergraph with Poisson hyperdegree distribution with an average hyperdegree $c=5$, cardinality distribution $Q(m)$ given by Eq. (\ref{eq:Qm}) and the regulatory interactions have a Poisson distribution with an average $c^+=5$ and $c^-$. }
  \label{fig:figure13}
 \end{figure}

It is thus instructive to investigate the behavior of IHOTP in this scenario (see Fig. \ref{fig:figure12}).
Recall that the dynamics of IHOTP encoded in the map $h_{m,p}(R)$ defined in Eq. (\ref{eq:h2}) is initialized as that at time $t=0$, when all hyperedges are active with probability $p_H^0$. When the percolation transition is discontinuous, there will be an interval of values of $R$ from $0$ to a finite value $\mathcal{R}$ that is not physical, i.e. cannot be achieved (see Fig. \ref{fig:figure12} (b) and (c)). Thus the corresponding map $h_{m,p}(R)$, is not defined in the entire range $[0,1]$ as it does not have compact support (see Fig. \ref{fig:figure12} (e) and (f)). 
This effect significantly changes the dynamical properties of IHOTP. In particular the map $h_{m,p}(R)$ might no longer display a local maximum.
As discussed in the case of HOTP, (see Appendix \ref{ApA}) in the situation in which $f_m(p_H)$ is a continuous function, i.e. when the percolation transition is continuous, the map $h_{m,p}(R)$ displays a local maximum for $R=\tilde{R}$ where $g_p^{\prime}(\tilde{R})=0$. Therefore the position of this maximum is independent of the percolation process under consideration.
Specifically, for Poisson positive and negative regulatory degree distributions with averages $c^+$ and $c^-$ we have that $\tilde{R}$ is given by Eq. (\ref{eq:tildeR}) as for HOTP.
When the considered percolation process displays instead a discontinuous phase transition at $p=p_c$ with a discontinuous jump $R=\mathcal{R}$ we cannot observe a local maximum of the map $h_{m,p}(R)$ as long as 
\bea
\mathcal{R}>\tilde{R}.
\label{eq:critical_r}
\eea
 In addition, to have a non-trivial dynamic, there must be at least one fixed point satisfying $R = h_{m,p}(R)$. Therefore $\mathcal{R}$ should further satisfy
\bea
\mathcal{R}<\max g_p^{-1}(p_c).
\eea
The critical value of $r$ at which we have $\mathcal{R}=\tilde{R}$ is here indicated as $r=r_c$.
Thus for $r<r_c$  we cannot observe a route to chaos of the order parameter of IHOTP and only stationary states and period-2 oscillations are allowed.
In Fig. \ref{fig:figure13} we show that $r_c$  is an increasing function of the average negative regulatory degree $c^-$. This can be explained as follows. First, we observe that  $\tilde{R}$ given by (Eq. (\ref{eq:tildeR})) is a decreasing function of  $c^-$. This indicates that a smaller discontinuity of the interdependent percolation order parameter will be enough to impede the route to chaos. Thus this is consistent with the monotonically increasing behavior of $r_c$ as a function of $c^-$.

\subsection{Generalization to regulated nodes}

\begin{figure}[!htb]
  \includegraphics[width=0.5\textwidth]{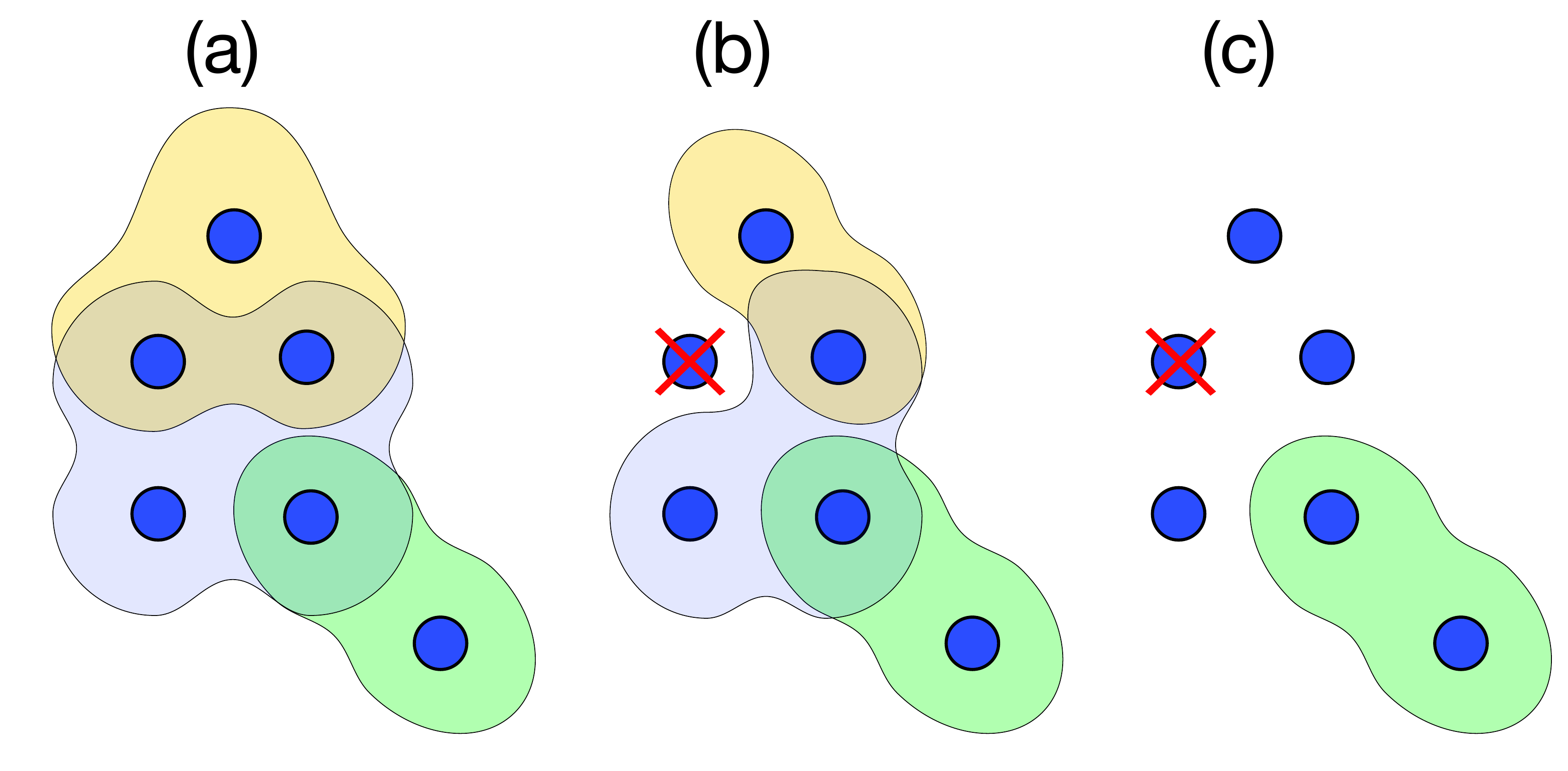}
  \caption{Schematic diagrams demonstrating the effect of node down-regulation in simple and cooperative node hyperedge percolation. Panel (a) shows a hypergraph. Panel (b) shows the effect of node deactivation in simple node hyperedge percolation. When a node is removed (deactivated), the cardinality of the hyperedges it belongs to decreases. Panel (c) shows the effect of node deactivation in cooperative node hyperedge percolation. When a node is deactivated, all hyperedges it belongs to are also deactivated. }
  \label{fig:figure14}
 \end{figure}

 An important generalization of HOTP is higher-order node dynamic percolation (HONDP) where the nodes are regulated instead of the hyperedges. {\color{black} The higher-order nature of this process is due to the fact that each node is typically regulated by several regulator nodes.} In the context of networks, the regulation of the nodes instead of the edges has been already considered in Ref. \cite{sun2023dynamic} where it was demonstrated that this generalization of triadic percolation still leads to a route to chaos of the order parameter.
Here we explore HONDP on hypergraphs emphasizing the peculiar properties of this dynamical percolation process and its relation to HOTP.
First of all, we observe a major difference between regulating hyperedges and regulating nodes: while hyperedges can be hierarchically regulated, nodes cannot. Thus we cannot define the node-regulation version of HHOTP.
The second observation that we make is that HONDP admits two formulations that depend on the higher-order nature of the hypergraph and have no correspondence in HOTP.  Indeed in hypergraphs,
 the effect of down-regulating a hyperedge is deterministic: it simply deactivates the hyperedge. Instead down-regulating a node can lead to two possible consequences. In hyperedge percolation \cite{sun2021higher} with random deactivation of the nodes, all the hyperedges involving the down-regulated nodes will reduce their cardinality by one. For instance, in social networks, a meeting can continue to take place also if one participant leaves. 
 Alternatively, down-regulating a node might lead to the deactivation of all the hyperedges the node belongs to \cite{bianconi2024theory}. This for instance occurs in chemical reactions where if a reactant is missing the reaction cannot take place. Here we call this process cooperative hypergraph percolation (CHP) (see Fig. \ref{fig:figure14} and Ref. \cite{bianconi2024theory} for further details). 
 According to these two definitions of node hypergraph percolation, we can define two versions of HONDP.

The dynamics of HONDP is defined as follows.
 At time $t=0$, every node is active with probability $p_H^0$. For $t \geq 1$:
\begin{itemize}
    \item Step 1: Given the set of active nodes at time $t-1$, a node is considered active at time $t$ if it is not deactivated at time $(t-1)$ and if it belongs to the giant component of (a) hyperedge percolation or (b) cooperative hyperedge percolation.
    \item Step 2: A node is deactivated if it is regulated by at least one active negative regulator and/or is not regulated by any active positive regulator and it is considered active otherwise. All other nodes are deactivated with probability $1-p$.
\end{itemize}
For completeness recall here the definition of the giant component in hyperedge percolation and in cooperative hypergraph percolation with regulation of the nodes.

The equations determining the size of the giant component $R^{(t)}$ given the probability $p_N^{(t-1)}$ that the nodes are active depend on the considered percolation process.
In the case (a), Step 1 implements hyperedge percolation with the regulation on the nodes. In this model nodes and hyperedges satisfy the following self-consistent and recursive conditions~\cite{sun2021higher}.
\begin{itemize}
\item
A node is in the giant component if it
\begin{itemize}\item[(i)] it is not down-regulated, \item[(ii)] belongs to at least one hyperedge that is in the hypergraph giant component.
\end{itemize}
 
\item A hyperedge is in the giant component if 
it includes at least one node that is in the hypergraph giant component.
\end{itemize}
The equations for hypergraphs~\cite{sun2021higher} can be used to obtain the size of the giant component $R^{(t)}$ of HONDP at time $t$ given the probability $p_N^{(t-1)}$ that nodes are not down-regulated at time $t-1$.
These equations are given by:
\bea
\hat{S}^{(t)}&=& \left[1-G_{1,m}(1-S^{(t)})\right], \nonumber\\
S^{(t)} &=& p_N^{(t-1)} \left[1-G_1(1-\hat{S}^{(t)})\right], \nonumber\\
R^{(t)} &=& p_N^{(t-1)} \left[1-G_0(1-\hat{S}^{(t)})\right].
\label{eq:simple_node_regulation}
\eea
In case (b) Step 1 implements cooperative hypergraph percolation \cite{bianconi2024theory}.
In cooperative hyperedge percolation,  nodes and hyperedges that are in the giant component satisfy the following self-consistent relationship:
\begin{itemize}
\item
A node is in the giant component if
\begin{itemize}
    \item[(i)] it is not down-regulated  with probability $p_N$, 
    \item[(ii)] it is connected to at least one hyperedge that is in the giant component.
\end{itemize}
\item A hyperedge is in the giant component if 
\begin{itemize}\item[(i)] {\em all its nodes} are not deactivated by regulatory interactions,
\item[(ii)] at least one of its nodes is in the giant component.
\end{itemize}
\end{itemize}
We thus have that  the equations determining the size of giant component $R^{(t)}$ of HONDP at time $t$ are given by \cite{bianconi2024theory}
\bea
\hat{S}^{(t)} &=& \sum_{m} \frac{m Q(m)}{\avg{m}} \left(p_N^{(t-1)}\right)^{m-1} \left[1-(1-S^{(t)})^{m-1}\right], \nonumber\\
S^{(t)} &=& 1-G_1(1-\hat{S}^{(t)}), \nonumber\\
R^{(t)} &=& p_N^{(t-1)} \left[1-G_0(1-\hat{S}^{(t)})\right].
\label{eq:cooperative_node_regulation}
\eea
In both cases, Step 2 is enforced by the equation implementing the regulation of the nodes, and thus expressing the probability $p_N^{(t)}$ that a node is active as a function of the probability $R^{(t)}$ that its regulator nodes are in the giant component at Step 1, which is given by,
\bea
p_N^{(t)} &=& p G_0^{-}\left(1-R^{(t)}\right) \left(1-G_0^{+}\left(1-R^{(t)}\right)\right).
\label{eq:node_reg}
\eea
If the hypergraph reduces to a network, i.e. if $Q(m)=\delta_{m, 2}$, both versions reduce to the dynamical percolation model with node regulation introduced in Ref. \cite{sun2023dynamic}.

The dynamics of HONDP,  as the dynamics of HOTP, is described by a percolation order parameter $R=R^{(t)}$ that, in the presence of both positive and negative regulatory interactions undergoes a route to chaos in the universality class of the logistic map, in both versions of the model.
To show this, let us adopt the same notation as for HOTP, and express 
the results of Step 1 given by the Eqs. (\ref{eq:simple_node_regulation}) in case (a) and Eqs. (\ref{eq:cooperative_node_regulation}) in case (b)
as 
\bea
R^{(t)}=f_m(p_N^{(t-1)}),
\eea
while we express the Eq. (\ref{eq:node_reg}) implementing Step 2 as
\bea
p_N^{(t)}=g_p(R^{(t)}).
\eea
With this notation, the dynamics of HONDP can be reduced to the one-dimensional map
\bea
R^{(t)} = h_{m,p}(R^{(t-1)}) = f_m\left(g_p\left(R^{(t-1)}\right)\right).
\eea
As for HOTP, also for HONDP, this map is continuous and since the function $f_m(R)$ is monotonous, the map $h_{m,p}(R)$ displays a maximum only when the function $p_N = g_p(R)$ displays a maximum.  This implies that only if both negative and positive regulatory interactions are present with finite probability, HONDP displays a route to chaos of its order parameter.

In order to illustrate some specific properties of HONDP we consider here the limiting case in which there are only positive regulations. In this scenario, similar to what we have previously discussed for HOTP, HONDP displays exclusively discontinuous hybrid transitions.

\begin{figure}[!htb]
  \includegraphics[width=0.45\textwidth]{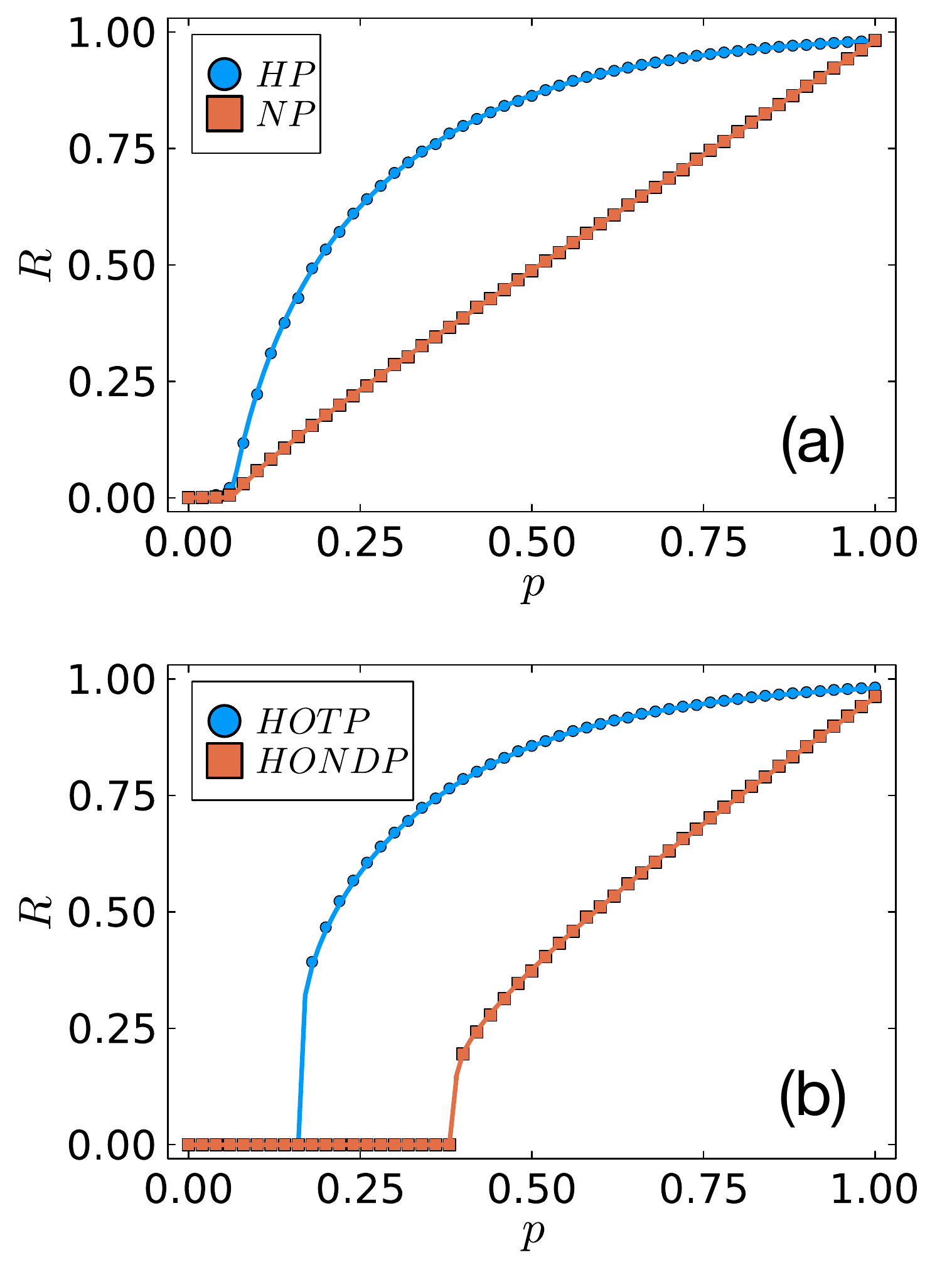}
  \caption{Panel (a) provides a comparison among the phase diagram of ordinary hypergraph percolation with node deactivation, indicated here as node percolation (NP), and with hyperedge deactivation, indicated here as hyperedge percolation  (HP). Panel (b)  provides a comparison among the phase diagrams of  HOTP and HONDP. Solid lines indicate theoretical predictions, symbols indicate Monte Carlo simulations conducted on hypergraphs of $N=10^4$ nodes. The structural hypergraph has a Poisson hyperdegree distribution $P(k)$ with an average hyperdegree $c=4$ and the hyperedges with fixed cardinality $m=5$. In panel (b) the positive regulatory interactions have a Poisson distribution with an average $c^+=4$ while negative regulatory interactions are absent.  }
  \label{fig:figure15}
 \end{figure}

We first compare case (i) of HONP with HOTP. In both cases Step 1, implements hypergraph percolation, however, in HONDP the nodes are upregulated with probability $p_N$, while in HOTP the hyperedges are upregulated with probability $p_H$.
Interestingly,  when regulatory interactions are not present, i.e. $p_N=p_H = p$, hypergraph percolation with the deactivation of nodes and hypergraph percolation with the deactivation of hyperedges displays a continuous phase transition at the same critical threshold \cite{sun2021higher}. However, in the presence of exclusive positive regulatory interactions, HONDP (i) and HOTP display discontinuous transitions and HONDP (i) has a larger critical threshold than HOTP, indicating a less robust giant component (see Fig. \ref{fig:figure15}).

\begin{figure}[!htb]
  \includegraphics[width=0.48\textwidth]{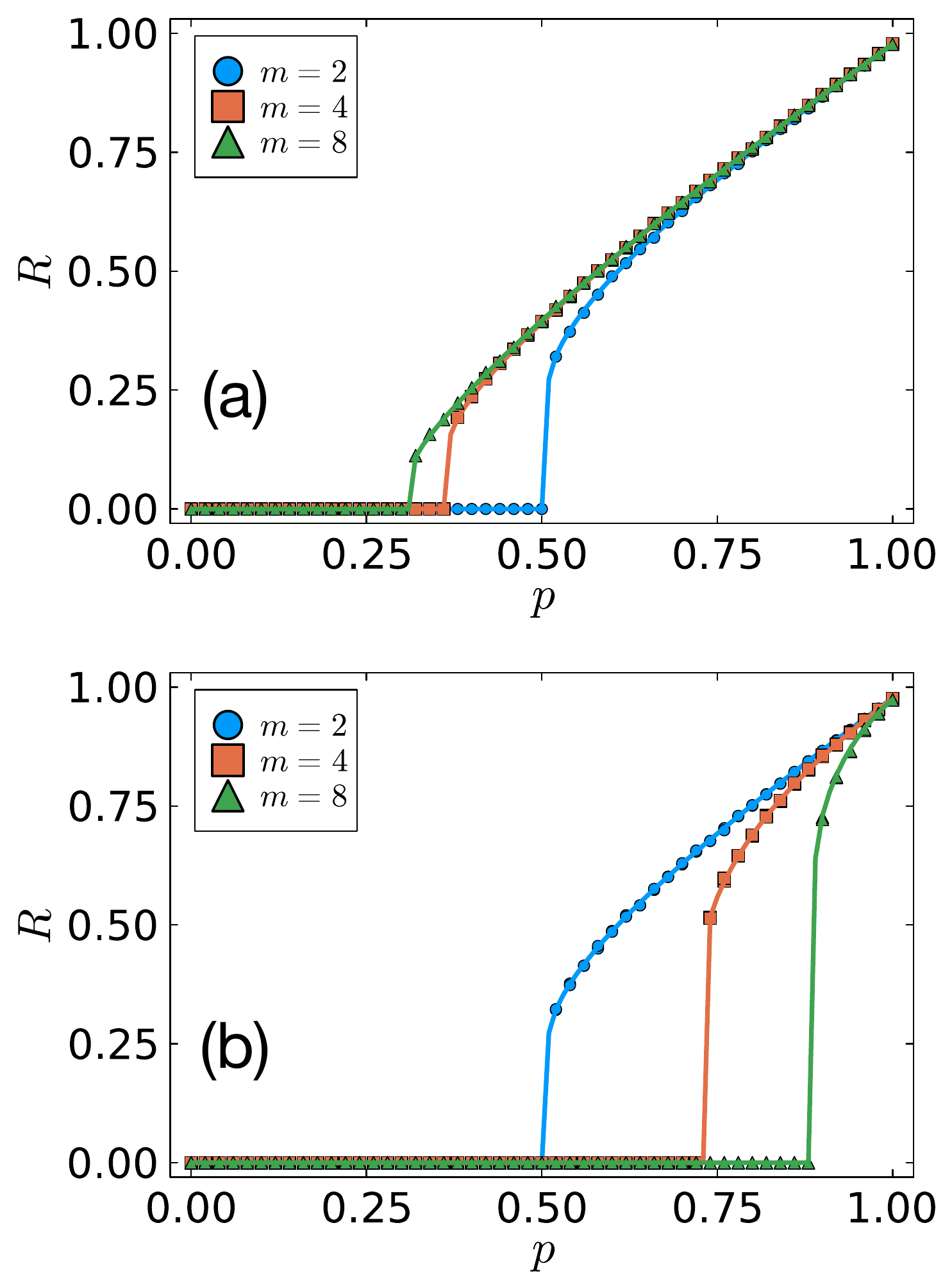}
  \caption{Phase diagrams of HONDP in case (a) implementing simple node hyperedge percolation (panel (a)) and in case (b) implementing cooperative node hyperedge percolation (panel (b)) for hypergraphs with hyperedges of fixed cardinality $m$ and in the absence of negative node regulations. The numerical predictions (solid lines) obtained from Eq. (\ref{eq:simple_node_regulation}) and Eq. (\ref{eq:cooperative_node_regulation})) are compared with Monte Carlo simulations (symbols). In both cases we observe a discontinuous hybrid transition, however in case (a) the percolation threshold decreases with increasing values of $m$ while in case (b) it increases. This demonstrates that while in case (a) HODNP the hypergraphs become more robust by increasing the value of $m$ in case (b) HODNP the hypergraphs become more fragile.   The considered structural hypergraph has $N=10^5$ nodes and a Poisson hyperdegree distribution $P(k)$ with an average hyperdegree $c=6$. The  Poisson degree distribution $\hat{P}^{+}(\kappa)$ has average positive regulatory degree $c^{+}=4$ while the  negative regulations are absent.  }
  \label{fig:figure16}
 \end{figure}

Secondly, we compare the properties of case (a) and case (b) HONDP and we observe that when the hyperedge cardinality is constant, and is given by $m$, these two models have different responses to the increase on $m$.  Both percolation processes display a discontinuous phase transition, but the critical threshold of monotonically decreasing with $m$ in case (a) where Step 1 is defined in terms of the hyperedge percolation process, while is an increasing function of $m$ in case (b) where Step 2 is defined in terms of the cooperative hyperedge percolation process (see Figure \ref{fig:figure16}). This different behavior is inherently related to the different nature of the two models as the rules of cooperative hyperedge percolation determine the increased fragility of hypergraphs with hyperedges of larger cardinality.

\section{Conclusion}

In this study, we propose a comprehensive theoretical framework that merges percolation theory with nonlinear dynamics to examine hypergraphs featuring a time-varying giant component.
In higher-order triadic percolation (HOTP) the dynamics is orchestrated by higher-order triadic interactions (HOTIs) that can upregulate or downregulate hyperedges turning percolation into a fully-fledged dynamical process. Specifically,  the order parameter of HOTP, indicating the fraction of nodes in the giant component, undergoes, in the most general scenario, period doubling and a route to chaos. Here we provide an in-depth study of the critical behavior of this model characterizing the emergence of discontinuous transition when only positive HOTIs are allowed; the emergence of both continuous and discontinuous phase transitions connected by tricritical points when only negative HOTIs are allowed; and proving that, in the presence of both positive and negative HOTIs, the order parameter of HOTP undergoes a route to chaos in the universality class of the logistic map.

Higher-order triadic interactions, however, can also be hierarchically nested, giving rise to hierarchical higher-order triadic interactions (HHOTIs) that have been observed for instance in ecological networks. In this scenario, it is important to consider a relevant variation of HOTP that we call hierarchical higher-order triadic percolation (HHOTP). Due to the improved combinatorial complexity of the regulatory interactions allowed in this scenario, we observe that the critical properties of HHOTP can be significantly more complex than those of HOTP. Among the most relevant differences between the two models, we highlight two of them. First, HHOTP can display a route to chaos of the order parameter also when only negative HHOTIs are allowed. Secondly, the route to chaos associated with HHOTP is not generally in the universality class of the logistic map.

We conclude our work by presenting two other generalizations of HOTP: interdependent higher-order triadic percolation (IHOTI) and higher-order node dynamic percolation (HONDP).

The underlying percolation process in IHOTI is the cooperative hypergraph percolation which already displays a discontinuous percolation transition in the absence of HOTIs. This impedes the route to chaos of the order parameter for certain parameter values, where only stationary states of period-2 oscillations of the order parameter can be observed.

In HONDP  we do not have HOTIs regulating the hyperedges, but only interactions up-regulating or down-regulating the nodes. In this case hierarchical regulation is not possible,  we can consider however different underlying percolation processes: the hypergraph percolation with random deactivation of the nodes (case(a)); and cooperative hypergraph percolation (case (b)) which imposes that all the nodes of a hyperedge must be intact for a hyperedge to be intact. The dynamics of HONDP displays interesting effects of the regulatory interactions that are peculiar to the model and distinct from that observed when the hyperedges are regulated by HOTIs.

In conclusion, this work provides a comprehensive framework for studying hypergraphs with time-varying connectivity, revealing the complex interplay between higher-order network structures and their dynamics. The results might shed light on the dynamics of climate networks, and biological and brain networks.

\bibliographystyle{apsrev4-1}
\bibliography{references}

\begin{thebibliography}{54}%
\makeatletter
\providecommand \@ifxundefined [1]{%
 \@ifx{#1\undefined}
}%
\providecommand \@ifnum [1]{%
 \ifnum #1\expandafter \@firstoftwo
 \else \expandafter \@secondoftwo
 \fi
}%
\providecommand \@ifx [1]{%
 \ifx #1\expandafter \@firstoftwo
 \else \expandafter \@secondoftwo
 \fi
}%
\providecommand \natexlab [1]{#1}%
\providecommand \enquote  [1]{``#1''}%
\providecommand \bibnamefont  [1]{#1}%
\providecommand \bibfnamefont [1]{#1}%
\providecommand \citenamefont [1]{#1}%
\providecommand \href@noop [0]{\@secondoftwo}%
\providecommand \href [0]{\begingroup \@sanitize@url \@href}%
\providecommand \@href[1]{\@@startlink{#1}\@@href}%
\providecommand \@@href[1]{\endgroup#1\@@endlink}%
\providecommand \@sanitize@url [0]{\catcode `\\12\catcode `\$12\catcode `\&12\catcode `\#12\catcode `\^12\catcode `\_12\catcode `\%12\relax}%
\providecommand \@@startlink[1]{}%
\providecommand \@@endlink[0]{}%
\providecommand \url  [0]{\begingroup\@sanitize@url \@url }%
\providecommand \@url [1]{\endgroup\@href {#1}{\urlprefix }}%
\providecommand \urlprefix  [0]{URL }%
\providecommand \Eprint [0]{\href }%
\providecommand \doibase [0]{http://dx.doi.org/}%
\providecommand \selectlanguage [0]{\@gobble}%
\providecommand \bibinfo  [0]{\@secondoftwo}%
\providecommand \bibfield  [0]{\@secondoftwo}%
\providecommand \translation [1]{[#1]}%
\providecommand \BibitemOpen [0]{}%
\providecommand \bibitemStop [0]{}%
\providecommand \bibitemNoStop [0]{.\EOS\space}%
\providecommand \EOS [0]{\spacefactor3000\relax}%
\providecommand \BibitemShut  [1]{\csname bibitem#1\endcsname}%
\let\auto@bib@innerbib\@empty
\bibitem [{\citenamefont {Battiston}\ \emph {et~al.}(2021)\citenamefont {Battiston}, \citenamefont {Amico}, \citenamefont {Barrat}, \citenamefont {Bianconi}, \citenamefont {Ferraz~de Arruda}, \citenamefont {Franceschiello}, \citenamefont {Iacopini}, \citenamefont {K{\'e}fi}, \citenamefont {Latora}, \citenamefont {Moreno} \emph {et~al.}}]{battiston2021physics}%
  \BibitemOpen
  \bibfield  {author} {\bibinfo {author} {\bibfnamefont {F.}~\bibnamefont {Battiston}}, \bibinfo {author} {\bibfnamefont {E.}~\bibnamefont {Amico}}, \bibinfo {author} {\bibfnamefont {A.}~\bibnamefont {Barrat}}, \bibinfo {author} {\bibfnamefont {G.}~\bibnamefont {Bianconi}}, \bibinfo {author} {\bibfnamefont {G.}~\bibnamefont {Ferraz~de Arruda}}, \bibinfo {author} {\bibfnamefont {B.}~\bibnamefont {Franceschiello}}, \bibinfo {author} {\bibfnamefont {I.}~\bibnamefont {Iacopini}}, \bibinfo {author} {\bibfnamefont {S.}~\bibnamefont {K{\'e}fi}}, \bibinfo {author} {\bibfnamefont {V.}~\bibnamefont {Latora}}, \bibinfo {author} {\bibfnamefont {Y.}~\bibnamefont {Moreno}},  \emph {et~al.},\ }\href@noop {} {\bibfield  {journal} {\bibinfo  {journal} {Nature Physics}\ }\textbf {\bibinfo {volume} {17}},\ \bibinfo {pages} {1093} (\bibinfo {year} {2021})}\BibitemShut {NoStop}%
\bibitem [{\citenamefont {Bianconi}(2021)}]{bianconi2021higher}%
  \BibitemOpen
  \bibfield  {author} {\bibinfo {author} {\bibfnamefont {G.}~\bibnamefont {Bianconi}},\ }\href@noop {} {\emph {\bibinfo {title} {Higher-order networks: An Introduction to Simplicial Complexes}}}\ (\bibinfo  {publisher} {Cambridge University Press},\ \bibinfo {year} {2021})\BibitemShut {NoStop}%
\bibitem [{\citenamefont {Boccaletti}\ \emph {et~al.}(2023)\citenamefont {Boccaletti}, \citenamefont {De~Lellis}, \citenamefont {Del~Genio}, \citenamefont {Alfaro-Bittner}, \citenamefont {Criado}, \citenamefont {Jalan},\ and\ \citenamefont {Romance}}]{boccaletti2023structure}%
  \BibitemOpen
  \bibfield  {author} {\bibinfo {author} {\bibfnamefont {S.}~\bibnamefont {Boccaletti}}, \bibinfo {author} {\bibfnamefont {P.}~\bibnamefont {De~Lellis}}, \bibinfo {author} {\bibfnamefont {C.}~\bibnamefont {Del~Genio}}, \bibinfo {author} {\bibfnamefont {K.}~\bibnamefont {Alfaro-Bittner}}, \bibinfo {author} {\bibfnamefont {R.}~\bibnamefont {Criado}}, \bibinfo {author} {\bibfnamefont {S.}~\bibnamefont {Jalan}}, \ and\ \bibinfo {author} {\bibfnamefont {M.}~\bibnamefont {Romance}},\ }\href@noop {} {\bibfield  {journal} {\bibinfo  {journal} {Physics Reports}\ }\textbf {\bibinfo {volume} {1018}},\ \bibinfo {pages} {1} (\bibinfo {year} {2023})}\BibitemShut {NoStop}%
\bibitem [{\citenamefont {Battiston}\ \emph {et~al.}(2020)\citenamefont {Battiston}, \citenamefont {Cencetti}, \citenamefont {Iacopini}, \citenamefont {Latora}, \citenamefont {Lucas}, \citenamefont {Patania}, \citenamefont {Young},\ and\ \citenamefont {Petri}}]{battiston2020networks}%
  \BibitemOpen
  \bibfield  {author} {\bibinfo {author} {\bibfnamefont {F.}~\bibnamefont {Battiston}}, \bibinfo {author} {\bibfnamefont {G.}~\bibnamefont {Cencetti}}, \bibinfo {author} {\bibfnamefont {I.}~\bibnamefont {Iacopini}}, \bibinfo {author} {\bibfnamefont {V.}~\bibnamefont {Latora}}, \bibinfo {author} {\bibfnamefont {M.}~\bibnamefont {Lucas}}, \bibinfo {author} {\bibfnamefont {A.}~\bibnamefont {Patania}}, \bibinfo {author} {\bibfnamefont {J.-G.}\ \bibnamefont {Young}}, \ and\ \bibinfo {author} {\bibfnamefont {G.}~\bibnamefont {Petri}},\ }\href@noop {} {\bibfield  {journal} {\bibinfo  {journal} {Physics Reports}\ }\textbf {\bibinfo {volume} {874}},\ \bibinfo {pages} {1} (\bibinfo {year} {2020})}\BibitemShut {NoStop}%
\bibitem [{\citenamefont {Salnikov}\ \emph {et~al.}(2018)\citenamefont {Salnikov}, \citenamefont {Cassese},\ and\ \citenamefont {Lambiotte}}]{salnikov2018simplicial}%
  \BibitemOpen
  \bibfield  {author} {\bibinfo {author} {\bibfnamefont {V.}~\bibnamefont {Salnikov}}, \bibinfo {author} {\bibfnamefont {D.}~\bibnamefont {Cassese}}, \ and\ \bibinfo {author} {\bibfnamefont {R.}~\bibnamefont {Lambiotte}},\ }\href@noop {} {\bibfield  {journal} {\bibinfo  {journal} {European Journal of Physics}\ }\textbf {\bibinfo {volume} {40}},\ \bibinfo {pages} {014001} (\bibinfo {year} {2018})}\BibitemShut {NoStop}%
\bibitem [{\citenamefont {Torres}\ \emph {et~al.}(2021)\citenamefont {Torres}, \citenamefont {Blevins}, \citenamefont {Bassett},\ and\ \citenamefont {Eliassi-Rad}}]{torres2021and}%
  \BibitemOpen
  \bibfield  {author} {\bibinfo {author} {\bibfnamefont {L.}~\bibnamefont {Torres}}, \bibinfo {author} {\bibfnamefont {A.~S.}\ \bibnamefont {Blevins}}, \bibinfo {author} {\bibfnamefont {D.}~\bibnamefont {Bassett}}, \ and\ \bibinfo {author} {\bibfnamefont {T.}~\bibnamefont {Eliassi-Rad}},\ }\href@noop {} {\bibfield  {journal} {\bibinfo  {journal} {SIAM Review}\ }\textbf {\bibinfo {volume} {63}},\ \bibinfo {pages} {435} (\bibinfo {year} {2021})}\BibitemShut {NoStop}%
\bibitem [{\citenamefont {Majhi}\ \emph {et~al.}(2022)\citenamefont {Majhi}, \citenamefont {Perc},\ and\ \citenamefont {Ghosh}}]{majhi2022dynamics}%
  \BibitemOpen
  \bibfield  {author} {\bibinfo {author} {\bibfnamefont {S.}~\bibnamefont {Majhi}}, \bibinfo {author} {\bibfnamefont {M.}~\bibnamefont {Perc}}, \ and\ \bibinfo {author} {\bibfnamefont {D.}~\bibnamefont {Ghosh}},\ }\href@noop {} {\bibfield  {journal} {\bibinfo  {journal} {Journal of the Royal Society Interface}\ }\textbf {\bibinfo {volume} {19}},\ \bibinfo {pages} {20220043} (\bibinfo {year} {2022})}\BibitemShut {NoStop}%
\bibitem [{\citenamefont {Mill{\'a}n}\ \emph {et~al.}(2020)\citenamefont {Mill{\'a}n}, \citenamefont {Torres},\ and\ \citenamefont {Bianconi}}]{millan2020explosive}%
  \BibitemOpen
  \bibfield  {author} {\bibinfo {author} {\bibfnamefont {A.~P.}\ \bibnamefont {Mill{\'a}n}}, \bibinfo {author} {\bibfnamefont {J.~J.}\ \bibnamefont {Torres}}, \ and\ \bibinfo {author} {\bibfnamefont {G.}~\bibnamefont {Bianconi}},\ }\href@noop {} {\bibfield  {journal} {\bibinfo  {journal} {Physical Review Letters}\ }\textbf {\bibinfo {volume} {124}},\ \bibinfo {pages} {218301} (\bibinfo {year} {2020})}\BibitemShut {NoStop}%
\bibitem [{\citenamefont {Ghorbanchian}\ \emph {et~al.}(2021)\citenamefont {Ghorbanchian}, \citenamefont {Restrepo}, \citenamefont {Torres},\ and\ \citenamefont {Bianconi}}]{ghorbanchian2021higher}%
  \BibitemOpen
  \bibfield  {author} {\bibinfo {author} {\bibfnamefont {R.}~\bibnamefont {Ghorbanchian}}, \bibinfo {author} {\bibfnamefont {J.~G.}\ \bibnamefont {Restrepo}}, \bibinfo {author} {\bibfnamefont {J.~J.}\ \bibnamefont {Torres}}, \ and\ \bibinfo {author} {\bibfnamefont {G.}~\bibnamefont {Bianconi}},\ }\href@noop {} {\bibfield  {journal} {\bibinfo  {journal} {Communications Physics}\ }\textbf {\bibinfo {volume} {4}},\ \bibinfo {pages} {120} (\bibinfo {year} {2021})}\BibitemShut {NoStop}%
\bibitem [{\citenamefont {Skardal}\ and\ \citenamefont {Arenas}(2019)}]{skardal2019abrupt}%
  \BibitemOpen
  \bibfield  {author} {\bibinfo {author} {\bibfnamefont {P.~S.}\ \bibnamefont {Skardal}}\ and\ \bibinfo {author} {\bibfnamefont {A.}~\bibnamefont {Arenas}},\ }\href@noop {} {\bibfield  {journal} {\bibinfo  {journal} {Physical Review Letters}\ }\textbf {\bibinfo {volume} {122}},\ \bibinfo {pages} {248301} (\bibinfo {year} {2019})}\BibitemShut {NoStop}%
\bibitem [{\citenamefont {Zhang}\ \emph {et~al.}(2021)\citenamefont {Zhang}, \citenamefont {Latora},\ and\ \citenamefont {Motter}}]{zhang2021unified}%
  \BibitemOpen
  \bibfield  {author} {\bibinfo {author} {\bibfnamefont {Y.}~\bibnamefont {Zhang}}, \bibinfo {author} {\bibfnamefont {V.}~\bibnamefont {Latora}}, \ and\ \bibinfo {author} {\bibfnamefont {A.~E.}\ \bibnamefont {Motter}},\ }\href@noop {} {\bibfield  {journal} {\bibinfo  {journal} {Communications Physics}\ }\textbf {\bibinfo {volume} {4}},\ \bibinfo {pages} {1} (\bibinfo {year} {2021})}\BibitemShut {NoStop}%
\bibitem [{\citenamefont {Gambuzza}\ \emph {et~al.}(2016)\citenamefont {Gambuzza}, \citenamefont {Frasca}, \citenamefont {Fortuna},\ and\ \citenamefont {Boccaletti}}]{gambuzza2016inhomogeneity}%
  \BibitemOpen
  \bibfield  {author} {\bibinfo {author} {\bibfnamefont {L.~V.}\ \bibnamefont {Gambuzza}}, \bibinfo {author} {\bibfnamefont {M.}~\bibnamefont {Frasca}}, \bibinfo {author} {\bibfnamefont {L.}~\bibnamefont {Fortuna}}, \ and\ \bibinfo {author} {\bibfnamefont {S.}~\bibnamefont {Boccaletti}},\ }\href@noop {} {\bibfield  {journal} {\bibinfo  {journal} {Physical Review E}\ }\textbf {\bibinfo {volume} {93}},\ \bibinfo {pages} {042203} (\bibinfo {year} {2016})}\BibitemShut {NoStop}%
\bibitem [{\citenamefont {Sun}\ \emph {et~al.}(2023)\citenamefont {Sun}, \citenamefont {Radicchi}, \citenamefont {Kurths},\ and\ \citenamefont {Bianconi}}]{sun2023dynamic}%
  \BibitemOpen
  \bibfield  {author} {\bibinfo {author} {\bibfnamefont {H.}~\bibnamefont {Sun}}, \bibinfo {author} {\bibfnamefont {F.}~\bibnamefont {Radicchi}}, \bibinfo {author} {\bibfnamefont {J.}~\bibnamefont {Kurths}}, \ and\ \bibinfo {author} {\bibfnamefont {G.}~\bibnamefont {Bianconi}},\ }\href@noop {} {\bibfield  {journal} {\bibinfo  {journal} {Nature Communications}\ }\textbf {\bibinfo {volume} {14}},\ \bibinfo {pages} {1308} (\bibinfo {year} {2023})}\BibitemShut {NoStop}%
\bibitem [{\citenamefont {Sun}\ and\ \citenamefont {Bianconi}(2021)}]{sun2021higher}%
  \BibitemOpen
  \bibfield  {author} {\bibinfo {author} {\bibfnamefont {H.}~\bibnamefont {Sun}}\ and\ \bibinfo {author} {\bibfnamefont {G.}~\bibnamefont {Bianconi}},\ }\href@noop {} {\bibfield  {journal} {\bibinfo  {journal} {Physical Review E}\ }\textbf {\bibinfo {volume} {104}},\ \bibinfo {pages} {034306} (\bibinfo {year} {2021})}\BibitemShut {NoStop}%
\bibitem [{\citenamefont {Liu}\ \emph {et~al.}(2023)\citenamefont {Liu}, \citenamefont {Chu},\ and\ \citenamefont {Meng}}]{liu2023higher}%
  \BibitemOpen
  \bibfield  {author} {\bibinfo {author} {\bibfnamefont {R.-R.}\ \bibnamefont {Liu}}, \bibinfo {author} {\bibfnamefont {C.}~\bibnamefont {Chu}}, \ and\ \bibinfo {author} {\bibfnamefont {F.}~\bibnamefont {Meng}},\ }\href@noop {} {\bibfield  {journal} {\bibinfo  {journal} {Chaos, Solitons \& Fractals}\ }\textbf {\bibinfo {volume} {177}},\ \bibinfo {pages} {114246} (\bibinfo {year} {2023})}\BibitemShut {NoStop}%
\bibitem [{\citenamefont {Kim}\ and\ \citenamefont {Goh}(2024)}]{kim2024higher}%
  \BibitemOpen
  \bibfield  {author} {\bibinfo {author} {\bibfnamefont {J.-H.}\ \bibnamefont {Kim}}\ and\ \bibinfo {author} {\bibfnamefont {K.-I.}\ \bibnamefont {Goh}},\ }\href@noop {} {\bibfield  {journal} {\bibinfo  {journal} {Physical Review Letters}\ }\textbf {\bibinfo {volume} {132}},\ \bibinfo {pages} {087401} (\bibinfo {year} {2024})}\BibitemShut {NoStop}%
\bibitem [{\citenamefont {Di~Gaetano}\ \emph {et~al.}(2024)\citenamefont {Di~Gaetano}, \citenamefont {Battiston},\ and\ \citenamefont {Starnini}}]{di2024percolation}%
  \BibitemOpen
  \bibfield  {author} {\bibinfo {author} {\bibfnamefont {L.}~\bibnamefont {Di~Gaetano}}, \bibinfo {author} {\bibfnamefont {F.}~\bibnamefont {Battiston}}, \ and\ \bibinfo {author} {\bibfnamefont {M.}~\bibnamefont {Starnini}},\ }\href@noop {} {\bibfield  {journal} {\bibinfo  {journal} {Physical Review Letters}\ }\textbf {\bibinfo {volume} {132}},\ \bibinfo {pages} {037401} (\bibinfo {year} {2024})}\BibitemShut {NoStop}%
\bibitem [{\citenamefont {Bianconi}\ and\ \citenamefont {Ziff}(2018)}]{bianconi2018topological}%
  \BibitemOpen
  \bibfield  {author} {\bibinfo {author} {\bibfnamefont {G.}~\bibnamefont {Bianconi}}\ and\ \bibinfo {author} {\bibfnamefont {R.~M.}\ \bibnamefont {Ziff}},\ }\href@noop {} {\bibfield  {journal} {\bibinfo  {journal} {Physical Review E}\ }\textbf {\bibinfo {volume} {98}},\ \bibinfo {pages} {052308} (\bibinfo {year} {2018})}\BibitemShut {NoStop}%
\bibitem [{\citenamefont {Bianconi}\ and\ \citenamefont {Dorogovtsev}(2024)}]{bianconi2024theory}%
  \BibitemOpen
  \bibfield  {author} {\bibinfo {author} {\bibfnamefont {G.}~\bibnamefont {Bianconi}}\ and\ \bibinfo {author} {\bibfnamefont {S.~N.}\ \bibnamefont {Dorogovtsev}},\ }\href@noop {} {\bibfield  {journal} {\bibinfo  {journal} {Physical Review E}\ }\textbf {\bibinfo {volume} {109}},\ \bibinfo {pages} {014306} (\bibinfo {year} {2024})}\BibitemShut {NoStop}%
\bibitem [{\citenamefont {Pan}\ \emph {et~al.}(2024)\citenamefont {Pan}, \citenamefont {Zhou}, \citenamefont {Zhou}, \citenamefont {Boccaletti},\ and\ \citenamefont {Bonamassa}}]{pan2024robustness}%
  \BibitemOpen
  \bibfield  {author} {\bibinfo {author} {\bibfnamefont {X.}~\bibnamefont {Pan}}, \bibinfo {author} {\bibfnamefont {J.}~\bibnamefont {Zhou}}, \bibinfo {author} {\bibfnamefont {Y.}~\bibnamefont {Zhou}}, \bibinfo {author} {\bibfnamefont {S.}~\bibnamefont {Boccaletti}}, \ and\ \bibinfo {author} {\bibfnamefont {I.}~\bibnamefont {Bonamassa}},\ }\href@noop {} {\bibfield  {journal} {\bibinfo  {journal} {Physical Review Research}\ }\textbf {\bibinfo {volume} {6}},\ \bibinfo {pages} {013049} (\bibinfo {year} {2024})}\BibitemShut {NoStop}%
\bibitem [{\citenamefont {Chen}\ \emph {et~al.}(2024)\citenamefont {Chen}, \citenamefont {Jia}, \citenamefont {Liu},\ and\ \citenamefont {Meng}}]{chen2024cascading}%
  \BibitemOpen
  \bibfield  {author} {\bibinfo {author} {\bibfnamefont {L.}~\bibnamefont {Chen}}, \bibinfo {author} {\bibfnamefont {C.}~\bibnamefont {Jia}}, \bibinfo {author} {\bibfnamefont {R.-R.}\ \bibnamefont {Liu}}, \ and\ \bibinfo {author} {\bibfnamefont {F.}~\bibnamefont {Meng}},\ }\href@noop {} {\bibfield  {journal} {\bibinfo  {journal} {arXiv preprint arXiv:2408.01172}\ } (\bibinfo {year} {2024})}\BibitemShut {NoStop}%
\bibitem [{\citenamefont {Landry}\ and\ \citenamefont {Restrepo}(2020)}]{landry2020effect}%
  \BibitemOpen
  \bibfield  {author} {\bibinfo {author} {\bibfnamefont {N.~W.}\ \bibnamefont {Landry}}\ and\ \bibinfo {author} {\bibfnamefont {J.~G.}\ \bibnamefont {Restrepo}},\ }\href@noop {} {\bibfield  {journal} {\bibinfo  {journal} {Chaos: An Interdisciplinary Journal of Nonlinear Science}\ }\textbf {\bibinfo {volume} {30}} (\bibinfo {year} {2020})}\BibitemShut {NoStop}%
\bibitem [{\citenamefont {St-Onge}\ \emph {et~al.}(2021)\citenamefont {St-Onge}, \citenamefont {Sun}, \citenamefont {Allard}, \citenamefont {H{\'e}bert-Dufresne},\ and\ \citenamefont {Bianconi}}]{st2021universal}%
  \BibitemOpen
  \bibfield  {author} {\bibinfo {author} {\bibfnamefont {G.}~\bibnamefont {St-Onge}}, \bibinfo {author} {\bibfnamefont {H.}~\bibnamefont {Sun}}, \bibinfo {author} {\bibfnamefont {A.}~\bibnamefont {Allard}}, \bibinfo {author} {\bibfnamefont {L.}~\bibnamefont {H{\'e}bert-Dufresne}}, \ and\ \bibinfo {author} {\bibfnamefont {G.}~\bibnamefont {Bianconi}},\ }\href@noop {} {\bibfield  {journal} {\bibinfo  {journal} {Physical Review Letters}\ }\textbf {\bibinfo {volume} {127}},\ \bibinfo {pages} {158301} (\bibinfo {year} {2021})}\BibitemShut {NoStop}%
\bibitem [{\citenamefont {Iacopini}\ \emph {et~al.}(2019)\citenamefont {Iacopini}, \citenamefont {Petri}, \citenamefont {Barrat},\ and\ \citenamefont {Latora}}]{iacopini2019simplicial}%
  \BibitemOpen
  \bibfield  {author} {\bibinfo {author} {\bibfnamefont {I.}~\bibnamefont {Iacopini}}, \bibinfo {author} {\bibfnamefont {G.}~\bibnamefont {Petri}}, \bibinfo {author} {\bibfnamefont {A.}~\bibnamefont {Barrat}}, \ and\ \bibinfo {author} {\bibfnamefont {V.}~\bibnamefont {Latora}},\ }\href@noop {} {\bibfield  {journal} {\bibinfo  {journal} {Nature Communications}\ }\textbf {\bibinfo {volume} {10}},\ \bibinfo {pages} {1} (\bibinfo {year} {2019})}\BibitemShut {NoStop}%
\bibitem [{\citenamefont {Carletti}\ \emph {et~al.}(2020)\citenamefont {Carletti}, \citenamefont {Battiston}, \citenamefont {Cencetti},\ and\ \citenamefont {Fanelli}}]{carletti2020random}%
  \BibitemOpen
  \bibfield  {author} {\bibinfo {author} {\bibfnamefont {T.}~\bibnamefont {Carletti}}, \bibinfo {author} {\bibfnamefont {F.}~\bibnamefont {Battiston}}, \bibinfo {author} {\bibfnamefont {G.}~\bibnamefont {Cencetti}}, \ and\ \bibinfo {author} {\bibfnamefont {D.}~\bibnamefont {Fanelli}},\ }\href@noop {} {\bibfield  {journal} {\bibinfo  {journal} {Physical Review E}\ }\textbf {\bibinfo {volume} {101}},\ \bibinfo {pages} {022308} (\bibinfo {year} {2020})}\BibitemShut {NoStop}%
\bibitem [{\citenamefont {Torres}\ and\ \citenamefont {Bianconi}(2020)}]{torres2020simplicial}%
  \BibitemOpen
  \bibfield  {author} {\bibinfo {author} {\bibfnamefont {J.~J.}\ \bibnamefont {Torres}}\ and\ \bibinfo {author} {\bibfnamefont {G.}~\bibnamefont {Bianconi}},\ }\href@noop {} {\bibfield  {journal} {\bibinfo  {journal} {Journal of Physics: Complexity}\ }\textbf {\bibinfo {volume} {1}},\ \bibinfo {pages} {015002} (\bibinfo {year} {2020})}\BibitemShut {NoStop}%
\bibitem [{\citenamefont {Cho}\ \emph {et~al.}(2016)\citenamefont {Cho}, \citenamefont {Barcelon},\ and\ \citenamefont {Lee}}]{cho2016optogenetic}%
  \BibitemOpen
  \bibfield  {author} {\bibinfo {author} {\bibfnamefont {W.-H.}\ \bibnamefont {Cho}}, \bibinfo {author} {\bibfnamefont {E.}~\bibnamefont {Barcelon}}, \ and\ \bibinfo {author} {\bibfnamefont {S.~J.}\ \bibnamefont {Lee}},\ }\href@noop {} {\bibfield  {journal} {\bibinfo  {journal} {Experimental neurobiology}\ }\textbf {\bibinfo {volume} {25}},\ \bibinfo {pages} {197} (\bibinfo {year} {2016})}\BibitemShut {NoStop}%
\bibitem [{\citenamefont {Grilli}\ \emph {et~al.}(2017)\citenamefont {Grilli}, \citenamefont {Barab{\'a}s}, \citenamefont {Michalska-Smith},\ and\ \citenamefont {Allesina}}]{grilli2017higher}%
  \BibitemOpen
  \bibfield  {author} {\bibinfo {author} {\bibfnamefont {J.}~\bibnamefont {Grilli}}, \bibinfo {author} {\bibfnamefont {G.}~\bibnamefont {Barab{\'a}s}}, \bibinfo {author} {\bibfnamefont {M.~J.}\ \bibnamefont {Michalska-Smith}}, \ and\ \bibinfo {author} {\bibfnamefont {S.}~\bibnamefont {Allesina}},\ }\href@noop {} {\bibfield  {journal} {\bibinfo  {journal} {Nature}\ }\textbf {\bibinfo {volume} {548}},\ \bibinfo {pages} {210} (\bibinfo {year} {2017})}\BibitemShut {NoStop}%
\bibitem [{\citenamefont {Bairey}\ \emph {et~al.}(2016)\citenamefont {Bairey}, \citenamefont {Kelsic},\ and\ \citenamefont {Kishony}}]{bairey2016high}%
  \BibitemOpen
  \bibfield  {author} {\bibinfo {author} {\bibfnamefont {E.}~\bibnamefont {Bairey}}, \bibinfo {author} {\bibfnamefont {E.~D.}\ \bibnamefont {Kelsic}}, \ and\ \bibinfo {author} {\bibfnamefont {R.}~\bibnamefont {Kishony}},\ }\href@noop {} {\bibfield  {journal} {\bibinfo  {journal} {Nature Communications}\ }\textbf {\bibinfo {volume} {7}},\ \bibinfo {pages} {12285} (\bibinfo {year} {2016})}\BibitemShut {NoStop}%
\bibitem [{\citenamefont {Chen}\ \emph {et~al.}(2023)\citenamefont {Chen}, \citenamefont {Liao},\ and\ \citenamefont {Liu}}]{chen2023teasing}%
  \BibitemOpen
  \bibfield  {author} {\bibinfo {author} {\bibfnamefont {C.}~\bibnamefont {Chen}}, \bibinfo {author} {\bibfnamefont {C.}~\bibnamefont {Liao}}, \ and\ \bibinfo {author} {\bibfnamefont {Y.-Y.}\ \bibnamefont {Liu}},\ }\href@noop {} {\bibfield  {journal} {\bibinfo  {journal} {Nature Communications}\ }\textbf {\bibinfo {volume} {14}},\ \bibinfo {pages} {2375} (\bibinfo {year} {2023})}\BibitemShut {NoStop}%
\bibitem [{\citenamefont {Boers}\ \emph {et~al.}(2019)\citenamefont {Boers}, \citenamefont {Goswami}, \citenamefont {Rheinwalt}, \citenamefont {Bookhagen}, \citenamefont {Hoskins},\ and\ \citenamefont {Kurths}}]{boers2019complex}%
  \BibitemOpen
  \bibfield  {author} {\bibinfo {author} {\bibfnamefont {N.}~\bibnamefont {Boers}}, \bibinfo {author} {\bibfnamefont {B.}~\bibnamefont {Goswami}}, \bibinfo {author} {\bibfnamefont {A.}~\bibnamefont {Rheinwalt}}, \bibinfo {author} {\bibfnamefont {B.}~\bibnamefont {Bookhagen}}, \bibinfo {author} {\bibfnamefont {B.}~\bibnamefont {Hoskins}}, \ and\ \bibinfo {author} {\bibfnamefont {J.}~\bibnamefont {Kurths}},\ }\href@noop {} {\bibfield  {journal} {\bibinfo  {journal} {Nature}\ }\textbf {\bibinfo {volume} {566}},\ \bibinfo {pages} {373} (\bibinfo {year} {2019})}\BibitemShut {NoStop}%
\bibitem [{\citenamefont {Mill{\'a}n}\ \emph {et~al.}(2023)\citenamefont {Mill{\'a}n}, \citenamefont {Sun}, \citenamefont {Torres},\ and\ \citenamefont {Bianconi}}]{millan2023triadic}%
  \BibitemOpen
  \bibfield  {author} {\bibinfo {author} {\bibfnamefont {A.~P.}\ \bibnamefont {Mill{\'a}n}}, \bibinfo {author} {\bibfnamefont {H.}~\bibnamefont {Sun}}, \bibinfo {author} {\bibfnamefont {J.~J.}\ \bibnamefont {Torres}}, \ and\ \bibinfo {author} {\bibfnamefont {G.}~\bibnamefont {Bianconi}},\ }\href@noop {} {\bibfield  {journal} {\bibinfo  {journal} {arXiv preprint arXiv:2311.14877}\ } (\bibinfo {year} {2023})}\BibitemShut {NoStop}%
\bibitem [{\citenamefont {Kozachkov}\ \emph {et~al.}(2023)\citenamefont {Kozachkov}, \citenamefont {Slotine},\ and\ \citenamefont {Krotov}}]{kozachkov2023neuron}%
  \BibitemOpen
  \bibfield  {author} {\bibinfo {author} {\bibfnamefont {L.}~\bibnamefont {Kozachkov}}, \bibinfo {author} {\bibfnamefont {J.-J.}\ \bibnamefont {Slotine}}, \ and\ \bibinfo {author} {\bibfnamefont {D.}~\bibnamefont {Krotov}},\ }\href@noop {} {\bibfield  {journal} {\bibinfo  {journal} {arXiv preprint arXiv:2311.08135}\ } (\bibinfo {year} {2023})}\BibitemShut {NoStop}%
\bibitem [{\citenamefont {Herron}\ \emph {et~al.}(2023)\citenamefont {Herron}, \citenamefont {Sartori},\ and\ \citenamefont {Xue}}]{herron2023robust}%
  \BibitemOpen
  \bibfield  {author} {\bibinfo {author} {\bibfnamefont {L.}~\bibnamefont {Herron}}, \bibinfo {author} {\bibfnamefont {P.}~\bibnamefont {Sartori}}, \ and\ \bibinfo {author} {\bibfnamefont {B.}~\bibnamefont {Xue}},\ }\href@noop {} {\bibfield  {journal} {\bibinfo  {journal} {PRX Life}\ }\textbf {\bibinfo {volume} {1}},\ \bibinfo {pages} {023012} (\bibinfo {year} {2023})}\BibitemShut {NoStop}%
\bibitem [{\citenamefont {Nicoletti}\ and\ \citenamefont {Busiello}(2024)}]{nicoletti2024information}%
  \BibitemOpen
  \bibfield  {author} {\bibinfo {author} {\bibfnamefont {G.}~\bibnamefont {Nicoletti}}\ and\ \bibinfo {author} {\bibfnamefont {D.~M.}\ \bibnamefont {Busiello}},\ }\href@noop {} {\bibfield  {journal} {\bibinfo  {journal} {Physical Review X}\ }\textbf {\bibinfo {volume} {14}},\ \bibinfo {pages} {021007} (\bibinfo {year} {2024})}\BibitemShut {NoStop}%
\bibitem [{\citenamefont {Baptista}\ \emph {et~al.}(2024)\citenamefont {Baptista}, \citenamefont {Niedostatek}, \citenamefont {Yamamoto}, \citenamefont {MacArthur}, \citenamefont {Kurths}, \citenamefont {Garcia},\ and\ \citenamefont {Bianconi}}]{baptista2024mining}%
  \BibitemOpen
  \bibfield  {author} {\bibinfo {author} {\bibfnamefont {A.}~\bibnamefont {Baptista}}, \bibinfo {author} {\bibfnamefont {M.}~\bibnamefont {Niedostatek}}, \bibinfo {author} {\bibfnamefont {J.}~\bibnamefont {Yamamoto}}, \bibinfo {author} {\bibfnamefont {B.}~\bibnamefont {MacArthur}}, \bibinfo {author} {\bibfnamefont {J.}~\bibnamefont {Kurths}}, \bibinfo {author} {\bibfnamefont {R.~S.}\ \bibnamefont {Garcia}}, \ and\ \bibinfo {author} {\bibfnamefont {G.}~\bibnamefont {Bianconi}},\ }\href@noop {} {\bibfield  {journal} {\bibinfo  {journal} {arXiv preprint arXiv:2404.14997}\ } (\bibinfo {year} {2024})}\BibitemShut {NoStop}%
\bibitem [{\citenamefont {Jost}\ and\ \citenamefont {Mulas}(2019)}]{jost2019hypergraph}%
  \BibitemOpen
  \bibfield  {author} {\bibinfo {author} {\bibfnamefont {J.}~\bibnamefont {Jost}}\ and\ \bibinfo {author} {\bibfnamefont {R.}~\bibnamefont {Mulas}},\ }\href@noop {} {\bibfield  {journal} {\bibinfo  {journal} {Advances in mathematics}\ }\textbf {\bibinfo {volume} {351}},\ \bibinfo {pages} {870} (\bibinfo {year} {2019})}\BibitemShut {NoStop}%
\bibitem [{\citenamefont {Dorogovtsev}\ \emph {et~al.}(2008)\citenamefont {Dorogovtsev}, \citenamefont {Goltsev},\ and\ \citenamefont {Mendes}}]{dorogovtsev2008critical}%
  \BibitemOpen
  \bibfield  {author} {\bibinfo {author} {\bibfnamefont {S.~N.}\ \bibnamefont {Dorogovtsev}}, \bibinfo {author} {\bibfnamefont {A.~V.}\ \bibnamefont {Goltsev}}, \ and\ \bibinfo {author} {\bibfnamefont {J.~F.}\ \bibnamefont {Mendes}},\ }\href@noop {} {\bibfield  {journal} {\bibinfo  {journal} {Reviews of Modern Physics}\ }\textbf {\bibinfo {volume} {80}},\ \bibinfo {pages} {1275} (\bibinfo {year} {2008})}\BibitemShut {NoStop}%
\bibitem [{\citenamefont {Bianconi}(2018)}]{bianconi2018multilayer}%
  \BibitemOpen
  \bibfield  {author} {\bibinfo {author} {\bibfnamefont {G.}~\bibnamefont {Bianconi}},\ }\href@noop {} {\emph {\bibinfo {title} {Multilayer networks: structure and function}}}\ (\bibinfo  {publisher} {Oxford university press},\ \bibinfo {year} {2018})\BibitemShut {NoStop}%
\bibitem [{\citenamefont {Lee}\ \emph {et~al.}(2018)\citenamefont {Lee}, \citenamefont {Kahng}, \citenamefont {Cho}, \citenamefont {Goh},\ and\ \citenamefont {Lee}}]{lee2018recent}%
  \BibitemOpen
  \bibfield  {author} {\bibinfo {author} {\bibfnamefont {D.}~\bibnamefont {Lee}}, \bibinfo {author} {\bibfnamefont {B.}~\bibnamefont {Kahng}}, \bibinfo {author} {\bibfnamefont {Y.}~\bibnamefont {Cho}}, \bibinfo {author} {\bibfnamefont {K.-I.}\ \bibnamefont {Goh}}, \ and\ \bibinfo {author} {\bibfnamefont {D.-S.}\ \bibnamefont {Lee}},\ }\href@noop {} {\bibfield  {journal} {\bibinfo  {journal} {Journal of the Korean Physical Society}\ }\textbf {\bibinfo {volume} {73}},\ \bibinfo {pages} {152} (\bibinfo {year} {2018})}\BibitemShut {NoStop}%
\bibitem [{\citenamefont {Li}\ \emph {et~al.}(2021)\citenamefont {Li}, \citenamefont {Liu}, \citenamefont {L{\"u}}, \citenamefont {Hu}, \citenamefont {Xu},\ and\ \citenamefont {Zhang}}]{li2021percolation}%
  \BibitemOpen
  \bibfield  {author} {\bibinfo {author} {\bibfnamefont {M.}~\bibnamefont {Li}}, \bibinfo {author} {\bibfnamefont {R.-R.}\ \bibnamefont {Liu}}, \bibinfo {author} {\bibfnamefont {L.}~\bibnamefont {L{\"u}}}, \bibinfo {author} {\bibfnamefont {M.-B.}\ \bibnamefont {Hu}}, \bibinfo {author} {\bibfnamefont {S.}~\bibnamefont {Xu}}, \ and\ \bibinfo {author} {\bibfnamefont {Y.-C.}\ \bibnamefont {Zhang}},\ }\href@noop {} {\bibfield  {journal} {\bibinfo  {journal} {Physics Reports}\ }\textbf {\bibinfo {volume} {907}},\ \bibinfo {pages} {1} (\bibinfo {year} {2021})}\BibitemShut {NoStop}%
\bibitem [{\citenamefont {Artime}\ \emph {et~al.}(2024)\citenamefont {Artime}, \citenamefont {Grassia}, \citenamefont {De~Domenico}, \citenamefont {Gleeson}, \citenamefont {Makse}, \citenamefont {Mangioni}, \citenamefont {Perc},\ and\ \citenamefont {Radicchi}}]{artime2024robustness}%
  \BibitemOpen
  \bibfield  {author} {\bibinfo {author} {\bibfnamefont {O.}~\bibnamefont {Artime}}, \bibinfo {author} {\bibfnamefont {M.}~\bibnamefont {Grassia}}, \bibinfo {author} {\bibfnamefont {M.}~\bibnamefont {De~Domenico}}, \bibinfo {author} {\bibfnamefont {J.~P.}\ \bibnamefont {Gleeson}}, \bibinfo {author} {\bibfnamefont {H.~A.}\ \bibnamefont {Makse}}, \bibinfo {author} {\bibfnamefont {G.}~\bibnamefont {Mangioni}}, \bibinfo {author} {\bibfnamefont {M.}~\bibnamefont {Perc}}, \ and\ \bibinfo {author} {\bibfnamefont {F.}~\bibnamefont {Radicchi}},\ }\href@noop {} {\bibfield  {journal} {\bibinfo  {journal} {Nature Reviews Physics}\ }\textbf {\bibinfo {volume} {6}},\ \bibinfo {pages} {114} (\bibinfo {year} {2024})}\BibitemShut {NoStop}%
\bibitem [{\citenamefont {Dorogovtsev}\ \emph {et~al.}(2006)\citenamefont {Dorogovtsev}, \citenamefont {Goltsev},\ and\ \citenamefont {Mendes}}]{dorogovtsev2006k}%
  \BibitemOpen
  \bibfield  {author} {\bibinfo {author} {\bibfnamefont {S.~N.}\ \bibnamefont {Dorogovtsev}}, \bibinfo {author} {\bibfnamefont {A.~V.}\ \bibnamefont {Goltsev}}, \ and\ \bibinfo {author} {\bibfnamefont {J.~F.~F.}\ \bibnamefont {Mendes}},\ }\href@noop {} {\bibfield  {journal} {\bibinfo  {journal} {Physical Review Letters}\ }\textbf {\bibinfo {volume} {96}},\ \bibinfo {pages} {040601} (\bibinfo {year} {2006})}\BibitemShut {NoStop}%
\bibitem [{\citenamefont {Buldyrev}\ \emph {et~al.}(2010)\citenamefont {Buldyrev}, \citenamefont {Parshani}, \citenamefont {Paul}, \citenamefont {Stanley},\ and\ \citenamefont {Havlin}}]{buldyrev2010catastrophic}%
  \BibitemOpen
  \bibfield  {author} {\bibinfo {author} {\bibfnamefont {S.~V.}\ \bibnamefont {Buldyrev}}, \bibinfo {author} {\bibfnamefont {R.}~\bibnamefont {Parshani}}, \bibinfo {author} {\bibfnamefont {G.}~\bibnamefont {Paul}}, \bibinfo {author} {\bibfnamefont {H.~E.}\ \bibnamefont {Stanley}}, \ and\ \bibinfo {author} {\bibfnamefont {S.}~\bibnamefont {Havlin}},\ }\href@noop {} {\bibfield  {journal} {\bibinfo  {journal} {Nature}\ }\textbf {\bibinfo {volume} {464}},\ \bibinfo {pages} {1025} (\bibinfo {year} {2010})}\BibitemShut {NoStop}%
\bibitem [{\citenamefont {Baxter}\ \emph {et~al.}(2012)\citenamefont {Baxter}, \citenamefont {Dorogovtsev}, \citenamefont {Goltsev},\ and\ \citenamefont {Mendes}}]{baxter2012avalanche}%
  \BibitemOpen
  \bibfield  {author} {\bibinfo {author} {\bibfnamefont {G.}~\bibnamefont {Baxter}}, \bibinfo {author} {\bibfnamefont {S.}~\bibnamefont {Dorogovtsev}}, \bibinfo {author} {\bibfnamefont {A.}~\bibnamefont {Goltsev}}, \ and\ \bibinfo {author} {\bibfnamefont {J.}~\bibnamefont {Mendes}},\ }\href@noop {} {\bibfield  {journal} {\bibinfo  {journal} {Physical Review Letters}\ }\textbf {\bibinfo {volume} {109}},\ \bibinfo {pages} {248701} (\bibinfo {year} {2012})}\BibitemShut {NoStop}%
\bibitem [{\citenamefont {Shekhtman}\ and\ \citenamefont {Havlin}(2018)}]{shekhtman2018percolation}%
  \BibitemOpen
  \bibfield  {author} {\bibinfo {author} {\bibfnamefont {L.~M.}\ \bibnamefont {Shekhtman}}\ and\ \bibinfo {author} {\bibfnamefont {S.}~\bibnamefont {Havlin}},\ }\href@noop {} {\bibfield  {journal} {\bibinfo  {journal} {Physical Review E}\ }\textbf {\bibinfo {volume} {98}},\ \bibinfo {pages} {052305} (\bibinfo {year} {2018})}\BibitemShut {NoStop}%
\bibitem [{\citenamefont {Del~Ferraro}\ \emph {et~al.}(2018)\citenamefont {Del~Ferraro}, \citenamefont {Moreno}, \citenamefont {Min}, \citenamefont {Morone}, \citenamefont {P{\'e}rez-Ram{\'\i}rez}, \citenamefont {P{\'e}rez-Cervera}, \citenamefont {Parra}, \citenamefont {Holodny}, \citenamefont {Canals},\ and\ \citenamefont {Makse}}]{del2018finding}%
  \BibitemOpen
  \bibfield  {author} {\bibinfo {author} {\bibfnamefont {G.}~\bibnamefont {Del~Ferraro}}, \bibinfo {author} {\bibfnamefont {A.}~\bibnamefont {Moreno}}, \bibinfo {author} {\bibfnamefont {B.}~\bibnamefont {Min}}, \bibinfo {author} {\bibfnamefont {F.}~\bibnamefont {Morone}}, \bibinfo {author} {\bibfnamefont {{\'U}.}~\bibnamefont {P{\'e}rez-Ram{\'\i}rez}}, \bibinfo {author} {\bibfnamefont {L.}~\bibnamefont {P{\'e}rez-Cervera}}, \bibinfo {author} {\bibfnamefont {L.~C.}\ \bibnamefont {Parra}}, \bibinfo {author} {\bibfnamefont {A.}~\bibnamefont {Holodny}}, \bibinfo {author} {\bibfnamefont {S.}~\bibnamefont {Canals}}, \ and\ \bibinfo {author} {\bibfnamefont {H.~A.}\ \bibnamefont {Makse}},\ }\href@noop {} {\bibfield  {journal} {\bibinfo  {journal} {Nature Communications}\ }\textbf {\bibinfo {volume} {9}},\ \bibinfo {pages} {2274} (\bibinfo {year} {2018})}\BibitemShut {NoStop}%
\bibitem [{\citenamefont {Son}\ \emph {et~al.}(2012)\citenamefont {Son}, \citenamefont {Bizhani}, \citenamefont {Christensen}, \citenamefont {Grassberger},\ and\ \citenamefont {Paczuski}}]{son2012percolation}%
  \BibitemOpen
  \bibfield  {author} {\bibinfo {author} {\bibfnamefont {S.-W.}\ \bibnamefont {Son}}, \bibinfo {author} {\bibfnamefont {G.}~\bibnamefont {Bizhani}}, \bibinfo {author} {\bibfnamefont {C.}~\bibnamefont {Christensen}}, \bibinfo {author} {\bibfnamefont {P.}~\bibnamefont {Grassberger}}, \ and\ \bibinfo {author} {\bibfnamefont {M.}~\bibnamefont {Paczuski}},\ }\href@noop {} {\bibfield  {journal} {\bibinfo  {journal} {Europhysics Letters}\ }\textbf {\bibinfo {volume} {97}},\ \bibinfo {pages} {16006} (\bibinfo {year} {2012})}\BibitemShut {NoStop}%
\bibitem [{\citenamefont {Cellai}\ \emph {et~al.}(2013)\citenamefont {Cellai}, \citenamefont {L{\'o}pez}, \citenamefont {Zhou}, \citenamefont {Gleeson},\ and\ \citenamefont {Bianconi}}]{cellai2013percolation}%
  \BibitemOpen
  \bibfield  {author} {\bibinfo {author} {\bibfnamefont {D.}~\bibnamefont {Cellai}}, \bibinfo {author} {\bibfnamefont {E.}~\bibnamefont {L{\'o}pez}}, \bibinfo {author} {\bibfnamefont {J.}~\bibnamefont {Zhou}}, \bibinfo {author} {\bibfnamefont {J.~P.}\ \bibnamefont {Gleeson}}, \ and\ \bibinfo {author} {\bibfnamefont {G.}~\bibnamefont {Bianconi}},\ }\href@noop {} {\bibfield  {journal} {\bibinfo  {journal} {Physical Review E}\ }\textbf {\bibinfo {volume} {88}},\ \bibinfo {pages} {052811} (\bibinfo {year} {2013})}\BibitemShut {NoStop}%
\bibitem [{\citenamefont {Baxter}\ \emph {et~al.}(2016)\citenamefont {Baxter}, \citenamefont {Bianconi}, \citenamefont {da~Costa}, \citenamefont {Dorogovtsev},\ and\ \citenamefont {Mendes}}]{baxter2016correlated}%
  \BibitemOpen
  \bibfield  {author} {\bibinfo {author} {\bibfnamefont {G.~J.}\ \bibnamefont {Baxter}}, \bibinfo {author} {\bibfnamefont {G.}~\bibnamefont {Bianconi}}, \bibinfo {author} {\bibfnamefont {R.~A.}\ \bibnamefont {da~Costa}}, \bibinfo {author} {\bibfnamefont {S.~N.}\ \bibnamefont {Dorogovtsev}}, \ and\ \bibinfo {author} {\bibfnamefont {J.~F.}\ \bibnamefont {Mendes}},\ }\href@noop {} {\bibfield  {journal} {\bibinfo  {journal} {Physical Review E}\ }\textbf {\bibinfo {volume} {94}},\ \bibinfo {pages} {012303} (\bibinfo {year} {2016})}\BibitemShut {NoStop}%
\bibitem [{\citenamefont {Feigenbaum}(1978)}]{feigenbaum1978quantitative}%
  \BibitemOpen
  \bibfield  {author} {\bibinfo {author} {\bibfnamefont {M.~J.}\ \bibnamefont {Feigenbaum}},\ }\href@noop {} {\bibfield  {journal} {\bibinfo  {journal} {Journal of statistical physics}\ }\textbf {\bibinfo {volume} {19}},\ \bibinfo {pages} {25} (\bibinfo {year} {1978})}\BibitemShut {NoStop}%
\bibitem [{\citenamefont {Strogatz}(2018)}]{strogatz2018nonlinear}%
  \BibitemOpen
  \bibfield  {author} {\bibinfo {author} {\bibfnamefont {S.~H.}\ \bibnamefont {Strogatz}},\ }\href@noop {} {\emph {\bibinfo {title} {Nonlinear dynamics and chaos: with applications to physics, biology, chemistry, and engineering}}}\ (\bibinfo  {publisher} {CRC press},\ \bibinfo {year} {2018})\BibitemShut {NoStop}%
\bibitem [{\citenamefont {Amburg}\ \emph {et~al.}(2020)\citenamefont {Amburg}, \citenamefont {Veldt},\ and\ \citenamefont {Benson}}]{landry_2023}%
  \BibitemOpen
  \bibfield  {author} {\bibinfo {author} {\bibfnamefont {I.}~\bibnamefont {Amburg}}, \bibinfo {author} {\bibfnamefont {N.}~\bibnamefont {Veldt}}, \ and\ \bibinfo {author} {\bibfnamefont {A.}~\bibnamefont {Benson}},\ }in\ \href@noop {} {\emph {\bibinfo {booktitle} {Proceedings of The Web Conference 2020}}}\ (\bibinfo {year} {2020})\ pp.\ \bibinfo {pages} {706--717},\ \bibinfo {note} {data available at \url{https://doi.org/10.5281/zenodo.10157609}}\BibitemShut {NoStop}%
\bibitem [{\citenamefont {Cirigliano}\ \emph {et~al.}(2024)\citenamefont {Cirigliano}, \citenamefont {Castellano},\ and\ \citenamefont {Bianconi}}]{cirigliano2024general}%
  \BibitemOpen
  \bibfield  {author} {\bibinfo {author} {\bibfnamefont {L.}~\bibnamefont {Cirigliano}}, \bibinfo {author} {\bibfnamefont {C.}~\bibnamefont {Castellano}}, \ and\ \bibinfo {author} {\bibfnamefont {G.}~\bibnamefont {Bianconi}},\ }\href@noop {} {\bibfield  {journal} {\bibinfo  {journal} {Physical Review E}\ }\textbf {\bibinfo {volume} {110}},\ \bibinfo {pages} {034302} (\bibinfo {year} {2024})}\BibitemShut {NoStop}%
\end{thebibliography}%
\appendix
\section{Universality class of the route to chaos associated with HOTP}
\label{ApA}

In this Appendix our aim is to determine the universality class associated with the route to chaos associated with HOTP.
A classic result of chaos theory is that an unimodal continuous one-dimensional map undergoes a route to chaos in the universality class of the logistic map if it has a uniquely differentiable quadratic maximum \cite{feigenbaum1978quantitative,strogatz2018nonlinear}. 

The map $h_{m,p}(R)$ determining the dynamics of HOTP  is unimodal and displays a maximum at $R=\tilde{R}$ obeying  \bea\left. \frac{\partial h_{m,p}(R)}{ \partial R} \right|_{R=\tilde{R}}=0.\eea Here we want to show that the map   $h_{m,p}(R)$  around $R=\tilde{R}$ is quadratic,  thus implying that the order parameter $R$ of HOTP undergoes a route to chaos in the universality class of the logistic map.
Specifically our aim is to show that for $|\delta R|\ll 1$ we have
\bea
h_{m,p}(\tilde{R}+\delta R) - h_{m,p}(\tilde{R}) \simeq \frac{1}{2} h_{m,p}^{\prime\prime}(\tilde{R}) (\delta R)^2.
\eea
In order to derive this scaling behavior, let us write the derivative explicitly. Starting from the self-consistent equations 
\bea
R^{(t)}&=&1-G_0(1-\hat{S}^{(t)}) = F_1\left(\hat{S}^{(t)}\right), \nonumber\\
\hat{S}^{(t)} &=& p_H^{(t-1)}\left[1-G_{1,m}(1-S^{(t)})\right] = F_2\left(p_H^{(t-1)}, S^{(t)}\right), \nonumber\\
S^{(t)} &=& 1-G_1(1-\hat{S}^{(t)}) = F_3\left(\hat{S}^{(t)}\right), \nonumber \\
p_H^{(t)}&=&p G_0^-\left(1-R^{(t)}\right)\left(1-G_0^+(1-R^{(t)}\right)=F_4(R^{(t-1)}).\nonumber 
\eea

Differentiating  the  equations  for $\hat{S}^{(t)}$ and $p_H^{(t-1)}$
 we obtain
 \bea
\frac{d \hat{S}^{(t)}}{d R^{(t-1)}}&=&\frac{\partial F_2}{\partial p_H^{(t-1)}} \frac{\partial F_4}{\partial R^{(t-1)}} + \frac{\partial F_2}{\partial S^{(t)}} \frac{\partial S^{(t)}}{\partial R^{(t-1)}},\nonumber \\
\frac{\partial S^{(t)}}{\partial R^{(t-1)}}&=&\frac{\partial F_3}{\partial \hat{S}^{(t)}} \frac{\partial \hat{S}^{(t)}}{\partial R^{(t-1)}}.\nonumber
\eea
Thus we can express ${d \hat{S}^{(t)}}/{d R^{(t-1)}}$ as
\bea
\frac{d \hat{S}^{(t)}}{d R^{(t-1)}} = \frac{\partial F_2}{\partial p_H^{(t-1)}} \frac{\partial F_4}{\partial R^{(t-1)}}\left(1-\frac{\partial F_2}{\partial S^{(t)}} \frac{\partial F_3}{\partial \hat{S}^{(t)}}\right)^{-1}.\nonumber
\eea
We observe that the derivative $h_{m,p}^{\prime}(R^{(t-1)})$ can be expressed in terms of ${d \hat{S}^{(t)}}/{d R^{(t-1)}}$ as 
\bea
h_{m,p}^{\prime}(R^{(t-1)})=\frac{d {R}^{(t)}}{d R^{(t-1)}}=\frac{d F_1}{d \hat{S}^{(t)}}\frac{d \hat{S}^{(t)}}{d R^{(t-1)}}.
\eea
Thus we obtain
\bea
h_{m,p}^{\prime}(R^{(t-1)})&=&\frac{d F_1}{d \hat{S}^{(t)}}\frac{\partial F_2}{\partial p_H^{(t-1)}} \frac{\partial F_4}{\partial R^{(t-1)}}\nonumber \\ &&\times \left(1-\frac{\partial F_2}{\partial S^{(t)}} \frac{\partial F_3}{\partial \hat{S}^{(t)}}\right)^{-1}.
\eea
From this equation, using the fact that
\bea
\frac{\partial F_1}{\partial \hat{S}^{(t)}} &=& G_0^\prime\left(1-\hat{S}^{(t)}\right)>0, \label{dis}\\
\frac{\partial F_2}{\partial p_H^{(t-1)}}&=&\frac{1-G_{1,m}(1-S^{(t)})}{1-p_H^{(t-1)} G^\prime_{1,m}(1-S^{(t)}) G_1^\prime(1-\hat{S}^{(t)})}>0,\nonumber
\eea
it follows that the map $R^{(t)}=h_{m,p}\Big(R^{t-1}\Big)$ has its maximum, at $R=\tilde{R}$  where  we have 
\bea
\left.\frac{\partial F_4}{\partial R^{(t-1)}}\right|_{R^{(t-1)}=\tilde{R}}=0.\eea
This implies that the map $h_{m,p}(R)$ admits a maximum only if the equations implementing the regulatory step are not monotonic, i.e.   $p_H^{(t)}$ is not a monotonic function of $R^{(t-1)}$. Thus this implies that both positive and negative HOTIs need to be present with a non-vanishing density.
In order to prove that the maximum $R=\tilde{R}$ is quadratic, we consider the second derivative of the map
\bea
h^{\prime\prime}_{m,p}(R^{(t)}) &=& \frac{d^2 R^{(t)}}{d (R^{(t-1)})^2}\nonumber\\
&=& \frac{d F_1}{d \hat{S}^{(t)}} \frac{d^2 \hat{S}^{(t)}}{d (R^{(t-1)})^2}+ \frac{d^2 F_1}{d (\hat{S}^{(t)})^2} \left(\frac{d \hat{S}^{(t)}}{d R^{(t-1)}}\right)^2.\nonumber
\eea
Specifically we are interest in proving that the sign of $h^{\prime\prime}_{m,p}(R^{(t)})$ is negative for $R=\tilde{R}$.
Since, for $R=\tilde{R}$ we have ${\partial F_4}/{\partial R^{(t-1)}}={d \hat{S}^{(t)}}/{d R^{(t-1)}}=0$ we obtain
\bea
h^{\prime\prime}_{m,p}(R^{(t)}) = \frac{d F_1}{d \hat{S}^{(t)}}\frac{\partial F_2}{\partial p_H^{(t-1)}} \frac{\partial^2 F_4}{\partial (R^{(t-1)})^2}\left(1-\frac{\partial F_2}{\partial S^{(t)}} \frac{\partial F_3}{\partial \hat{S}^{(t)}}\right)^{-1},\nonumber 
\eea
where we have used that 
\bea
\frac{d^2 \hat{S}^{(t)}}{d (R^{(t-1)})^2} = \frac{\partial F_2}{\partial p_H^{(t-1)}} \frac{\partial^2 F_4}{\partial (R^{(t-1)})^2}\left(1-\frac{\partial F_2}{\partial S^{(t)}} \frac{\partial F_3}{\partial \hat{S}^{(t)}}\right)^{-1}.\nonumber
\eea

This equation implies that, given the inequalities (\ref{dis}), the sign of 
$h^{\prime\prime}_{m,p}(\tilde{R})$ depends only on the form of ${\partial^2 F_4}/{\partial (R^{(t-1)})^2}$. In particular,  the map $h_{m,p}(R)$ will always have a quadratic maximum. 
This demonstrates that the map $h_{m,p}(R)$ will always have a quadratic maximum, if $p_H^{(t)}=F_4(R^{(t-1)})$ displays a quadratic maximum.
Thus in such general hypotheses the order parameter $R=R^{(t)}$ of the HOTP displays a route to chaos in the universality class of the logistic map.
Note that this implies that the universality class of HOTP is independent of the hypergraph structure and that also changing the regulatory degree distribution might not affect this result as long as $p_H^{(t)}=F_4(R^{(t-1)})$ displays a quadratic maximum at a non-zero value of $R$.

In this work, we focus mostly on the relevant case in which the regulatory (positive and negative) degree distributions are Poisson distributions with average degrees $c^\pm$ respectively. This case satisfies the hypothesis for observing a route to chaos of the order parameter of HOTP in the universality class of the logistic map. In fact, we  have 
\bea
\left.\frac{\partial^2 p_H^{(t-1)}}{\partial (R^{(t-1)})^2}\right|_{R^{(t-1)}=\tilde{R}} = -p \tilde{c} c^{+} e^{-\tilde{c} \tilde{R}} < 0,
\eea
where $\tilde{c} = c^{-}+c^{+}$, thus
\bea
\left.h_{m,p}^{\prime\prime}(R)\right|_{R=\tilde{R}} < 0.
\eea
\begin{figure}[!htb!]
  \includegraphics[width=0.45\textwidth]{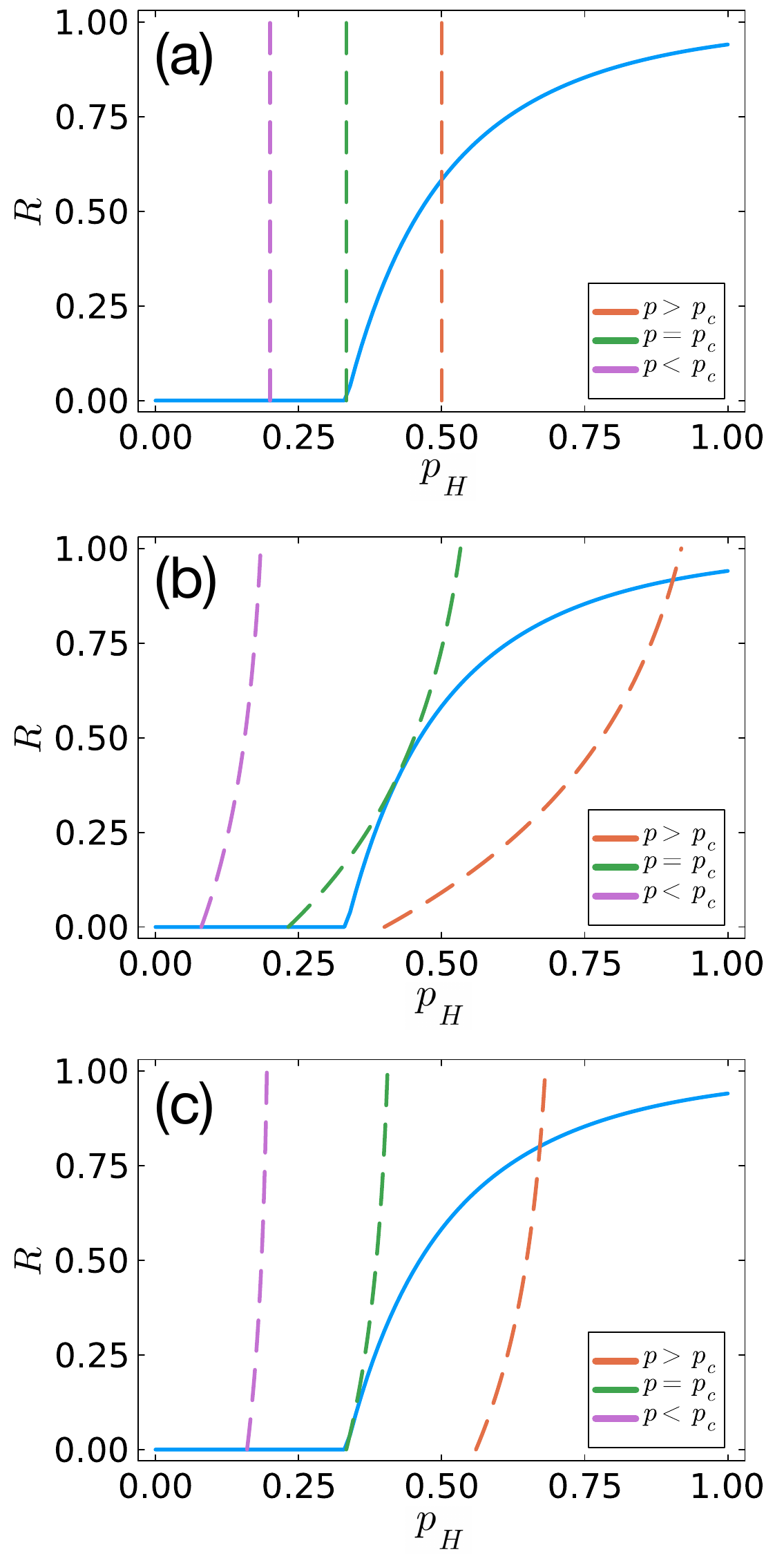}
  \caption{Illustration of the functions $f_{m}(p_H)$ (solid line) and $g_p(R)$ (dashed lines) determining the critical properties of HOTP when hyperedges are not regulated with probability $\rho_0$.   In panel (a) we consider values of $\rho_0$ above the tricritical point, where a continuous phase transition is observed.  In panel (b) we consider values of $\rho_0$ below the tricritical point where a discontinuous phase transition is observed. In panel (c) we consider values of $\rho_0$ at the tricritical point. In each panel, $p_c$ indicates the critical threshold for discontinuous transition (panel (a)) continuous transition (panel (b)), and tricritical phase transition (panel (c)). The structural hypergraph has fixed hyperedge cardinality $m=2$ and a Poisson hyperdegree distribution with an average $c=3$. The regulatory network is formed exclusively by positive regulations with a Poisson regulatory degree distribution with an average $c^+=2$.}
  \label{fig:hier_dependency}
 \end{figure}
\section{HOTP with partial regulation}
\label{ApB}
In the main text, we have shown that  HOTP in the absence of negative regulations can only display a discontinuous hybrid phase transition, while HOTP in the absence of positive regulation can display either continuous transition or period-2 oscillations. In order to reveal the nature of the difference between these two scenarios, we discuss here a general scenario of HOTP with partial regulation where hyperedges are not regulated with probability $\rho_0$. Finally, we discuss also the effect of partial regulation in HONDP.
\subsubsection{HOTP in the absence of negative regulations}
Let us consider HOTP where hyperedges are not regulated with probability $\rho_0$, i.e. the HOTIs are present with a probability $1-\rho_0$.  the probability of retaining a hyperedge $p_H^{(t)}$ at time $t$ follow the following generalization of Eq. (\ref{eq:hypergraph_triadic_pL}), 
\bea
p_H^{(t)} = p \left[\rho_0 + (1-\rho_0) \left(1-G_0^+\left(1-R^{(t)}\right)\right)\right].
\label{eq:gp_dependency}
\eea
When $\rho_0=1$, we recover ordinary hypergraph percolation with random hyperedge deactivation that displays a second-order continuous phase transition. When $\rho_0=0$, we recover the HOTP in the absence of negative regulation which displays a discontinuous hybrid phase transition. 

Let us investigate the general scenario when $0<\rho_0<1$. First, we observe that when $R=0$, $p_H=\rho_0 p$. Let us denote the percolation threshold for 
for simple hyperedge percolation as $p_H^c$ (the blue line in Fig. \ref{fig:hier_dependency}) and let us adopt the map notation for the dynamics, 
\bea
R^{(t)}=f_m\left(p_H^{(t-1)}\right), \quad p_H^{(t)}=g_p\left(R^{(t)}\right)
\eea
and
\bea
R^{(t)} = h_{m, p}\left(R^{(t-1)}\right) = f_m\left(g_p\left(R^{(t-1)}\right)\right).
\eea

If $\rho_0 p \geq p_H^c$, $R^{\star}=h_{m,p}(R^{\star})$ has a unique stable fixed point (see Fig. \ref{fig:hier_dependency} (a)) and if $\rho_0 p < p_H^c$, $R^{\star}=h_{m,p}(R^{\star})$ could have three fixed points, while the smallest ($R=0$) and the largest are stable (see Fig. \ref{fig:hier_dependency}(b)). Therefore, the discontinuous transition takes place when $y=h_{m,p}(R)$ is tangent to $y=R$ at $p_H=p_H^{\star}$, $R=R^{\star}$ (see Fig. \ref{fig:hier_dependency} (b) when $p=p_c$), i.e.
\bea
J = h_{m,p}^\prime(R^{\star}) = \left.\frac{\partial f_m}{\partial p_H^{(t)}}\right|_{p_H=p_H^{\star}} \left. \frac{\partial g_p}{\partial R^{(t)}}\right|_{R = R^{\star}}=1.
\eea
If $R^{\star}=h_{m,p}(R^{\star})$ always has a unique stable fixed point, the transition is always continuous. The continuous transition takes place when the fixed point decreases to zero at $p \rho_0 = p_H^c$. In addition, to guarantee the fixed point is stable, it is necessary to have
\bea
\left.\frac{\partial f_m}{\partial p_H^{(t)}}\right|_{p_H=p_H^c} \left. \frac{\partial g_p}{\partial R^{(t)}}\right|_{R = 0}<1.
\eea
Thus, the tricritical point separating continuous and discontinuous transition is reached when
\bea
\left.\frac{\partial f_m}{\partial p_H^{(t)}}\right|_{p_H=p_H^c} \left. \frac{\partial g_p}{\partial R^{(t)}}\right|_{R = 0}=1 ,\quad p \rho_0 = p_H^c.
\eea
To illustrate this critical behavior, let us assume that the structural hypergraph has a Poisson hyperdegree distribution with an average $c$ and the hyperedge cardinality $m$. Thus $p_H^c = 1/c(m-1)$ \cite{sun2021higher} and 
\bea
\left.\frac{\partial f_m}{\partial p_H^{(t)}}\right|_{p_H=p_H^c}=2c, \quad \left. \frac{\partial g_p}{\partial R^{(t)}}\right|_{R = 0} = p(1-\rho_0)c^+.
\eea
Thus the tricritical point is reached when (see Fig. \ref{fig:hier_dependency} (c))
\bea
\rho_0 = \frac{2c^+}{2c^++m-1}, \quad p = \frac{2c^++m-1}{2cc^+(m-1)}.
\eea

We observe that $\rho_0>0$ for any $c^+$ and $m$. Therefore in the case of $HOTP$ in the absence of negative regulations, $\rho_0=0$, the transition will always be discontinuous.

\subsubsection{HHOTP in the absence of positive regulations}
When only negative regulations are present,  the probability of retaining a hyperedge is given by
\bea
p_H^{(t)} = p \left[\rho_0 + (1-\rho_0) \left(G_0^-\left(1-R^{(t)}\right)\right)\right]
\label{eq:gp_dependency_neg}
\eea
Similarly, we observe that when $R=0$, we have $p_H=p$. Thus the continuous transition is observed when $p=p_H^c$ together with $|J|<1$, where $p_H^c$ denotes the critical threshold of hyperedge percolation. The period-2 bifurcation happens when the non-trivial fixed point loses its stability when $p>p_H^c$ and $|J|=1$. Thus the tricritical point is reached when
\bea
J = -1 ,\quad p_H^c = p
\eea
Assuming the hypergraph has a Poisson structural hyperdegree distribution with an average $c$, a Poisson regulatory degree distribution with an average $c^-$, and a fixed hyperedge cardinality $m$,
\bea
J = -2c(1-\rho_0)pc^- = -1, \quad p = 1/c(m-1).
\eea
Thus the tricritical point is reached when
\bea
\rho_0 = 1-\frac{m-1}{2c^-}, \quad p = \frac{1}{c(m-1)}
\eea
Thus,  when $\rho_0=0$, we have $c^-=(m-1)/2$. Therefore, in the case of HOTP in the absence of positive regulation, as long as $c^-<(m-1)/2$, the transition is always continuous.

\subsubsection{HONDP with partial regulation}
We discuss also HONP when hyperedges are not regulated with probability $\rho_0$ observing a critical behavior similar to the one of HOTP  with partial regulation discussed in the previous paragraphs of this Appendix. The major difference between HONP and HOTP is that the map $R=f_m(p)$ implements hypergraph percolation (case (a)) or cooperative hypergraph percolation (case (b)) with deactivation of the nodes instead of hypergraph percolation with hyperedge deactivation. Let us discuss the case (a) as an example. In this case, we have
\bea
\left.\frac{\partial f_m}{\partial p_N^{(t)}} \right|_{p_N = p_N^c}= \frac{2c(m-1)}{c(m-1)+(m-2)}.
\eea
Thus for HONDP in the absence of negative regulations, the tricritical point is given by
\bea
\rho_0 &=& \frac{2c^+}{2c^+ + c(m-1) + (m-2)}, \nonumber\\ p &=& \frac{c(m-1) + (m-2) + 2c^+}{2c^+ c (m-1)}.
\eea
For HONDP in the absence of positive regulation, we have instead that the tricritical point is given by
\bea
\rho_0 = 1-\frac{c(m-1)+(m-2)}{2c^-}, ~p = \frac{1}{c(m-1)}.
\eea

\end{document}